\documentclass{iopart}
\usepackage{graphicx}
\usepackage{exscale}
\usepackage{iopams}
\eqnobysec

\newcommand{\text}[1]{\ensuremath \mbox{\textrm #1}}
\newcommand{\vect}[1]{\mathbf{#1}}

\begin{document}

\review{The solar magnetic field}{\RPP {\bf 69} (2006) 563--668}%
{doi:10.1088/0034-4885/69/3/R02}

\author{Sami K. Solanki%
, Bernd Inhester, and Manfred Sch\"ussler}

\address{Max-Planck-Institut f\"ur Sonnensystemforschung,
37191 Katlenburg-Lindau, Germany}

\ead{solanki@mps.mpg.de}

\begin{abstract}
The magnetic field of the Sun is the underlying cause of the many
diverse phenomena combined under the heading of solar
activity. Here we describe the magnetic field as it threads its way from
the bottom of the convection zone, where it is built up by the solar
dynamo, to the solar surface, where it manifests itself in the form of
sunspots and faculae, and beyond into the outer solar atmosphere and,
finally, into the heliosphere. On the way it, transports energy from the
surface and the subsurface layers into the solar corona, where it heats
the gas and accelerates the solar wind.
\end{abstract}





\section{Introduction}
\label{sec_intro}

\begin{quote}
{\em After 5 billion years, the Sun is still popping and boiling, unable to
settle down into the decadent middle age that simple theoretical
considerations would suggest. (...) It appears that the radical element
for the continuing thread of cosmic unrest is the magnetic field.}

(E.N.\ Parker, Cosmical Magnetic Fields, 1979)
\end{quote}

What Eugene Parker \cite{Parker:1979d} wrote for the Universe is
certainly true for the Sun. In the absence of a magnetic field, the Sun
would be a considerably simpler object: it would show only minor deviations
from spherical symmetry due to differential rotation and meridional
circulation, while convection and the oscillations and waves
that it excites would provide some background noise. Although these
processes themselves provide a rich field of study, they pale in
comparison with the overwhelming variety of phenomena and processes
existing and acting on the real Sun. Without a magnetic field, phenomena
as diverse as sunspots and coronal loops, faculae and solar
flares, the solar wind and prominences, the solar cycle and irradiance
variability, to name but a few, would be unknown to us.

It is therefore not surprising that the magnetic field holds a central
position within solar research, although this is sometimes masked by the
fact that the indirect manifestations of the Sun's field (e.g.\ the
relatively cool gas within sunspots or the very hot gas in the corona)
are much easier to detect than the magnetic field itself. Direct in-situ
measurements of the Sun's magnetic field are restricted to locations
accessible to spacecraft, i.e.\ up to now to distances greater than
$60\, {\rm R}_\odot$ (sampled by the Helios space probes
\cite{Schwenn:Marsch:1990,Schwenn:Marsch:1991}).  Only a small fraction
of the Sun's total magnetic flux reaches out to this distance. We
therefore rely on remote sensing in order to detect more than just the
tip of the proverbial iceberg. The magnetic field near the solar surface
is usually measured using the Zeeman effect, which is best observed in
spectral lines formed in the solar photosphere. Although observations
based on cyclotron resonance, the Hanle effect and Faraday rotation, as
well as the recording of the Zeeman effect in non-photospheric spectral
lines are becoming increasingly important, the vast majority of all
recordings of the magnetic field still refer to measurements of the
Zeeman effect in the photosphere.

Hence the photospheric magnetic configuration of, e.g., sunspots is
known with considerable accuracy, while our knowledge of the strength
and structure of the magnetic field in the corona owes as much to the
extrapolation (using potential or force-free fields) from photospheric
measurements and the use of proxies (e.g., EUV images of coronal loops)
as to direct measurement. This is extremely unfortunate since the
magnetic field plays a comparatively minor role in the photosphere, but
completely dominates proceedings in the corona. The absence of sensitive
and high-resolution coronal magnetic field measurements may be the
largest single factor blocking progress in coronal physics and
frustrating our attempts to answer questions related to coronal heating,
the triggering of flares and coronal mass ejections, as well as the
acceleration of the fast and slow solar wind.

Our conception of the Sun's magnetic field is a prototype for the
magnetism of other cool stars. Many of these very common stars show
signs of magnetic activity, such as X-ray and EUV emission typical of
stellar coronae, or the brightness modulations due to the passage of
starspots across the stellar disc \cite{Schrijver:Zwaan:2000}.  The sum
of all cool stars in the solar neighborhood can supply information on
the generation and manifestation of stellar magnetic fields which cannot
be obtained from the Sun alone, so that their investigation indirectly
provides an enrichment of our knowledge of the Sun. Examples are the
dependence of activity level, of the surface magnetic flux and activity
cycle parameters or of starspot latitudes on stellar rotation rate, mass
and evolutionary state. The investigation of stellar magnetism is,
however, strongly hampered by the inability to resolve stellar surfaces
at the spatial scales at which many of the physical processes take
place, not withstanding the success of techniques such as (Zeeman)
Doppler imaging
\cite{Strassmeier:etal:2002,Berdyugina:2004,Donati:2001}. For example,
it is estimated that Doppler imaging resolves only a small fraction of
the starspots on active stars \cite{Solanki:Unruh:2004}. Only on the Sun
do we (nearly) achieve the necessary spatial resolution to get at the
heart of the physics. Thus it is not surprising that both physical
mechanisms proposed to explain the presence of high-latitude spots on
rapid rotators
\cite{Schuessler:Solanki:1992,Schrijver:Title:2001} are extrapolations
to the relevant stellar parameters of mechanisms known to act on the Sun.
Magnetically driven processes occur also in systems such as accretion
disks around evolved stars or supermassive black holes at the centres
of galaxies and other astrophysical systems. Once more, the Sun provides
access to similar processes at the fine spatial scales at which many of
them occur.

In addition to being a normal star the Sun is also the central star of
the solar system and the main source of energy for the Earth.  The
magnetically variable Sun can influence the Earth in multiple
ways. Changes in the radiative output of the Sun affect the Earth's
energy balance and influence stratospheric chemistry. The cyclic
variations of the Sun's open magnetic flux modulates the cosmic ray flux
reaching the Earth.  These and other solar variables have been proposed
as drivers of global climate change. The evidence for such an influence
is increasing \cite{Bond:etal:2001,Haigh:1999}. In addition, coronal
mass ejections, large eruptions that fling coronal gas into
interplanetary space, inject particles and energy into the
magnetosphere, ionosphere and upper atmosphere. There they produce
aurorae, substorms and other phenomena combined under the heading of
space weather. Since all the relevant solar phenomena (e.g. irradiance
variations and coronal mass ejections) are magnetically driven, a better
understanding of Sun-Earth relations also requires a good knowledge of
the Sun's magnetic field and its evolution on timescales from minutes to
millenia.

Finally, the Sun's magnetic field and in particular its interaction with
solar convection and rotation also presents a variety of problems of
basic physical interest (e.g., self-excited dynamos, magnetoconvection,
interaction of radiation with magnetized gas, magnetic
reconnection). Indeed, in a diagram of the location of laboratory and
natural plasmas in the $n$-$T$ plane (where $n$ is the particle density
and $T$ the temperature) the plasma in the solar core, corona and
heliosphere occupies parameter regimes that other plasmas in the
laboratory or elsewhere in the solar system do not.  Thus, the Sun is a
unique laboratory of magnetized plasmas.

This review is aimed both at the young solar researcher entering the
field, as well as physicists working in other, possibly related
fields. We have reduced mathematical descriptions to a minimum and have
not aimed at completeness in the references. The review is structured as
follows. We start by giving a very brief overview of the Sun's magnetic
field and introducing a few relevant parameters
(Sect.~\ref{sec_basic}). This is followed by a description of the large
scale structure of the field and the solar cycle
(Sect.~\ref{sec_largescale}).  After that, the physics of the solar
magnetic field is described step by step. We start in the solar interior
(Sect.~\ref{sec_convzone}), discuss the dynamo (Sect.~\ref{sec_dynamo}),
and then move upward to the solar photosphere
(Sects.~\ref{sec_sunspots} and \ref{sec_small-scale}), for which the
largest amount of empirical data is available. Proceeding further
outward, we discuss the magnetic structure of the solar corona
(Sect.~\ref{sec_corona}) and end in the heliosphere
(Sect.~\ref{sec_heliospheric}).


\section{Overview and basic considerations}
\label{sec_basic}

A dynamo mechanism operating in the lower part of the solar convection
zone is generally considered to be the source of the Sun's magnetic
field.  Current models (see Sec.~\ref{sec_dynamo}) place the dynamo at
the interface between the convection zone and the radiative core, a
layer marked by convective overshooting and a strong radial shear in the
Sun's differential rotation. Strands of strong azimuthal (toroidal)
magnetic field generated by the dynamo break out from the overshoot
layer owing to magnetic buoyancy and related instabilities, rise through
the convection zone and emerge at the visible solar surface in the form
of bipolar magnetic regions, finally forming loop structures whose tops
usually lie in the corona. The locations where the legs of a loop
intersect the solar surface are visible as either sunspots or faculae.
Higher in the atmosphere, in the chromosphere, the latter appear as
bright plages.  The upper layers of a loop radiate at EUV and X-ray
wavelengths.



The structure of the field and its properties undergo a remarkable
transition with height. In the photosphere the field is highly
filamented. Most of the magnetic energy resides in magnetic flux tubes,
concentrations of magnetic field that can be roughly described as
bundles of nearly parallel field lines with a relatively sharp boundary,
although these cover less than 5\% of the solar surface. The flux tubes
visible at the surface range from the very small and bright (magnetic
elements) to the very large and dark (sunspots).  The strength of the
flux tube fields at the solar surface is remarkably homogeneous,
however, being around 1--2 kG\footnote[1]{%
  \parbox[t]{0.95\textwidth}{Throughout this review, we write all equations
  according to the SI unit
 system. However, when giving values of the
  magnetic field strength and
 the magnetic flux, we mostly follow the
  practice in the astrophysical
 literature and use cgs units. The
  relationship between both systems of
 units is: $1\,{\rm Tesla} =
  10^4\,{\rm Gauss}$ (magnetic flux density)
 and $1\,{\rm Weber} =
  10^8\,{\rm Maxwell}$ (magnetic flux).}}
when averaged over their cross sections.  Owing to the pressure exerted
by the magnetic field, radial force balance across the tube boundary
implies that the flux tubes are strongly evacuated. This evacuation
leads to considerable buoyancy, which ensures that, on average, the
magnetic field remains nearly vertical.

Because of the evacuation, the plasma $\beta$ $(= 2\mu_0 p/B^2)$ at the
solar surface is in the range 0.2--0.4, if the gas pressure ${p}$ is
taken within a flux tube.  Thus, locally the magnetic energy density can
exceed the thermal energy density, although, because of the small area
coverage by strong fields, globally most of the energy is in the form of
the thermal energy of the gas. Below the solar surface the plasma
$\beta$ increases rapidly with depth, reaching values estimated to be
10$^5$ or higher near the bottom of the convection zone.

Another relevant ratio is that of magnetic to kinetic energy
density, $B^2/\mu_0 \rho v^2$, which indicates whether the magnetic field
locally dominates the bulk motions or vice versa. Taking values for
velocity, $v$, and density, $\rho$, typical of the surface convection
(granulation) and $B$ typical of the flux tubes, one obtains a value of
the order of 10 (while it can become of order unity for the supersonic
peaks of the granular flow). Globally, the energy in the granulation is
of the same order of magnitude as the energy in the magnetic field. In
the convection zone, the value of $B^2/\mu_0\rho v^2$ is more uncertain,
but there are indications for values significantly above unity in
intense flux tubes located in the lower convection zone.

The situation changes rapidly as one moves higher into the atmosphere.
Initially, both the gas pressure and the field strength decrease
exponentially with height, so that $\beta$ remains approximately
constant, but above a certain height, which typically lies in the
chromosphere, the expanding magnetic flux tubes fill the whole
volume. Beyond this height, the field strength drops more sedately,
although still with a large negative power of $r$ (the radial distance
from the solar center). It drops much more rapidly than a locally
monopolar (${\propto r^{-2}}$) or dipolar (${\propto r^{-3}}$)
configuration. This comes from the fact that, in addition to the
increasing volume available to be filled, the field lines loop back to
the solar surface. Hence, with increasing $r$ one moves beyond the tops
of an increasing number of loops. Alternatively, the magnetic field in
the corona, if described as a linear combination of multipoles, requires
that very high orders are included, which drop with a correspondingly
high power of $r$. Nonetheless, the gas pressure and density drop even
more rapidly with height than the field (exponential vs.\ power law
decrease). Consequently, we have $\beta\ll 1$ and $B^2/\mu_0\rho v^2
\gg 1$ in the corona, so that the energetics and dynamics are dominated
by the magnetic field.

As a consequence of these conditions, the magnetic Lorentz force cannot
be balanced by any other force, so that the coronal magnetic field has
to arrange itself into a force-free configuration, which is rather
homogeneous in strength (compared to the photosphere or convection
zone), but not so in direction.  Owing to the presence of loops reaching
up to different altitudes, all values of the field inclination can be
found at a given height in the corona, so that the situation is exactly
the opposite to the photosphere, where the field is inhomogeneous in
strength, but mainly vertically oriented.

At $r\gtrsim$ 2--3 $R_\odot$, most of the remaining field lines are
`open', i.e., they reach out into the heliosphere. Only a small fraction
(a few percent) of the total flux that emerges from the solar surface is
in the form of such open field lines.  Finally, even further out, in the
regime of the solar wind the relative importance of the various energy
densities changes again, with ${\rho v^2/2} > B^2/2\mu_0>P$ beyond the
point where the kinetic energy density of the solar wind equals the
magnetic energy density (the Alfv\'en radius). This point is located
10--20 $R_\odot$ from the solar surface, for the fast and slow wind,
respectively.  The kinetic energy of the solar wind dominates and forces
the magnetic field lines to follow the radially directed wind. In
combination with solar rotation this results in a spiral pattern of the
magnetic field.


\section{Large-scale structure and solar cycle}
\label{sec_largescale}

While the solar magnetic field is strongly structured down to scales
at the limit of observational resolution, it shows, at the same time,
a remarkable degree of large-scale spatio-temporal order and
organization. Properties of the bipolar magnetic regions on the solar
surface (like their mean emergence latitude and the spatial
orientation of their polarities) and the direction of the global
dipole field vary systematically in the course of the solar activity
cycle. Magnetic flux is organized in network structures defined by
convection patterns and becomes globally redistributed by large-scale
flows.

\subsection{Flux distribution on the solar surface}
\label{subsec_distribution}

\begin{figure}
\centering
\resizebox{1.0\hsize}{!}
{\includegraphics[width=0.49\hsize]{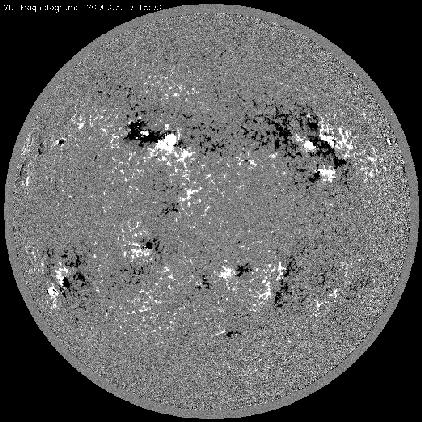}
 \hglue 3mm
 \includegraphics[width=0.49\hsize]{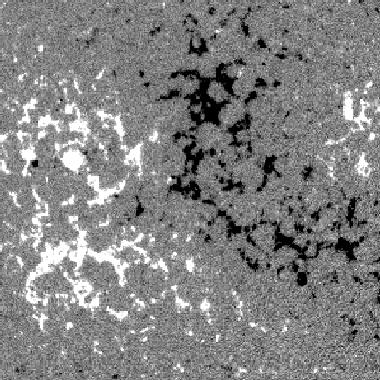}}
\caption{\label{fig:magnetograms}
 Distribution of magnetic flux in the visible solar surface
layers. The circular polarization of a spectral line due to the Zeeman
effect serves as a quantitative measure of magnetic flux. Black and
white indicate positive and negative magnetic polarity, respectively,
while grey signifies low magnetic flux levels. {\sl Left}: Magnetic
image of the visible hemisphere on July 31, 2000, near the maximum of
the current activity cycle. The map is dominated by large bipolar
magnetic regions and extended unipolar domains. The Sun rotates from
left to right.  {\sl Right}: Close-up of a decaying bipolar region of
about $400\times 400\,$Mm$^2$ size, showing the network structure
outlining the convective flow pattern of supergranulation (Images taken
with the {\sl Michelson Doppler Imager (MDI)} on board the {\sl Solar
and Heliospheric Observer (SOHO)}, a joint ESA/NASA mission).}
\end{figure}

Maps of the magnetic field on the visible solar surface (the thin layer
where the solar plasma becomes transparent for light in the visible
wavelength range) displays structures on a wide range of scales. During
periods of high solar activity, large bipolar magnetic regions indicate
the locations of recent magnetic flux emergence from the deep solar
interior (see Figure~\ref{fig:magnetograms}, left panel). Large areas of
apparently unipolar field result from the decay and spread of the
bipolar regions. During most of the time, such unipolar regions can be
found around the solar poles; they define the global dipole component of
the solar magnetic field, which is roughly aligned with the rotation
axis during most of the time.

\subsubsection{Bipolar regions.}
\label{subsubsec_bipolar}

Magnetic flux appears at the solar surface in the form of bipolar
magnetic regions with a wide range of values for the (unsigned) magnetic
flux and lifetimes \cite{Harvey:Zwaan:1993,Hagenaar:etal:2003}. While
the largest {\em active regions} reach fluxes of nearly $10^{23}\,$Mx
($10^{15}\,$Wb) and lifetimes of months, the smallest {\em ephemeral
regions} contain less than $10^{19}\,$Mx and live less than a day before
their magnetic flux cancels or merges with the pre-existing background
flux.  Large bipolar regions form conspicuous sunspot groups while the
smaller regions can only be detected through magnetic field
measurements.

\begin{figure}
\centering
\includegraphics[width=0.6\hsize]{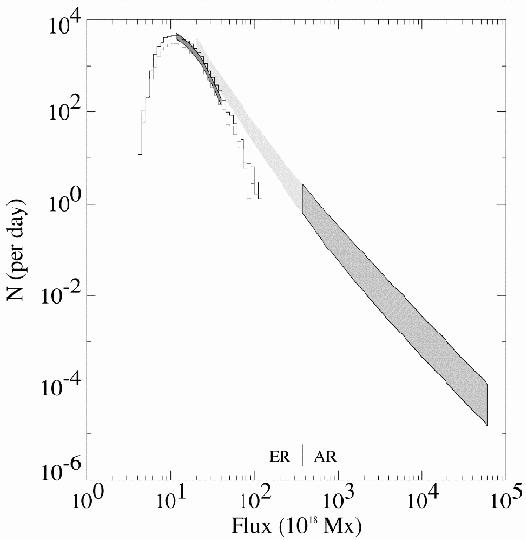}
\caption{\label{fig:bipolar}
 Emergence rate of bipolar regions per (unsigned) flux interval
of $10^{18}$~Mx. The dark grey band shows the distribution for larger
regions (AR: active regions \cite{Harvey:Zwaan:1993}); the variation
by about a factor of 8 through a typical activity cycle is indicated
by the width of the band. The variation for the smallest ephemeral
regions (ER) is much smaller and possibly in antiphase with the
sunspot cycle (dark shading); full histograms are shown for 1997
October (black, solar minimum) and 2000 August (gray, solar
maximum). The turnover below $10^{18}$~Mx reflects the detection
threshold of the instrument. The lightly shaded area between the
smallest ephemeral regions and the active regions is conjectural (from
\cite{Hagenaar:etal:2003}).}
\end{figure}

Figure~\ref{fig:bipolar} shows that the emergence rate of bipolar regions
as a function of flux (or size) roughly follows a power law, which
possibly extends down to the ephemeral regions, indicating a common
origin of all emerging flux. On the other hand, the variation of the
emergence rate in the course of the solar cycle is different: large
active regions vary by about a factor 8, while ephemeral regions hardly
vary at all and may actually show a slight variation in antiphase
\cite{Hagenaar:etal:2003}.

Bipolar active regions exhibit a number of systematic properties,
providing information on the magnetic structure of the source
region near the bottom of the convection zone from where the surface
flux emerges (see Sect.~\ref{subsubsec_rules}). Active regions (a) are
generally orientated roughly in the East-West direction, (b) are
systematically tilted in latitude, (c) appear only in low and middle
latitudes (below 45 degrees), (d) show different proper motions of
the two opposite-polarity patches with respect to the surrounding
plasma. During and shortly after their emergence at the surface, these
properties are displayed by the bipolar regions in a statistical
sense; smaller regions show a larger scatter
\cite{Hagenaar:etal:2003,Schrijver:Zwaan:2000}.

\subsubsection{Magnetic network.}
\label{subsubsec_network}

The relative importance of induction versus diffusion effects in the
temporal evolution of the magnetic field in a plasma with bulk flow is
described by the {\em magnetic Reynolds number}, $R_{\rm m} =
U\,L/\eta$, where $U$ is a typical flow velocity, $L$ the length scale
of the flow, and $\eta$ is the magnetic diffusivity (inversely
proportional to the electrical conductivity). For all relevant flow
patterns in the Sun we have $R_{\rm m} \gg 1$, so that the magnetic
field evolution is governed by induction, with the magnetic field lines
being fixed to the fluid elements ({\em Alfv\'en's theorem}, `flux
freezing' \cite{Choudhuri:1998}). As a consequence, the magnetic flux
distribution at the solar surface reflects the dominant patterns of
convection: magnetic flux is transported by the horizontal flows to the
convective down drafts and forms a network pattern on both the scales of
granulation (the dominant scale of convective energy transport with
$L\simeq 1-2\,$Mm, see Sect.~\ref{sec_small-scale}) and supergranulation
($L\simeq 30\,$Mm, see Figure~\ref{fig:magnetograms}, right panel).

\subsubsection{Magnetic flux transport.}
\label{subsubsec_transport}

The dominance of induction effects has the consequence that the
evolution of the magnetic flux through the solar surface after its
emergence is largely governed by the large-scale velocity
fields. These are (a) differential rotation (the equator rotating
about 30\% faster than the polar regions), (b) a meridional flow of
about 10--25 m$\,$s$^{-1}$ from the equator towards the poles, and
(c) supergranulation. At large length scales, the effect of the latter
can be described by a random-walk process with an effective
(`turbulent') magnetic diffusivity of the order of
$5\,10^{8}\,$m$^2\,$s$^{-1}$ \cite{Leighton:1964}.  The main
consequences of this transport of magnetic flux on the solar surface
are: (a) the spreading of bipolar regions in time, (b) the development
of quasi-rigidly rotating magnetic patterns, and (c) the transport of
magnetic flux to the poles. In connection with the tilt of the bipolar
region with respect to the East-West direction (see
Sect.~\ref{subsubsec_rules}) the latter process leads to reversals of
the polar fields: the westward polarity patch of a bipolar region is
at somewhat lower latitudes and thus preferentially cancels with its
opposite-polarity counterparts on the other hemisphere, so that a net
flux of eastward polarity is transported to the poles by the
meridional flow.

The observed evolution of the large-scale solar magnetic field on a
time scale of months to years, including the polarity reversals of the
polar fields, is remarkably well reproduced by two-dimensional
flux-transport models
\cite{Sheeley:1992,Wang:Sheeley:1994,Wang:1998,Baumann:etal:2004},
which describe the passive transport of a purely vertical (radial)
magnetic field by differential rotation, meridional flow, and
turbulent diffusion. The input required for this kind of models is the
temporal and spatial distribution of magnetic flux emergence in
bipolar magnetic regions. The results indicate that the evolution of
the observable magnetic flux after emergence is largely a surface
phenomenon, so that inferences about the generation of magnetic flux
and the dynamo in the solar interior can only be drawn from the
properties of bipolar regions in their early phases.

\subsection{The 11(22)-year cycle} 
\label{subsec_cycle}

The magnetic activity of the Sun, as most apparent by the coming and
going of dark sunspots (strongly magnetized regions on the visible
surface, see Sect.~\ref{sec_sunspots}) and the eruptive phenomena in the
upper atmosphere and corona of the Sun, shows a cyclic (but not strictly
periodic) variation with a mean period of about 11 years. There is a
remarkable degree of regularity in the large-scale properties of this
{\em solar cycle}, which underlies the seemingly random appearance and
evolution of the magnetic features in the turbulent surface layers of
the Sun. The systematic reversals of the polar magnetic fields and the
polarity orientation of sunspot groups show that a complete magnetic cycle
covers two 11-year activity cycles.

\subsubsection{Sunspot numbers.}
\label{subsubsec_ssn}

\begin{figure}
\centering
\includegraphics[width=0.9\hsize]{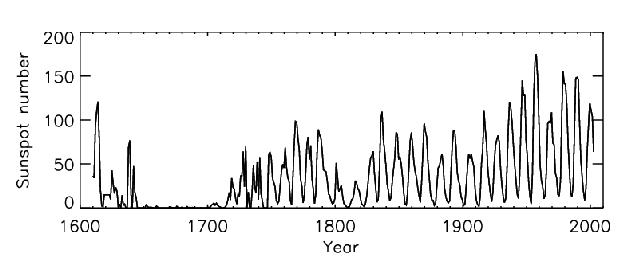}
\vskip -5mm
\caption{\label{fig:gsn}
The record of yearly averaged group sunspot numbers
\cite{Hoyt:Schatten:1998}.  Besides the dominant 11-year cycle, there is
a long-term modulation of sunspot activity with extended periods of
almost vanishing sunspots (1640-1700, Maunder minimum) or rather few
sunspots (1800-1820, Dalton minimum).}
\end{figure}

The longest systematic direct record of solar activity is the series
of sunspot numbers, which starts soon after the invention of the
telescope in the beginning of the 17th century. By compilation of the
records of many observers it was possible to derive a series of
sunspot numbers which reaches back to 1611
\cite{Hoyt:Schatten:1998}.

\begin{figure}
\centering
\includegraphics[width=\hsize,angle=0]{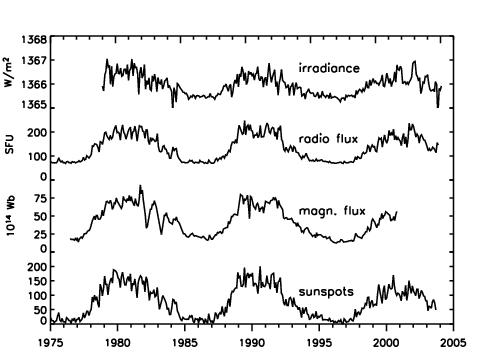}
\vskip -0mm
\caption{\label{fig:flux_comb}
 Various quantities varying in phase with the 11-year cycle of
sunspots (lowest curve: monthly sunspot number provided by the Royal
Observatory of Belgium, Brussels): total solar irradiance at the top of
the Earth's atmosphere in W/m$^2$ (source:
\cite{Frohlich:Lean:1998,Froehlich:2004}), 10.7 cm (2800 MHz) solar
radio flux in `solar flux units' (SFU) of $10^{-22}$ J$\,{\rm s}^{-1}
\,{\rm m}^{-2}\,{\rm Hz}^{-1}$ (source: Radio Astrophysical Observatory,
Penticton, Canada) and the total (unsigned) magnetic flux in the solar
photosphere (source: Wilcox Solar Observatory, Stanford, USA
\cite{Arge:etal:2002}).  }
\end{figure}

Figure~\ref{fig:gsn} shows the existing record of (yearly averaged) sunspot
numbers. While the 11-year cycle is obvious, there is also a modulation
of the sunspot activity on longer time scales. Most remarkable is the
second half of the 17th century (the Maunder minimum \cite{Eddy:1976}),
when sunspots were almost completely absent. We now know that this is just
one case of the occasional `grand minima' of solar activity (see
Sect.~\ref{subsec_longterm}); a less marked example is the Dalton
minimum at the beginning of the 19th century.

Although the definition of the sunspot number appears to be somewhat
arbitrary, it is in fact very well correlated with more objective
measures like the total solar radio flux at 10.7~cm wavelength or the
total (unsigned) surface magnetic flux and thus represents a
quantitatively valid measure of solar magnetic activity
\cite{Tapping:2000} (see Figure~\ref{fig:flux_comb}).

\subsubsection{Sunspot latitudes, polarity rules, and tilt angle of 
bipolar regions.}
\label{subsubsec_rules}

\begin{figure}
\centering
\includegraphics[width=\hsize]{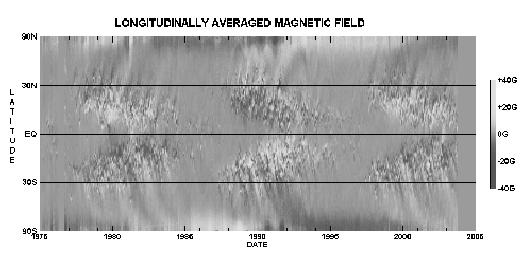}
\\[-5mm]
\caption{\label{fig:herringbone}
Time-latitude diagram of the longitudinally averaged magnetic
field in the solar photosphere for the last three activity cycles. The
emergence of magnetic flux in active regions leads to characteristic
structures in the lower latitudes, which have been first described in
similar `butterfly diagrams' based on sunspot observations
\cite{Maunder:1913}. The combined effects of convection, and meridional
circulation lead to the magnetic flux transport to high latitudes and
thus cause reversals of the polar magnetic fields in phase with the
activity cycle (courtesy D. Hathaway, NASA).}
\end{figure}

The average latitudes of newly appearing sunspots show a systematic
variation in the course of the 11-year activity cycle. Sunspots emerge
in two broad latitude belts, which are roughly symmetric with respect to
the solar equator. At the beginning of a cycle, these belts are centered
around $\pm 30{^\circ}$ latitude. In the course of the cycle, the sunspot
belts migrate towards the equator and reach about $\pm 5{^\circ}$
average latitude towards the end of the cycle. In a latitude-time
diagram of sunspot occurrence, this drift leads to a characteristic
butterfly-shaped pattern, which is also clearly visible in the related
diagram of the surface magnetic field (see Figure~\ref{fig:herringbone}). 

Sunspots typically appear in groups embedded in magnetically bipolar
regions, whose local magnetic polarity they share in most cases. The
bipolar regions are roughly oriented in East-West-direction (the
direction of solar rotation, see Figure~\ref{fig:magnetograms}) and their
polarities are arranged according to the following set of {\em polarity
rules\/} ({\em Hale's law,} first formulated on the basis of magnetic
field measurements in sunspot groups \cite{Hale:Nicholson:1925}):
\begin{enumerate}
\item[1)] The magnetic orientation of bipolar regions remains the same in
each hemisphere during an 11-year activity cycle.
\item[2)] The bipolar regions in the Northern and Southern hemispheres
have opposite magnetic orientation.
\item[3)] The magnetic orientation of the bipolar regions reverses from one
cycle to the next.
\end{enumerate}
As a consequence of these rules, the pattern of magnetic orientations
repeats itself with a period of two activity cycles, i.e., the magnetic
cycle of the Sun has a duration of about 22 years.

Another systematic property of bipolar regions is their deviation from a
precise East-West orientation: on both hemispheres, the more westward
located polarity (leading with respect to the direction of rotation) is
nearer to the equator than the following polarity. On average, the
corresponding tilt angle with respect to the East-West direction,
$\gamma$, is proportional to the mean heliographic latitude, $\lambda$,
of the bipolar region: $\gamma \simeq 0.5 \lambda$ (Joy's law
\cite{Howard:1991}).

Larger bipolar regions obey the polarity rules and Joy's law more
strictly than smaller regions (without sunspots), which are probably
more strongly affected by disturbances and deformation of the underlying
magnetic structure by convective motions.

The systematic properties of large bipolar regions are important
constraints for models of solar magnetic activity and the solar
cycle. Together with the polarity reversals of the global magnetic
field, they indicate a remarkable degree of large-scale order and
self-organization that is not obvious considering the non-stationary
nature of the convective motions, from which the solar magnetic field
must ultimately derive its energy (see Sect.~\ref{sec_dynamo}).

\subsubsection{Effect on solar rotation}
\label{subsubsec_torsional}

Magnetic activity affects the differential rotation of the solar plasma
only very slightly. Helioseismological measurements have not revealed a
systematic change of the shear layer (tachocline) at the bottom of the
convection zone in the course of the 11-year cycle, but there are
indications of a time variation of the rotation near the bottom and
below the solar convection zone with a period of about $1.3\,$years
\cite{Howe:etal:2000a}. It is presently unknown whether this oscillation
is related to the magnetic field.  A clearer relationship exists in
the case of the (somewhat misleadingly named) `torsional
oscillations', a pattern of slightly slower and faster rotating bands
differing from the average rotation by well below 1\%.  The bands
propagate in latitude toward the equator and are related, at low
latitudes, to the migration of the sunspot zones as well as to
variations in the pattern of the meridional flow on the solar surface
\cite{Beck:etal:2002}. The torsional oscillation was first observed by 
Doppler measurements of surface flows \cite{Howard:Labonte:1981} and
later found to penetrate at least a third of the convection zone
depth \cite{Howe:etal:2000b,Vorontsov:etal:2002}. Suggestions
concerning the physical origin of the alternating bands include
driving by the magnetic Lorentz force \cite{Schuessler:1981} and
geostrophic flows driven by surface cooling in active regions
\cite{Spruit:2003}.

\subsubsection{Irradiance variation and other impacts of the solar cycle}
\label{subsubsec_irradiance}

The variation of magnetic activity in the course of the 11-year solar
cycle affects not only `magnetic' quantities like magnetic flux or
sunspot number; the radiative output of the Sun is also modulated in
phase with the solar cycle. This variation is particularly pronounced
for the radiation originating from the hot plasma in the upper layers of
the solar atmosphere (chromosphere to corona). Here, the dominating
magnetic forces and the dissipation of magnetic energy determine the
atmospheric structure, which therefore strongly changes in the course of
the activity cycle (see Sect.~\ref{sec_corona}). As a consequence, the
radiation at UV, EUV and radio wavelengths varies by factors in the
range $2 \dots 10$ between solar activity minimum and maximum while the
soft X-ray flux changes by a factor of order 100
\cite{White:1994}. Moreover, the total (frequency-integrated) solar
irradiance at Earth's orbit as measured with bolometric instruments in
space also changes slightly (of the order of 0.1\%) in phase with the
solar cycle (see Figure~\ref{fig:flux_comb}, \cite{Froehlich:2004}). The
fact that the Sun brightens during activity maximum in spite of the
larger fraction of the surface covered by dark sunspots is due
to the effect of small-scale magnetic fields, which locally enhance the
radiative flux in so-called faculae (see Sect.~\ref{sec_small-scale}).

The variability of the solar irradiance has potential effects on
terrestrial climate, both on time scales of the solar cycle and,
probably even more importantly, on centennial and longer time scales
\cite{Reid:2000}. While changes in the total irradiance concern the
basic energy input into the climate system, the (relatively much
stronger) variability of the UV radiation affects the temperature
structure of the stratosphere through photochemical reactions
involving ozone \cite{Larkin:etal:2000}.  A third possible route by
which solar variability may affect the terrestrial climate is via the
modulation of the galactic cosmic ray flux by the varying heliospheric
magnetic field \cite{Belov:2000} (see Sect.~\ref{sec_heliospheric}). It
has been suggested that cosmic rays trigger cloud formation, so that
their variation could possibly affect the total cloud cover and thus
climate \cite{Marsh:Svensmark:2000}.

\subsection{Long-term modulation of magnetic activity}
\label{subsec_longterm}

\begin{figure}
\centering
\includegraphics[width=\hsize,angle=0]{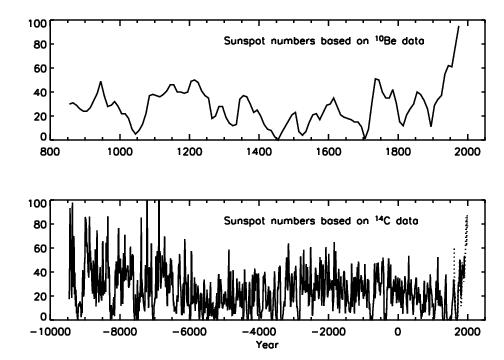}
\caption{\label{fig:SN_recons}
Reconstructions of the (10-year averaged) sunspot number from
the measured concentrations of cosmogenic isotopes. {\em Upper panel:}
Sunspot number since AD 850 based on $^{10}$Be data
\cite{Usoskin:etal:2003, Usoskin:etal:2004a}. {\em Lower panel:}
Reconstruction from 9455 BC until AD 1900 based on $^{14}$C data from
tree rings \cite{Solanki:etal:2004} (full line). After $\simeq 1900$,
the $^{14}$C record is strongly contaminated by the burning of fossil fuels
(Suess effect) and cannot be used for the reconstruction.  The dotted
line gives the 10-year averaged actual group sunspot number from 1611 AD
onward.}
\end{figure}

As already apparent from the record of sunspot numbers shown in
Figure~\ref{fig:gsn}, there is a long-term amplitude modulation of the
11-year sunspot cycle, including periods of low or even almost vanishing
sunspot activity. The detection of further regularities or periodicities
is hampered by the short length of the directly measured sunspot
record. Quantitative information about the solar activity before 1611
can be obtained through proxies like the concentrations of the
`cosmogenic' isotopes $^{10}$Be and $^{14}$C \cite{Beer:2000}. These
isotopes are produced from atmospheric oxygen and nitrogen by spallation
reactions caused by galactic cosmic rays (mainly protons and
$\alpha$-particles). The modulation of the cosmic ray flux (in antiphase
with the solar cycle) by the varying heliospheric magnetic field
\cite{Bazilevskaya:2000} imprints the signature of the solar cycle and
the strength of solar activity upon the production rate of the
cosmogenic isotopes: the higher the solar activity, the lower the
production rate, and vice versa. The isotopes are subsequently removed
from the atmosphere and incorporated in `archives' of past solar
magnetic activity in the form of yearly layers of polar ice shields
($^{10}$Be, by precipitation) or tree rings ($^{14}$C, by plant
metabolism).

By inverting the various components of the chain of processes
connecting the measured isotope concentration with characteristic
measures of solar activity (like the sunspot number), it is possible
to reconstruct the latter for periods greatly exceeding the length of
the directly measured record. As an example, Figure~\ref{fig:SN_recons}
shows the (11-year averaged) reconstructed sunpot number based upon
$^{10}$Be data from Greenland and Antarctica (from AD 850 onward
\cite{Usoskin:etal:2003}) and a reconstruction from $^{14}$C in tree
rings (for the whole holocene from the year 9400 BC onward
\cite{Solanki:etal:2004}).  Both reconstructions demonstrate that
periods of very low activity like the Maunder minimum in the 17th
century are not uncommon. Furthermore, they also show that episodes of
consistently high average activity (like the current period since
about 1940) are much more rare; in fact, similarly long periods of
comparable activity levels can only be found more than 8000 years
before present.

The cosmogenic isotope records permit also the search for other
periodicities than the 11/22-year basic sunspot cycle.  Spectral
analysis of the isotope data in fact reveals significant signals at
periods around 90 years (the Gleissberg cycle, which is also found in the
sunspot record \cite{Gleissberg:1939}), 210 years, and 2200 years
\cite{Beer:2000,Wagner:etal:2001,Peristykh:Damon:2003}.

\subsection{A `fossil' magnetic field in the radiative interior?}
\label{subsec_fossil}

Owing to the skin effect, the oscillating magnetic field of the solar
cycle can penetrate the radiative interior by only a few km. In the
absence of turbulent convection, the skin depth is determined by the
molecular magnetic diffusivity, $\eta=(\mu_0\sigma)^{-1}$, which is in
the range $10^{-2} \dots 1\,{\rm m}^2\,{\rm s}^{-1}$ ($\sigma$ is the
electrical conductivity).

On the other hand, such values of the diffusivity in the interior
together with the large size of the system lead to a very long diffusive
decay time for a large-scale magnetic field, $\tau_d
\simeq R_\odot^2/\eta$, exceeding the age of the Sun.  It is
therefore conceivable that a slowly decaying `fossil' magnetic field
resides in the solar interior. This field could be a remnant from the
magnetization of the interstellar cloud out of which the Sun formed or
it could be a trapped field from a dynamo acting during the very early
evolutionary phase of the Sun when it was fully convective
\cite{Schuessler:1975,Kitchatinov:etal:2001}. It is also conceivable
that the  combination of differential rotation and magnetic
instability of a toroidal magnetic field leads to dynamo generation of
magnetic field in the radiative interior of the Sun \cite{Spruit:2002}.

However, there is no direct observational evidence for such a magnetic
field in the solar interior. In fact, the existence of a sharp
transition between the differentially rotating convection zone and the
almost uniformly rotating radiative interior in form of a narrow radial
shear layer (the solar `tachocline' \cite{Thompson:etal:2003}) indicates
that any fossil magnetic field is closely confined to the radiative part
and thus has no surface manifestations. On the other hand, at least a weak
internal magnetic field seems to be required to maintain both the
uniform rotation of the solar interior
\cite{Mestel:Weiss:1987,Charbonneau:MacGregor:1993a} and the sharpness
of the tachocline
\cite{Ruediger:Kitchatinov:1997,Macgregor:Charbonneau:1999,Garaud:2002}.


\section{Magnetic fields in the convection zone}
\label{sec_convzone}

In a convecting medium with large magnetic Reynolds number, magnetic
flux is expelled from regions of closed streamlines and assembled in
flux concentrations between the convection cells
\cite{Parker:1963,Weiss:1966}. This process of flux expulsion, which can
be directly observed in the flux distribution on the solar surface (see
Sect.~\ref{sec_small-scale}), leads to a strongly intermittent
distribution of magnetic flux. By analogy as well as through qualitative
minimum-energy arguments \cite{Parker:1984} and numerical simulations
\cite{Brandenburg:etal:1995}, such intermittency of the magnetic field
is also suggested to prevail throughout the whole solar convection
zone.  The question is whether intermittent structures generated by
flux expulsion can become sufficiently strong and large to decouple
themselves from the convective velocity field to be governed by their
internal dynamics. Sunspots are an example of such an `autonomous'
structure, but observations show that they are not formed by flux
expulsion by surface flows but emerge as coherent (albeit initially
fragmented) entities from below.  In fact, the sunspot polarity rules
and other systematic features indicate that the magnetic flux
responsible for the formation of sunspots and (large) bipolar regions
is not dominated by non-stationary convective motion but originates from
a source region of largely ordered and azimuthally oriented magnetic
flux in the deep convection zone \cite{Schuessler:1984,
Schuessler:1996b}. These observations are in accordance with the
picture of a `rising tree' \cite{Zwaan:1978, Zwaan:1992}: strands of
magnetic flux detach from the source region, rise through the
convection zone and emerge at the surface in a dynamically active way
to form bipolar magnetic regions and sunspots.  Only after the initial
stage of flux emergence and after fragmenting into small-scale flux
concentrations, does the surface field come progressively under the
influence of convective flow patterns and large-scale surface flows,
so that its large-scale evolution can be well described by flux
transport models (see Sect.~\ref{subsubsec_transport}). Consequently,
it is necessary to consider the dynamics of magnetic structures in the
convection zone in order to make the connection between the observed
properties of the surface fields and the dynamo process generating the
magnetic flux in the first place (see Sect.~\ref{sec_dynamo}).

\subsection{Magnetic flux tubes and flux storage}
\label{subsec_fluxtubes}

The dynamics of magnetic structures can be conveniently described using
the concept of {\it isolated magnetic flux tubes.}  These are defined as
bundles of magnetic field lines (comprising constant magnetic flux),
which are separated from their non-magnetic environment by a tangential
discontinuity (surface current) in ideal MHD or a narrow resistive boundary
layer. As a consequence, the coupling between an isolated flux tube and
its environment is purely hydrodynamic and mediated by pressure forces.

If the diameter of a flux tube is small compared to all other relevant
length scales (scale heights, wavelengths, radius of curvature, etc.)
the {\it thin flux tube approximation\/} can be employed, a quasi-1D
description that greatly simplifies the mathematical treatment
\cite{Spruit:1981,FerrizMas:Schuessler:1993}.  The forces which are most
important for the dynamics of a magnetic flux tube are the buoyancy
force, the magnetic curvature force, the Coriolis force (in a rotating
system), and the aerodynamic drag force (for motion relative to the
surrounding plasma). Very thin flux tubes are effectively coupled to
the motion of the surrounding plasma by the drag force, while larger
(but still `thin' in the above sense) tubes can move relative to the
surrounding plasma owing to the action of the other forces, similar to
a flexible solid body immersed in a fluid.

An upward directed buoyancy force results from the density deficit in
a flux tube arising from the necessity to compensate the magnetic
pressure by a reduced gas pressure \cite{Parker:1955a,Parker:1975}.
Such {\em magnetic buoyancy\/} has important consequences for dynamo
models. It is the obvious mechanism to drive the rise and eruption at
the surface of magnetic flux generated by the dynamo, but it is so
efficient that fields of equipartition strength are lost from the
convection zone in a time much shorter than the 11-year time scale for
field generation and amplification by the dynamo \cite{Parker:1975}.
This problems is actually made even more severe by the unstable
superadiabatic stratification of the convection zone, which leads to
convective buoyancy and even faster flux loss
\cite{Moreno-Insertis:1983}, and by the intrinsic instability of
magnetic flux tubes in an unstably stratified medium
\cite{Spruit:Ballegooijen:1982,Ferriz-Mas:Schuessler:1993,
Ferriz-Mas:Schuessler:1995}.

The buoyancy problem and the resulting `storage problem' could
possibly be somewhat alleviated by `convective pumping'
\cite{Tobias:etal:1998,Dorch:Nordlund:2001,Ziegler:Ruediger:2003}. 
This effect provides a net transport of magnetic flux toward the bottom
of a strongly stratified convecting layer or into an underlying stable
layer \cite{Abbett:etal:2004}. The cause of this process is the strong
asymmetry between up- and downflows in stratified convection: while most
of the rising fluid has to turn horizontal owing to the decreasing
density after ascending for about a scale height, the downflows
accelerate, converge and merge, and can traverse many scale heights. The
magnetic buoyancy is therefore reduced by the `pummeling' from above by
strong downflow plumes. It is not clear, however, whether the numerical
simulation results which prompted this suggestion are applicable to the
realistic solar situation. The effects of limited grid resolution and
numerical diffusion may well lead to an overestimate of the flux
transport by pumping \cite{Schuessler:Rempel:2002}.

In any case, the storage problem for magnetic flux resulting from
magnetic buoyancy in the convection zone can be resolved by assuming
that the azimuthal magnetic flux which, upon emergence at the surface,
leads to large active regions and sunspot groups is generated and
stored in a stably (subadiabatically) stratified layer of overshooting
convection below the convection zone proper
\cite{Galloway:Weiss:1981,Ballegooijen:1982b,Schuessler:1983,Rempel:2004}, 
where it can reach a stable equilibrium configuration
\cite{Moreno-Insertis:etal:1992, Ferriz-Mas:Schuessler:1995}.
This layer overlaps with the region of strong radial differential
rotation (tachocline) and thus also provides the rotational shear
required for building up a strong azimuthal magnetic field.

There are two classes of possible magnetic configurations in the
overshoot layer: a continuous layer of magnetic flux and an ensemble
of isolated magnetic flux tubes.  In the case of a layer of azimuthal
magnetic field in mechanical equilibrium, the magnetic Lorentz force
is balanced by a combination of gas pressure gradient and Coriolis
force due to a field-aligned flow \cite{Rempel:etal:2000}. The
relative importance of both forces for the balance of the magnetic
curvature force depends on the degree of subadiabaticity of the
stratification as measured by the quantity $\delta = \nabla -
\nabla_{\rm ad}$, where $\nabla$ and $\nabla_{\rm ad} = \left(
\hbox{d} \, \ln T / \hbox{d} \ln p \right)_{\rm ad}$ represent the
actual and the adiabatic logarithmic temperature gradient (i.e., the
logarithmic temperature gradient in a homentropic stratification),
respectively.  In a strongly subadiabatic region (like the radiative
core of the Sun with $\delta < - 0.1$), the latitudinal pressure
gradient is dominant for the magnetic equilibrium, while the
contribution of the Coriolis force is small. For $\delta
\simeq -10^{-3}$, both contributions are similar, while for a value of
$\delta \simeq -10^{-6}$, as is probably realistic for a layer of
convective overshoot, the magnetic curvature force is balanced
practically by the Coriolis force alone. For a field strength of the
order of $10^5\,$~G ($10\,$Tesla), an azimuthal velocity of the order
of $100\,\hbox{m}\,\hbox{s}^{-1}$ relative to the background rotation
is required, so that the profile of differential rotation would be
significantly modified.

Another important point is the stability of a magnetic layer. In the
simplest case, the decrease of the field strength at the top of such a
layer drives an instability of Rayleigh-Taylor type, whose nonlinear
evolution leads to the formation of tubular magnetic structures
\cite{Wissink:etal:2000,Fan:2001}. In the absence of diffusion
effects, a magnetic layer in the overshoot region becomes unstable to
flux tube formation when the field strength exceeds a value of the
order of the equipartition field strength with respect to the kinetic
energy density of the convective velocities
\cite{Schuessler:Ferriz-Mas:2003}.  The limiting field strength for
stability may be even smaller if doubly-diffusive instability becomes
relevant in the case of a dominating thermal diffusion due to radiation
\cite{Acheson:1979}. Another road to the fragmentation of a magnetic layer
into flux tubes is the Kelvin-Helmholtz-type instability due to the shear
flow \cite{Cline:etal:2003b}.

As long as they remain within a stably stratified region, the flux
tubes formed by the instability of a magnetic layer can find a new
equilibrium governed by the balance of curvature force and Coriolis
force \cite{Moreno-Insertis:etal:1992}, which is similar to the
equilibrium of a magnetic layer in a slightly subadiabatic region
\cite{Rempel:etal:2000}. To obtain mechanical equilibrium in the
idealized case of an azimuthal flux tube, i.e. a flux ring contained in a
plane parallel to the equator, the buoyancy force must vanish since its
component parallel to the axis of rotation cannot be balanced by any
other force. In the direction perpendicular to the axis of rotation, the
magnetic curvature force is balanced by the Coriolis force due to a
faster rotation of the plasma within the flux ring compared to its
nonmagnetic environment.

\subsection{Flux-tube instability}
\label{subsec_instab}

If the dynamo-generated azimuthal magnetic flux is stored in the
convective overshoot layer in the form of a non-buoyant equilibrium
configuration, a trigger mechanism is required to initiate the rise of
flux tubes towards the surface.  One possibility is radiative heating
\cite{Fan:Fisher:1996, Moreno-Insertis:etal:2002}, which leads to a slow
rise of flux tubes through the overshoot layer until they enter the
convection zone proper, whereupon they rapidly rise to the surface owing
to convective and magnetic buoyancy \cite{Moreno-Insertis:1983}.

\begin{figure}
\centering
\resizebox{0.9\hsize}{!}
{\includegraphics[scale=1.5]{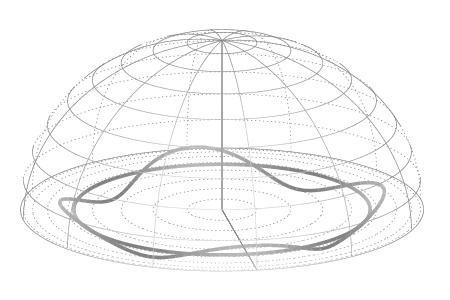}
 \includegraphics[scale=1.2]{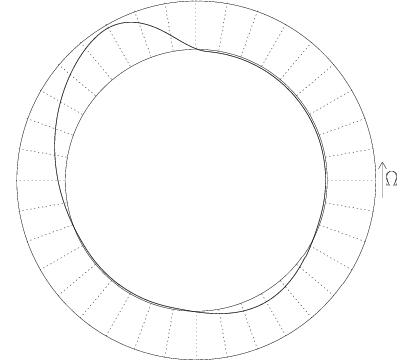}}
\vskip 3mm
\resizebox{\hsize}{!}
{\includegraphics[scale=1.]{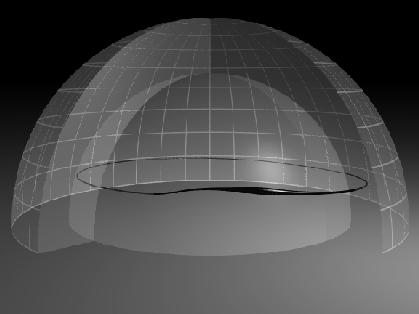}
 \includegraphics[scale=1.]{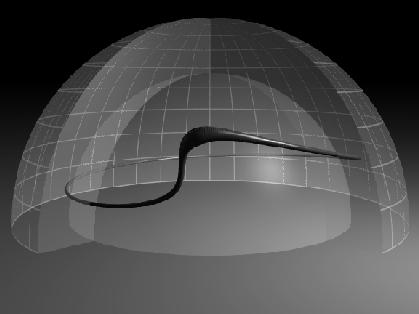}}
\caption{\label{fig:instability}
Instability and rise of magnetic flux tubes. {\em Upper left:}
Sketch of the undulatory instability. An initially axisymmetric
magnetic flux tube (a flux ring) in force equilibrium at the bottom of
the convection zone is perturbed by a displacement with azimuthal
wavenumber $m=4$. A downflow of plasma from the crests into the
troughs lets the summits rise while the valleys sink.  {\em Upper
right:} Polar view of an emerging loop resulting from the nonlinear
development of the undulatory instability (azimuthal wave number
$m=2$) of a flux tube with an initial field strength of about
12~Tesla. Shown is the projection of the tube onto the equatorial
plane. Note the distinct geometric asymmetry between the two legs of
the emerging loop resulting from the Coriolis force. {\em Lower left:}
Three-dimensional view of the rising flux tube shown in the upper
right panel. The two transparent half-spheres correspond to the solar
surface and the bottom of the convection zone, respectively. The
geometry of the rising loop is in accordance with the observed
properties of bipolar regions: low emergence latitude, positive tilt
angle with respect to the azimuthal direction, and geometric asymmetry
between the two legs of the loop.  {\em Lower right:} Emerging flux
loop with initial equipartition field of 1~Tesla carrying the same
magnetic flux as the tube on the left side. The loop emerges at much
too high latitude and the tilt angle even has the wrong sign (from
\cite{Caligari:etal:1995}).}
\end{figure}

As the field strength of a flux tube is intensified by differential
rotation, at some stage it exceeds the threshold for the onset of the
{\it undulatory instability\/} \cite{Spruit:Ballegooijen:1982,
Ferriz-Mas:Schuessler:1993}. This instability in most cases sets in for
non-axisymmetric perturbations of an equilibrium flux ring as sketched
in the left panel of Figure~\ref{fig:instability}. A downflow of plasma
along the field lines within the flux tube leads to an upward buoyancy
force acting on the outward displaced parts and a downward force on the
troughs, so that the perturbation grows. As a consequence, flux loops
form, rise through the convection zone, and finally emerge at the
surface to form sunspot groups and active regions.

A detailed linear stability analysis \cite{Ferriz-Mas:Schuessler:1995}
shows that flux tubes in the solar convective overshoot layer are
strongly stabilized by the combined effects of stratification, rotation,
and magnetic curvature forces. It requires a field strength of the order
of 10~Tesla ($10^5$~G) for the formation of a rising loop that
eventually emerges at the solar surface. This threshold value is
about an order of magnitude larger than the equipartition field
strength with respect to the kinetic energy density of the convective
motions in the deep convection zone.

\subsection{Numerical simulations of rising flux tubes}
\label{subsec_rising}

Simulations of the nonlinear development of the undulatory instability
on the basis of the thin flux tube approximation have revealed the
important role played by the Coriolis force (angular momentum
conservation) for the evolution of a rising flux tube in the solar
convection zone \cite{Fan:2004}:

\begin{enumerate} 
\item[1)] The expanding flux loop experiences a retarding Coriolis force
directed perpendicular to the rotation axis. For sufficiently weak
field, the Coriolis force can balance the corresponding component of
the buoyancy force. The unbalanced axial component of the buoyancy
force then leads to a motion of the flux loop parallel to the axis of
rotation, deflecting its motion to high latitudes. It demands a
sufficiently strong buoyancy force to let the loop emerge in low
latitudes, as observed on the Sun; this requires initial field
strengths of the order of 10~Tesla ($10^5\,$G)
\cite{Choudhuri:Gilman:1987, Choudhuri:1989, Fan:etal:1994,
Schuessler:etal:1994}. The lower part of Figure~\ref{fig:instability}
shows the quite different behavior of emerging flux loops with
initial fields of 12~Tesla (left panel) and 1~Tesla (equipartition
field strength, right panel).
\item[2)] The strong stratification of the convection zone leads to a
draining of mass from the rising part by a flow along the loop. The
Coriolis force on the horizontal component of this flow (corresponding
to an expansion) leads to a twist of the rising loop (clockwise in the
northern and anti-clockwise in the southern hemisphere). Upon
emergence of the flux loop at the surface, this twist manifests itself
as a tilt angle of the leading and the following parts of the
newly-formed bipolar region with respect to the azimuthal direction
(see Sect.~\ref{subsubsec_rules}). A quantitative agreement with the
observed tilt angles requires a strength of the dynamo-generated
magnetic field at the bottom of the convection zone in the range of
3--10~Tesla
\cite{Dsilva:Choudhuri:1993, Caligari:etal:1995}.
\item[3)] Angular momentum conservation retards the azimuthal motion of
the upper parts of a rising loop and leads to a geometric asymmetry:
the preceding part in the direction of rotation is more inclined with
respect to the vertical than the following part
\cite{Moreno-Insertis:etal:1994, Caligari:etal:1998}. The asymmetry can
clearly be seen in the upper right panel of
Figure~\ref{fig:instability}.  The rise of such an asymmetric structure
leads to the characteristic proper motions and geometrical asymmetries
observed in young sunspot groups \cite{Gilman:Howard:1985}.
\end{enumerate}

\noindent 2D and 3D simulations of buoyantly magnetic flux tubes confirm
the basic results based on the thin-tube approximation
\cite{Fisher:etal:2000, Fan:2004}. Moreover, they show that a rising
magnetic flux tube is liable to fragmentation \cite{Schuessler:1979b}
unless it is sufficiently strongly twisted, i.e., has
an azimuthal field component in addition to the axial field
\cite{Emonet:Moreno-Insertis:1998, Fan:etal:1998b, Dorch:etal:1999}.  At
a later stage of the rise, the twist can lead to kink instability, the
signature of which is indicated in X-ray images \cite{Fan:etal:1999}.
The effects of rotation and convection further complicate the picture
\cite{Abbett:etal:2001,Fan:etal:2003}.

\subsection{Origin of super-equipartition fields}
\label{subsec_strong}

The requirement of a field strength of the order of 10~Tesla for the
azimuthal magnetic field in the source region of large bipolar active
regions and sunspot groups is based on a number of reasons: (1) the
criterion for undular instability, (2) the low emergence latitudes of
active regions, (3) the tilt angles of sunspot groups, and (4) the
preservation of the coherence of the flux tube during its rise in view
of the surrounding turbulent convection \cite{Moreno-Insertis:1992}. 

It is unclear, however, how such a strong field is generated in the
Sun. The magnetic energy density of a field of 10~Tesla is two orders of
magnitude larger than the kinetic energy density of the convective
motions in the lower solar convection zone. This largely excludes flux
expulsion by convection as a mechanism. It is conceivable that
convective flows could {\it locally\/} be much stronger (for instance,
in concentrated downflows) and compress the field; however, such local
concentrations would correspond to large azimuthal wave numbers ($m>10$,
say), for which the undulatory instability requires a field strength
well in excess of $10^6$~G (100~Tesla). Since the velocity differences
over the tachocline are of the same order as the convective velocities,
the same argument applies to stretching by differential rotation as a
mechanism for the generation of the strong field: the back-reaction via
the Lorentz force of the growing magnetic field upon the shear flow
limits the field strength to about equipartition values. Moreover, this
would lead to a strong variation of the tachocline flows during the solar
cycle, which is not observed \cite{Eff-Darwich:Korzennik:2003}.

A possible field intensification mechanism that does not rely on
mechanical stress is related to the sudden weakening of the field
strength at the apex of a slowly rising flux loop that remains in
approximate hydrostatic equilibrium along the magnetic field lines
\cite{Moreno-Insertis:etal:1995,Schuessler:Rempel:2002}. 
Since the plasma in the flux loop is almost isentropic while the
surrounding convection zone shows a decreasing entropy with height, the
pressure within the flux tube decreases less rapidly than the external
pressure. At some critical height in the convection zone, both pressures
become equal, so that the magnetic pressure formally has to vanish: the
flux tube `explodes' at its apex and a region of weak field develops
between the two remaining `stumps' of the loop.  After such an
explosion, the high-entropy material within the tube streams out of the
stumps owing to its buoyancy. Numerical simulations indicate that this
outflow continues for a sufficiently long time, so that lateral pressure
balance leads to a significant field amplification in the non-exploded
sections of the loop, including the deep, still stored part of the flux
tube \cite{Rempel:Schuessler:2001, Ossendrijver:2005}. This represents
an amplification process for the magnetic field that does not rely on
the mechanical energy of the convective or rotational motions but
directly utilizes the huge amount of potential energy residing in the
superadiabatic stratification of the convection zone.


\section{The solar dynamo}
\label{sec_dynamo}

The regular reversals of the global magnetic field in the course of
the solar cycle indicate that induction effects by bulk flows of the
electrically conducting solar plasma dominate over diffusion effects
in determining the evolution of the solar magnetic field. In the
radiative interior, radially overturning motions are strongly
suppressed by the very stable stratification and the rotation is
almost rigid, so that it can harbor only a slowly decaying fossil
field (see, however, \cite{Spruit:2002}). The magnetic fields
responsible for the solar cycle therefore most probably originates in
the overlying convection zone, where the convective motions and
differential rotation lead to strong induction effects. The field
reversals in the 11-year time frame, which is extremely short compared
to the diffusion time scales, indicate that these flows have to
provide mechanisms for both the fast generation and for the efficient
removal and dissipation of magnetic flux. While the latter process can
be ascribed to the drastic enhancement of the effective magnetic
diffusivity for a large-scale magnetic field by turbulent flows in the
convection zone (`turbulent diffusivity'), the mechanisms for flux
generation starting from a small initial seed field is the crucial
issue of solar dynamo theory. Various possibilities have been
considered and a number of models have been constructed, but a fully
consistent and predictive theory is still lacking.  In what follows, a
brief sketch of the basic concepts and the present state of research
on the solar dynamo is given. Recent comprehensive reviews can be
found in \cite{Petrovay:2000,Ossendrijver:2003,Charbonneau:2005}.

\subsection{Basic concepts}
\label{subsec_basic}

In the framework of magneto-hydrodynamics (MHD), the dynamo problem can
be stated as the search for solutions of the MHD equations (the
combination of the non-relativistic Maxwell equations, Ohm's law, and
the hydrodynamic equations \cite{Choudhuri:1998}) with non-vanishing
magnetic energy as time goes to infinity. In order to exclude trivial or
unrealistic cases, the flow providing the induction effects is restricted to
a compact volume and has to be regular (finite kinetic energy and
gradients). Moreover, the magnetic field must not be maintained by
sources (currents) at infinity \cite{Moffatt:1978,
Krause:Raedler:1980}. We can define a self-excited dynamo by the
instability of the solution with vanishing magnetic field, a trivial
solution of the MHD equations.  The evolution of a magnetic field, {\bf
B}, under the influence of a velocity field, {\bf u}, and a magnetic
diffusivity, $\eta$, is described by the {\it induction equation}, viz.
\begin{equation} 
   \frac{\partial{\bf B}}{\partial t} = \nabla\times
   \left({\bf u}\times{\bf B}\right) - 
   \nabla\times\left(\eta\nabla\times{\bf B}\right)\,.
\label{eq:induction}
\end{equation}
For dynamo action, the induction term (first term on the right-hand
side) has to overcome (in a global sense) the diffusion term (the second
term). Consequently, a necessary requirement for dynamo action is that
the order-of-magnitude ratio of both terms, the magnetic Reynolds
number, is larger than unity.

Another necessary requirement for a working dynamo results from {\it
Cowling's theorem\/}, which states that a rotationally symmetric
magnetic field (like a dipole field) cannot be maintained by dynamo
action under rather general circumstances
\cite{Cowling:1933,Nunez:1996}. Accordingly, the solar magnetic field is
clearly non-axisymmetric. Another important such `anti-dynamo' theorem
is due to Elsasser \cite{Elsasser:1946} and states that differential
rotation alone (or, more generally, a purely toroidal motion in an
expansion of the velocity field into spherical harmonics) cannot drive
a dynamo in a sphere, provided that the magnetic diffusivity is
constant on spherical surfaces \cite{Ivers:James:1988,Proctor:2004}.
As a consequence of these restrictions, a working dynamo requires a
complex three-dimensional and non-axisymmetric structure of both the
generated magnetic field and the driving velocity field.

\begin{figure}
\centering
\parbox{0.44\hsize}{\includegraphics[width=1.0\hsize]{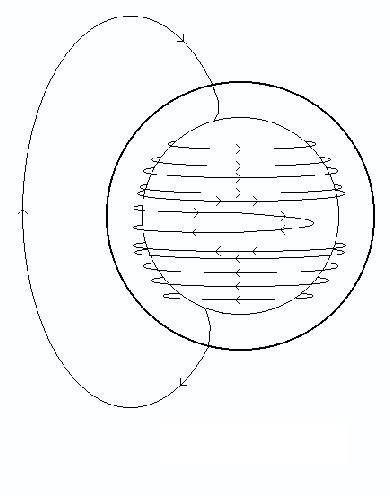}}
\parbox{0.54\hsize}{\includegraphics[width=1.0\hsize]{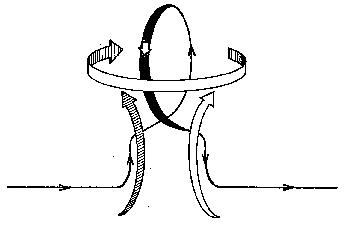}}
\vglue -5mm 
\caption{\label{fig:dyn_processes}
 {\em Left:} Generation of an azimuthal magnetic field through
winding of meridional magnetic field lines by differential
rotation. Sketched here is the case of a latitude-dependent angular
velocity with a faster-rotating equatorial region (courtesy D. Hathaway,
NASA). {\em Right:} Sketch of the Parker loop. An expanding convective
upflow comes into vortical motion owing to the action of the Coriolis
force in a rotating system (clockwise in the Northern hemisphere, where
the rotation vector has an upward directed component, counter-clockwise
in the Southern hemisphere). A magnetic field line frozen into the
plasma is twisted into a loop with magnetic field components
perpendicular to the plane of projection (after \cite{Parker:1955}).}
\end{figure}

In the case of the Sun, differential rotation and convective flows are
considered to be the most important ingredients of self-excited dynamo
action. For high magnetic Reynolds numbers, the magnetic field lines
are fixed to the fluid elements, so that the shearing effect of the
latitudinal and radial gradients of the angular velocity of rotation
leads to the generation and amplification of an azimuthal magnetic
field (aligned with the direction of rotation) from a meridional field
with radial and latitudinal components (see left panel of
Figure~\ref{fig:dyn_processes}).  Elsasser's theorem shows that
differential rotation alone is not sufficient to maintain the solar
magnetic field; a flow regenerating the meridional part of the field
is required in addition. The conceptual basis for most
convection-based models of this process is the {\em Parker loop}
\cite{Parker:1955}, which describes how convective up- and downflows
in a rotating system generate meridional flux loops through twisting
of the field lines due to the action of the Coriolis force on the flow
(see right panel of Figure~\ref{fig:dyn_processes}). Upflows expand and
downflows contract owing to the pressure stratification of the
convection zone. This leads to the same sense of twisting by the
Coriolis force in one hemisphere and, consistently, to the opposite
sense in the other hemisphere. For an azimuthal field of opposite
polarity in both hemispheres as generated from a dipolar field by
differential rotation (cf. left panel of
Figure~\ref{fig:dyn_processes}), the superposition of many such
meridional loops forms a large-scale meridional field of reversed
polarity with respect to the original field from which the azimuthal
field had been wound up in the first place. Therefore, the Parker loop
provides a simple explanation for the polarity reversals of the
large-scale meridional field of the Sun. In connection with the
winding by differential rotation it leads to periodic reversals of
both meridional and azimuthal magnetic field and thus accounts for the
observed basic features of the solar cycle: large-scale field
reversals and the polarity rules of sunspots groups (assuming that the
latter originate from the azimuthal field component, see
Sect.~\ref{sec_convzone}).

The Parker mechanism escapes Cowling's prohibition of a dynamo-generated
axisymmetric field through the intrinsic non-axisymmetry of the loop
formation while the spatially averaged generated field may well have a
dominating axisymmetric component.

\subsection{Mean-field theory}
\label{subsec_meanfield}

The Parker mechanism has been mathematically formalized and greatly
generalized in the theory of `mean-field electrodynamics' first
developed by Krause, R\"adler and Steenbeck \cite{Krause:Raedler:1980},
which describes the evolution of a suitably averaged magnetic field in a
turbulent flow of an electrically conducting fluid. Magnetic field and
velocity are written as sums of their mean (denoted by angular brackets
in what follows) and fluctuating parts (with zero mean, denoted by
primes). Under certain conditions (e.g., a clear separation of scales),
a spatial, temporal or azimuthal average can be considered, but
mathematically more convenient are ensemble averages, which always
satisfy the Reynolds rules (like interchangeability of averaging and
differentiation \cite{Ossendrijver:2003}).

Averaging the induction equation (\ref{eq:induction}), one obtains an
equation for the time evolution of the mean magnetic field, viz.
\begin{equation} 
   \frac{\partial\langle{\bf B}\rangle}{\partial t} = \nabla\times
   \left( \langle{\bf u}\rangle\times\langle{\bf B}\rangle
   + \langle{\bf u}^{\prime}\times{\bf B}^{\prime}\rangle\right)
   - \nabla\times\left(\eta\nabla\times\langle{\bf B}\rangle\right)\,.
\label{eq:mean_induction}
\end{equation}
The term $\langle{\bf u}^{\prime}\times{\bf B}^{\prime}\rangle$
describes the effect on the mean field of the correlations between the
fluctuating quantities. A corresponding equation for ${\bf
B}^{\prime}$ is obtained by subtracting Eq.~(\ref{eq:mean_induction})
from Equation~(\ref{eq:induction}). Under certain conditions (e.g., if the
magnitude of the fluctuations is small compared to the mean field), an
approximate solution of this equation in terms of the mean field can
be found.  In the case of (locally) isotropic and homogeneous
turbulence, a series expansion in terms of the spatial derivatives of
$\langle{\bf B}\rangle$ then leads to
\begin{equation} 
  \langle{\bf u}^{\prime}\times{\bf B}^{\prime}\rangle = 
         \alpha \langle{\bf B}\rangle 
       + \beta \nabla\times \langle{\bf B}\rangle + \dots,
\label{eq:emf}
\end{equation}
where 
\begin{equation} 
   \alpha = -{\tau_c\over 3} \langle{\bf u}\cdot(\nabla\times{\bf u})
   \rangle  \quad {\rm and} \quad 
   \beta = {\tau_c\over 3} \langle{\bf u}\cdot{\bf u}\rangle \;.
 \label{eq:alpha_beta}
\end{equation}
Here, $\tau_c$ denotes the correlation time of the fluctuating part of the
velocity field. Inserting Equation~(\ref{eq:emf}) into
Equation~(\ref{eq:mean_induction}) yields
\begin{equation} 
   \frac{\partial\langle{\bf B}\rangle}{\partial t} = \nabla\times
   \left( \langle{\bf u}\rangle\times\langle{\bf B}\rangle
   + \alpha\langle{\bf B}\rangle\right)
   - \nabla\times\left[(\eta+\beta) 
     \nabla\times\langle{\bf B}\rangle\right]\,.
\label{eq:mean_induction_emf}
\end{equation}
If non-vanishing, the contribution $\alpha \langle{\bf B}\rangle$ to
the mean electric field (the so-called {\em $\alpha$-effect\/}) drives
a mean current parallel to or anti-parallel to the mean magnetic
field. The $\alpha$-effect term therefore generates a meridional field
from an azimuthal field, and vice versa.  The pseudo-scalar $\alpha$
is only non-vanishing for flows lacking mirror symmetry, i.e., flows
possessing a finite average helicity, $\langle{\bf
u}\cdot(\nabla\times{\bf u})\rangle$, resulting from a net correlation
between velocity and vorticity. These conditions are fulfilled in the
case of convection in a rotating stratified medium (`cyclonic
convection'), since the Coriolis force leads to the required
correlation between velocity and vorticity. Therefore, the mean-field
approach yields a formalization of the dynamo mechanism illustrated by
the Parker loop.

Equation~(\ref{eq:mean_induction_emf}) shows that the positive quantity
$\beta$ formally is equivalent to an additional magnetic diffusivity,
the {\em turbulent diffusivity.} It describes the enhanced effective
diffusion of the mean magnetic field due to the (random-walk type)
transport of magnetic field lines by the fluctuating velocity field
and the creation of small-scale structure and enhanced dissipation
through the development of a turbulent cascade. The turbulent
diffusivity in the deep solar convection zone is many orders of
magnitude larger than the molecular diffusivity $\eta \simeq 1
... 10\,{\rm m}^2\,{\rm s}^{-1}$: with typical values of $10\,{\rm
m}\,{\rm s}^{-1}$ for the fluctuating convective velocity and about a
month for the correlation time, we find $\beta\simeq 10^{8}\,{\rm
m}^2\,{\rm s}^{-1}$, so that the molecular magnetic diffusivity can be
neglected for the evolution of the mean magnetic field. In fact, such
values of the turbulent diffusivity lead to typical decay times of the
order of decades, which is the right order of magnitude for the solar
cycle.

Inserting differential rotation for the mean flow, $\langle{\bf
u}\rangle$, in Equation~(\ref{eq:mean_induction_emf}) and using simple
estimates for $\alpha$ and $\beta$ leads to excited dynamo solutions,
i.e., a mean field growing from a small seed field. If the induction
effect of the $\alpha$-term is much smaller than that of differential
rotation (also called the {\em $\Omega$-effect}), the solutions are
oscillatory with periodic reversals of both the meridional and the
azimuthal field components. Moreover, the solutions of such
$\alpha\Omega$-dynamos represent latitudinally propagating waves for
predominantly depth-dependent differential rotation. Whether such dynamo
waves propagate equatorward or poleward depends on the sign of  $\alpha
\partial\Omega / \partial r$, where $\Omega$ is the angular velocity and
$r$ the radial coordinate \cite{Yoshimura:1975, Kitchatinov:2002}.  The
period of the oscillatory solutions is basically determined by the
diffusion time based upon the turbulent diffusivity, which yields the
value for the solar cycle to within an order of magnitude.

The presentation above covers only the most basic aspects of mean
field theory. In general, the coefficients of the expansion given in
Equation~(\ref{eq:emf}) are (pseudo)tensors, so that, depending on the
preferred directions in the system (due to stratification and
rotation), anisotropic turbulent diffusion and transport as well as
anisotropic dynamo coefficients result \cite{Krause:Raedler:1980,
Ruediger:Kitchatinov:1993}. While the symmetric part of the general
$\alpha$-tensor represents the dynamo effect, the antisymmetric part
describes a turbulent transport of the mean field.  The simplest of
such effects is {\sl turbulent diamagnetism\/}
\cite{Zeldovich:1957,Kitchatinov:Ruediger:1992}, the expulsion of
magnetic field from regions of intense turbulence. In fact, numerical
simulations of convection in rotating systems indicate that the
$\alpha$-tensor is highly anisotropic, with the term describing the
generation of meridional field from azimuthal field generally dominating
\cite{Brandenburg:etal:1990,Ossendrijver:etal:2001, Ossendrijver:etal:2002}.

So far we have only considered the linear (kinematic) aspects of the
dynamo process: magnetic field lines are passively carried and twisted
by the convective flows. As the field strengths grows in the course of
the dynamo process, the back-reaction of the magnetic field on the
generating velocity fields through the Lorenz force becomes
important. This limits the field strength that can be reached in the
course of the dynamo process (at least roughly to order of magnitude) to
the `equipartition field' $B_{\rm eq}$, i.e., the field strength for
which the magnetic energy density equals the kinetic energy density of
the generating motions:
\begin{equation}
\frac{B^2_{\rm eq}}{2\mu_0}\; \simeq \; \frac{1}{2} \rho v^2\,,
\end{equation}
where $\rho$ is the density and $v$ the velocity amplitude. For the
lower half of the solar convection zone we find $B_{\rm eq} \simeq
10^4$~G if we use current estimates of the convective velocity.

There are basically two non-linear effects to be considered in
mean-field theory. One is the back-reaction of the magnetic field on
the statistical properties of the velocity field (like the flow
helicity in cyclonic convection) that give rise to dynamo action in
the first place. This leads to a dependence of the $\alpha$-effect on
the magnetic field (also called {\em $\alpha$-quenching}). Simple
models give just an algebraic relationship (like $\alpha\propto
B^{-2}$), but the incorporation of effects with long time scales in
situations characterized by large magnetic Reynolds numbers (like the
evolution of the spectrum of the current helicity, the product
of the magnetic field with the electrical current density
\cite{Field:etal:1999}), leads to a dynamic (differential) equation
for $\alpha$ \cite{Kleeorin:etal:2003}.  Under certain circumstances,
$\alpha$-quenching can become `catastrophic' in the sense that
$\alpha$ becomes inversely proportional to the magnetic Reynolds
number and the magnetic field saturates at very low levels in systems
characterized by large magnetic Reybolds number
\cite{Vainshtein:Cattaneo:1992}.  The second non-linear effect results
from the mean Lorentz force due to the mean field, driving a mean
flow. The induction effect of these flows, in turn, reacts back and
contributes to limiting the growth of the mean field
\cite{Malkus:Proctor:1975, Schuessler:1979b}. This effect includes also
the modification of the differential rotation in $\alpha\Omega$-dynamos
\cite{Phillips:etal:2002, Brooke:etal:2002}.

Mean-field $\alpha\Omega$-dynamo models are capable of reproducing
many of the key features of the solar cycle, namely, the periodic
field reversals, the polarity rules of the sunspot groups, and the
equatorward drift of the activity zones -- given that the parameters
($\alpha$-effect, magnetic diffusivity, and differential rotation) are
suitably chosen. Elementary considerations on the basis of the Parker
loop already show that the (pseudo-scalar) $\alpha$ generated by
convection should be positive in the northern hemisphere and negative
in the southern hemisphere of the Sun. Equatorward propagation of
dynamo waves would then require that the dominant component of
differential rotation should be a radially inward increase of the
angular velocity. This is in clear contrast to the results of
helioseismology, which show almost no radial differential rotation in
the convection zone and an inward {\em decrease} of the angular
velocity in the low-latitude tachocline \cite{Thompson:etal:2003}. On
the other hand, the velocity correlations leading to the
$\alpha$-effect reverse near the bottom of the convection zone, so
that the sign of $\alpha$ reverses also and the correct sense of
dynamo wave propagation ensues in low latitudes.

\subsection{Dynamo simulations and fast dynamos}
\label{subsec_simulations}

Local simulations of compressible magneto-convection describing small
parts of the solar convection zone have been carried out mostly in
cartesian geometry \cite{Nordlund:etal:1992,
Brandenburg:etal:1996}. These simulations show excited dynamo action,
albeit they produce only little `mean' field on the scale of the
computational box.  While the total magnetic energy saturates at a small
fraction of the kinetic energy in the flow, the magnetic field is
locally intensified in flux-tube-like structures (`flux cigars') with
about equipartition field strength.

Self-consistent simulations of the generation of differential rotation
and dynamo action by convection in a spherical shell were already
attempted in the 1980s \cite{Gilman:Miller:1981, Gilman:1983,
Glatzmaier:1985}. Excited dynamo action was found, but the generated
fields indicated poleward dynamo wave propagation in disagreement with
the empirical facts. More recent attempts
(\cite{Brun:2004,Brun:etal:2004}, see also \cite{Miesch:2005}) showed
the generation of an intermittent `turbulent' magnetic field but failed
to reproduce the large-scale ordered fields responsible for the 22-year
solar cycle.

Numerical simulations have also provided a few examples of `fast
dynamos' \cite{Childress:Gilbert:1995}, i.e., dynamos whose growth rate
remains finite in the limit of infinite magnetic Reynolds number,
$R_m$. This is an important issue since the very large value of $R_m$
for the flows in the solar convection zone must not restrict the growth
time of the dynamo to time scales exceeding the solar age. Typically,
fast dynamo action appears for flows with chaotic stream lines,
corresponding to positive Lyapunov exponents, so that the magnetic field
lines are repeatedly stretched, twisted, and folded. In fact, numerical
simulations of Boussinesq convection in a closed box show fast dynamo
action and strong intermittency of the generated magnetic field
\cite{Cattaneo:1999}. At present it is unclear, however, how these
concepts can be applied to the study of dynamo action by solar
convection and rotational shear or which fraction of the small-scale
field at the solar surface could possibly be generated by such local
fast dynamo action. Another topic of considerable current interest is
the dependence of small-scale dynamo action on the value of the magnetic
Prandtl number, $\nu/\eta$, the ratio of kinematic viscosity and
magnetic diffusivity of the fluid, which is a small number in many
astrophysical contexts \cite{Boldyrev:Cattaneo:2004,
Schekochihin:etal:2005, Ponty:etal:2005}.  

\subsection{Magnetic helicity conservation}
\label{subsec_helicity}

A conceptually important quantity in the study of dynamo action in flows
with large magnetic Reynolds number is the {\em magnetic helicity,}
viz. the quantity
\begin{equation}
H = \int_V {\bf A}\cdot {\bf B}\,{\rm d}V\;,
\label{eq:helicity}
\end{equation}
where {\bf A} is the vector potential of the magnetic field and the
volume $V$ contains the magnetic field, so that the normal component of
{\bf B} vanishes at its boundary. $H$ is gauge-invariant under these
conditions and its evolution equation is given by
\begin{equation}
\frac{{\rm d}H}{{\rm d}t} = -2\eta\int_V {\bf j}\cdot {\bf B}\,{\rm d}V\;,
\end{equation}
which shows that the magnetic helicity is a conserved quantity under
ideal conditions ($\eta=0$), which is also true in the limit $\eta\to 0$
\cite{Brandenburg:etal:2002}. Note that the velocity field does not
explicitly appear in the evolution equation for $H$, so that there is
no turbulent diffusion of $H$: the net magnetic helicity in a flow
with large magnetic Reynolds number changes only on the very long time
scale of molecular diffusion.  This represents a relevant constraint
on dynamo models for the Sun, since the 11/22-year period of the solar
cycle is very much shorter than the resistive time scale based upon
(microscopic) diffusion
\cite{Brandenburg:2001,Brandenburg:etal:2003}. For dynamos based upon
helicity and/or shear, the build-up of a large-scale magnetic field is
necessarily connected with the build-up of large-scale magnetic
helicity. This growth can happen in a time significantly smaller than
the resistive time scale only if (approximately) the same amount of
magnetic helicity of the opposite sign resides in a location separated
either spatially (e.g., on the other hemisphere, as for helicity
generation by differential rotation) or in wavenumber (i.e., in small
scales, as in the case of the $\alpha$-effect due to helical flows). In
any case, the Sun has to shed magnetic helicity in the course of the
solar cycle \cite{Brandenburg:Subramanian:2005}, either in the form of
resistive dissipation at very small scales, by cross-hemispheric
transport, or by losses into outer space (possibly in the form of
coronal mass ejections
\cite{Low:2001}).

\subsection{Current models}
\label{subsec_dynmodels}

\begin{figure}
\centering
\includegraphics[width=0.3\hsize]{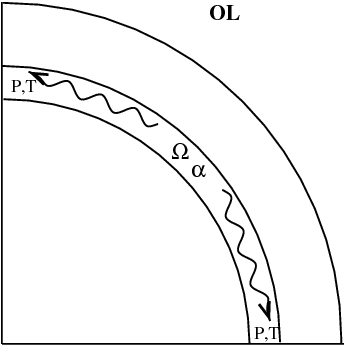}
\includegraphics[width=0.3\hsize]{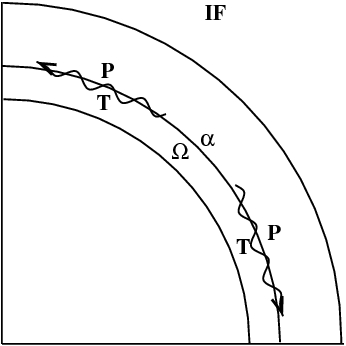}
\includegraphics[width=0.3\hsize]{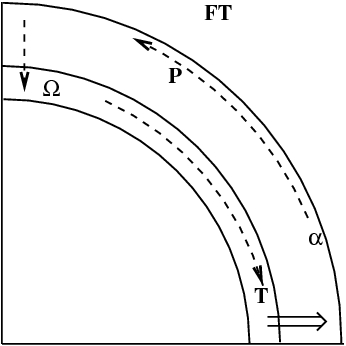}
\caption{\label{fig:dynamo_models}
Schematic illustration of the key features of various classes of dynamo
models. The images represent a quadrant from a meridional cut through
the Sun, indicating the radiative core (inner part), the layer of
convective overshoot and radial shear (inner shell, not to scale), and
the convection zone proper (outer shell). {\em Left panel:} overshoot
layer dynamo (OL). Differential rotation ($\Omega$) and $\alpha$-effect
are both confined to the overshoot layer; dynamo waves (wavy line) of
poloidal (P) and toroidal (T) magnetic field propagate in latitude,
according to the sign combination of the radial shear and the
$\alpha$-effect.  {\em Middle panel:} interface dynamo (IF). The
$\alpha$-effect acts in the convection zone proper while the radial
shear is confined to the overshoot layer (tachocline). Both regions are
connected by (turbulent) diffusion. The dynamo waves take on the
character of a surface wave along the interface between overshoot layer
and convection zone. {\em Right panel:} flux transport dynamo (FT). The
poloidal field generated by an $\alpha$-effect in the convection zone
(for instance, from the tilt induced near the surface by the Coriolis
force on rising flux tubes) is transported by a meridional circulation
(dashed lines) poleward and down to the tachocline, where the rotational
shear generates a toroidal field. This field is transported equatorward
by the return flow of the meridional surface flow and, owing to magnetic
instability, erupts at low latitudes (double-lined arrow) (adapted from
\cite{Petrovay:2000}.)  }
\end{figure}

Conventional models of $\alpha\Omega$-dynamos working in the bulk of
the convection zone have difficulties to reproduce the latitudinal
migration of the sunspot zone since the radial rotational shear is
much smaller than the latitudinal shear in the convection zone and
dynamo waves propagate along isolines of angular velocity
\cite{Yoshimura:1975} (see, however, \cite{Kitchatinov:2002} and a
recent re-appraisal of distributed dynamo action in the convection zone,
\cite{Brandenburg:2005}). 
Moreover, such models are neither capable of producing and storing
strongly super-equipartition azimuthal fields nor can the turbulent
$\alpha$-effect of cyclonic convection act efficiently upon such strong
fields. Various possibilities to overcome these problems have been
suggested. In most of these models, it is assumed that a layer of
overshooting convection overlapping with the tachocline is the location
where a strong azimuthal magnetic field is generated and stored. The
models can be classified into the following three types (for detailed
reviews see \cite{ Ossendrijver:2003,Petrovay:2000, Ruediger:Arlt:2003,
Schuessler:Schmitt:2004}):
\begin{enumerate}
\item[1)] \textit{overshoot layer dynamos:} $\alpha$-effect is restricted
to the overshoot region;
\item[2)] \textit{interface dynamos:} $\Omega$-effect is dominant in the
overshoot region and a conventional  $\alpha$-effect operates in the 
convection zone above, both regions being coupled by magnetic diffusion;
\item[3)] \textit{flux transport dynamos:} the radial transport of
magnetic flux into the overshoot layer and the latitudinal migration of
the magnetic field are dominated by advection by a large-scale meridional
flow.
\end{enumerate}
The basic concepts of these three classes of models are illustrated in
Fig~\ref{fig:dynamo_models}. In the first case of a dynamo working
completely in the overshoot region, the regeneration of the meridional
(poloidal) field has to be reconsidered because the strong fields
suppress the turbulent flows and the kinematic $\alpha$-effect would
hardly work. Possible alternative mechanisms rely on instabilities of
strong magnetic fields driven by buoyancy \cite{Schmitt:1987,
Ferriz-Mas:etal:1994, Brandenburg:Schmitt:1998, Schmitt:2002}, radial
shear flow in the tachocline \cite{Cline:etal:2003a}, or latitudinal
differential rotation
\cite{Dikpati:Gilman:2001a}.  In contrast to the conventional kinematic
dynamo models, the velocity field is not prescribed but consistently
determined from the (linearized) MHD equations. Super-equipartition
fields pose no problem but are actually required in some models for the
mechanism to operate, since the instability must be excited.  On the
other hand, in most of these models, the dynamo is not truly
self-excited since the instability requires the field strength to exceed
a certain threshold value. Consequently, a turbulent dynamo as a
`starter' is required to work in the background.  A frequent problem of
overshoot dynamo models is the considerable overlap of the individual
cycles in a time-latitude diagram of the azimuthal field
\cite{Ruediger:Brandenburg:1995}, which is in disagreement with the
corresponding observational diagrams of active-region occurrence at the
solar surface.

The interface dynamos circumvent the problem of the suppression of the
$\alpha$-effect by a spatial separation of the generation region for
the azimuthal and the meridional magnetic fields
\cite{Parker:1993}. Radial differential rotation builds up a strong
azimuthal field in the overshoot region, while the regeneration of the
meridional field is achieved by the `classical' $\alpha$-effect
operating in the bulk of the convection zone. The connection between the
two regions is accomplished via (turbulent) diffusion, with the magnetic
diffusivity in the overshoot layer being reduced by a factor $\sim
10^2...  10^3$.  As a result, the magnetic field is strong below the
interface and sufficiently weak above for the $\alpha$-effect to work.
The resulting dynamo wave takes on the character of a surface wave
propagating along the interface between the convection zone and the
overshoot layer \cite{Tobias:1996, Charbonneau:MacGregor:1997,
Markiel:Thomas:1999}

\begin{figure}
\centering
\includegraphics[width=.65\hsize]{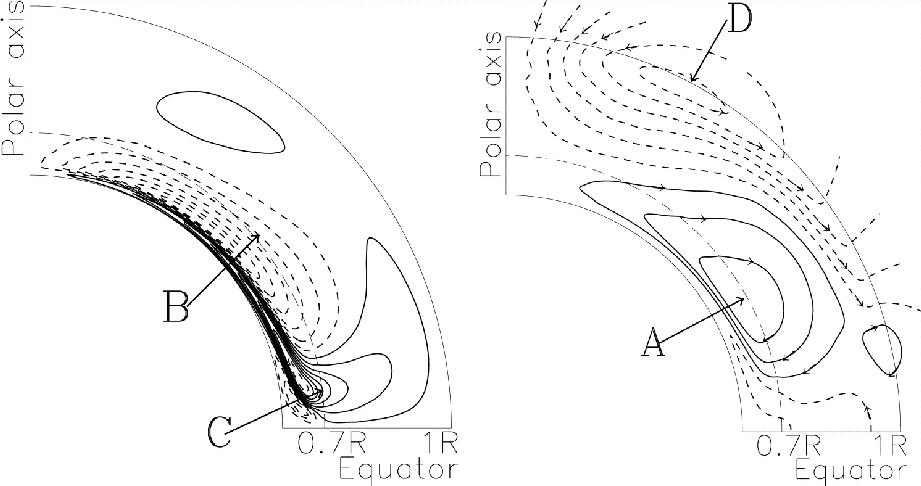}
\caption{\label{fig:flux_transport_dynamo}
Snapshot from a simulation of a Babcock-Leighton type flux-transport
dynamo with a realistic profile of angular velocity at a time close to
`sunspot minimum', after reversal of the high-latitude polar magnetic
field. The meridional circulation (not shown in the figure) is poleward
in the upper part of the convection zone and equatorward in the lower
part. Contour lines of the azimuthal magnetic field strength (left) and
field lines of the meridional magnetic field component (right) are
plotted in a meridional quadrant, with the long-dashed line indicating
the interface between the radial shear layer (overshoot layer) and the
convection zone proper. Solid contours correspond to positive
(clockwise-oriented) azimuthal field (meridional field line), with
opposite polarity (orientation) denoted by dashed contours.  The
meridional flux system located around A is acted upon mostly by the
(positive) latitudinal differential rotation, since the radial shear
vanishes at mid-latitudes.  This leads to the growth of the azimuthal
flux system at B. The azimuthal flux system near C is the (decaying)
remnant of the cycle just completed, and that at B represents the new
cycle just beginning. The latter is already contributing to the
generation of a surface meridional field at D, of opposite polarity to
the meridional flux system in A.  The finite time required for
meridional circulation to advect the magnetic field from D to A
introduces a time lag in the dynamo process (from
\cite{Charbonneau:Dikpati:2000}).  }
\end{figure}

The most often studied examples of flux transport dynamos are the
`Babcock-Leighton dynamos' \cite{Wang:etal:1991,Dikpati:2005}, named so
owing to some similarities with early models for the solar cycle
\cite{Babcock:1961, Leighton:1969}.  In these models, the regeneration
of the meridional field from the azimuthal field component is assumed to
originate from the twist imparted by the Coriolis force on azimuthal
flux tubes rising through the upper layers of the convective
envelope. This twist becomes apparent as the latitudinal tilt of active
regions at the solar surface.  If the radial shear in the tachocline
dominates the $\Omega$-effect, the two induction effects are widely
separated in space. It is assumed that the generated poloidal field is
transported by a large-scale meridional circulation in the convection
zone, a combination of the observed poleward surface flow and a
conjectured equatorward subsurface return flow
\cite{Choudhuri:etal:1995,Durney:1995, Dikpati:Charbonneau:1999b,
Nandy:Choudhuri:2001,Kueker:etal:2001, Guerrero:Munoz:2004,
Charbonneau:etal:2005}.  The deep return flow controls the migration of
the dynamo wave such that it can propagate equatorward (in accordance
with the observed latitude drift of the sunspot zone) even if the sign
combination of $\alpha$-effect and angular velocity gradient would lead
to poleward propagation in the absence of a meridional flow.
Furthermore, the resulting cycle period is largely determined by the
flow speed of the circulation. An example of such a model is shown in
Figure~\ref{fig:flux_transport_dynamo}. Flux transport by meridional
flow has also been considered in some models of overshoot-layer dynamos
\cite{Dikpati:Gilman:2001b} and interface dynamos
\cite{Petrovay:Kerekes:2004, Dikpati:etal:2005}.

\subsubsection{Long-term modulation and grand minima}
\label{subsubsec_longterm}

In order to understand the origin of the long-term modulation of the
solar cycle in terms of dynamo theory, various mechanisms have been
proposed:
\begin{enumerate}
\item[1)] a modulation of the differential rotation through the nonlinear
back-reaction of the magnetic field,
\item[2)] a stochastic fluctuation of the $\alpha$-effect,
\item[3)] a variation in the meridional circulation, and
\item[4)] on-off intermittency owing to a threshold field strength for dynamo
action.
\end{enumerate}

The back-reaction of the large-scale magnetic field on the differential
rotation in the tachocline can lead to complicated nonlinear behaviour,
including long-term amplitude modulations, intermittency and chaos as
well as symmetry-breaking bifurcations, so that the generated field may
flip between dipole and quadrupole states in a grand minimum
\cite{Tobias:1996a, Beer:etal:1998, Kueker:etal:1999, Moss:Brooke:2000}.

A different approach to account for the modulation of the solar cycle is
the stochastic behaviour of the dynamo itself
\cite{Hoyng:1987b,Hoyng:1988}.  Defining averages over longitude in a
mean-field dynamo model and considering a finite number of large
convective cells, all mean quantities, including the $\alpha$-effect,
retain a stochastic component \cite{Hoyng:1993,Hoyng:1996b}. Such a
variation can lead to the occasional excitation of higher dynamo modes
as well as to long intervals of low activity
\cite{Ossendrijver:etal:1996b}.

Fluctuations of both the meridional flow and the $\alpha$-effect have been
studied in connection with flux-transport dynamos
\cite{Charbonneau:Dikpati:2000}.  The results suggest that the
meridional circulation speed, the primary determinant of the cycle
period, acts as a clock and thus maintains the phase stability of the
cycle \cite{Charbonneau:etal:2004}.  The model also exhibits a clear
correlation between the azimuthal field strength of a given cycle and
the strength of the high-latitude surface magnetic field of the
preceding cycle, which is in qualitative agreement with observational
inferences \cite{Legrand:Simon:1991}.  Producing extended periods of
reduced activity, however, turns out to be rather difficult in the
framework of these models.

The dynamic $\alpha$-effect due to magnetic buoyancy only sets in beyond
a threshold field strength \cite{Ferriz-Mas:etal:1994,Cline:etal:2003a}.
Therefore it requires a starting mechanism in the form of fluctuating
magnetic fields transported by downdrafts from the turbulent convection
zone into the overshoot region.  At the same time, such fluctuations,
when destructive, can lead to a sequence of low-amplitude cycles or even
drive the dynamo subcritical until another, constructive magnetic
fluctuation restarts the dynamo again. This leads to on-off intermittent
solutions, which can be related to the occurrence of grand minima
\cite{Schmitt:etal:1996} .

\subsection{Magnetic activity and dynamos of other stars}
\label{subsec_stars}

While the proximity of the Sun allows us to study its magnetic activity
and the underlying dynamo process in detail, observations of other stars
open up the possibility to investigate the dependence of magnetic activity
on stellar parameters like age, rotation rate, and depth of the
convection zone, thus providing opportunities to test dynamo theories
\cite{Dobler:2005}.

Magnetic fields have been detected on many stars of various types
\cite{Mathys:etal:2001}. However, solar-like magnetic activity
characterized by strong surface inhomogeneity, hot chromospheres and
coronae, rapid time variability and flaring, etc., seems to be
restricted to stars with outer convection zones. These are
comparatively cool stars in various evolutionary stages ranging from
very young stars still accreting matter over hydrogen-burning stars to
evolved giant stars \cite{Schrijver:Zwaan:2000}. In some cases, the
magnetic field can be directly measured through the Zeeman effect, but
most often `proxies' of magnetic activity like photometric variations,
emission in chromospheric lines, and coronal X-ray emission are used
to infer the stellar activity. Large starspots can also be detected by
the spectroscopic techniques of (Zeeman) Doppler imaging
\cite{Vogt:Penrod:1983,Semel:1989}.

The magnetic activity of cool stars is related to their rotation rate:
faster spinning stars are more active. This is in general accordance
with turbulent dynamo theory, which predicts an increase of the
$\alpha$-effect with rotation rate.  For a number of stars there are
detections of cyclic activity variations with periods between 7 and 14
years, while others exhibit either irregular variations on a high
activity level or have a flat low activity level, possibly indicating a
grand minimum \cite{Baliunas:etal:1995}. Typically, stars with a cyclic
or flat activity level rotate slowly, while those with a irregular
variations are rapid rotators \cite{Montesinos:Jordan:1993}.


\section{Sunspots}
\label{sec_sunspots}

Sunspots are the most readily visible signs of the interaction between
concentrated solar magnetic fields and the solar plasma, with the
largest sunspots being visible to the (suitably protected) naked eye.
Although sunspots have been extensively studied for almost 400 years
and their magnetic nature has been known for practically a century,
our understanding of a number of their basic properties is still
evolving.

The first telescopic observations of sunspots by Galilei, Scheiner and
others around 1611 marked the beginning of the systematic study of the
Sun in the western world and heralded the dawn of research into the
Sun's physical character.  Over the ages the prevailing view on the
nature of sunspots has undergone major revisions. The breakthrough
came in 1908 when G.E. Hale \cite{Hale:1908} first measured a magnetic
field in sunspots. This was the first time that a magnetic field had
been measured outside the Earth. Since then the magnetic field has
become firmly established as the root cause of the sunspot phenomenon.

Recent overviews of the structure and physics of sunspots are given in
the proceedings edited by Thomas and Weiss \cite{Thomas:Weiss:1992},
Schmieder et al.\ \cite{Schmieder:etal:1997}, Strassmeier et al.\
\cite{Strassmeier:etal:2002}, as well as in the
review article by Solanki \cite{Solanki:2003}.

\subsection{Brightness and thermal structures}
\label{subsec_brightness}

\begin{figure}
\centering
\includegraphics[width=.6\hsize]{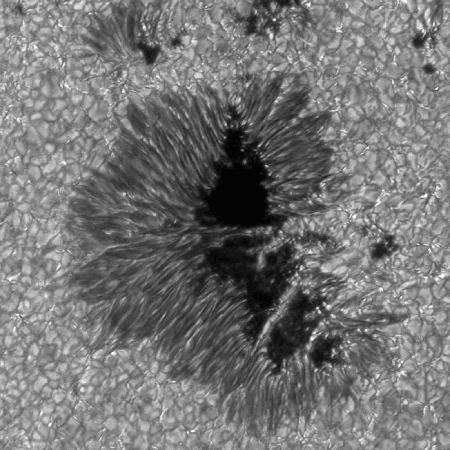}
\caption{\label{fig:1}A white-light image of a sunspot. The dark core is the
umbra, the radially striated part the penumbra. The surrounding bright
cells with dark boundaries are granular convection (Image obtained
with the German Vacuum Tower Telescope on Tenerife, courtesy
Kiepenheuer Institut f\"ur Sonnenphysik).}
\end{figure}

An image of a sunspot is shown in Figure~\ref{fig:1}. Each sunspot is
characterized by a dark core, the umbra, and a less dark halo, the
penumbra. The presence of a penumbra distinguishes sunspots from the
usually smaller pores. In addition, some sunspots contain light
bridges, i.e. bright bands crossing the umbra (two bright light
bridges and a fainter one are visible in Figure~\ref{fig:1}).

Integrated over wavelength, the intensity of the radiation coming from
the umbra is approximately 20\% of that of the quiet Sun, that of the
penumbra is approximately 75\%. Since the penumbral area is roughly four
to five times as large as the umbral area, the averaged, wavelength
integrated intensity of a sunspot is roughly 60--70\% of that of the
quiet Sun.

\begin{figure}
\centering
\includegraphics[height=.6\hsize,angle=90]{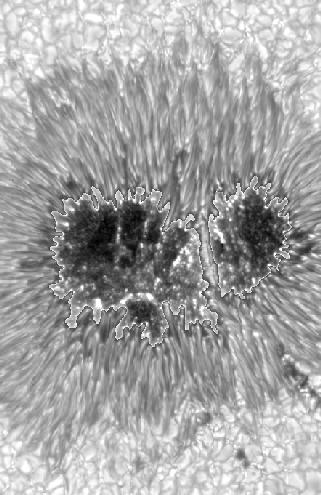}
\caption{\label{fig:2}Image of a sunspot made in the band head of the
TiO molecule at 705.5~nm wavelength with the Swedish Solar Telescope
on La Palma by V. Zakharov and A. Gandorfer. The umbral brightness is
artificially enhanced in order to show the numerous umbral dots more
clearly.}
\end{figure}

The brightness and thus the temperature of a sunspot are functions of
spatial position within the spot. They change on large scales, e.g.\
gradually in the umbra, but small-scale structure is also prominent.
The umbra harbors small bright structures, primarily umbral dots. An
image of a sunspot umbra taken in the TiO band-head at 705.5~nm is
shown in Figure~\ref{fig:2}. It reveals the presence of many umbral
dots, which have diameters ranging from the spatial resolution of the
observations to roughly 600~km, with the number decreasing
quadratically with increasing size
\cite{Sobotka:etal:1997a,Tritschler:Schmidt:2002b}. This
suggests that many umbral dots are spatially unresolved. The exact
brightness of umbral dots is not established beyond doubt, but the
majority is probably significantly darker than the undisturbed
photosphere outside sunspots \cite{Grossmann:Schmidt:1986,
Sobotka:etal:1997, Socas-Navarro:etal:2004}.

The sunspot penumbra is dominated by small-scale structure, most
prominently the elongated bright and dark penumbral filaments, but
also the point-like penumbral grains, which have similarities to the
umbral dots. At the highest currently reachable resolutions the bright
penumbral filaments or fibrils show a sharp dark lane running inside
them \cite{Scharmer:etal:2002} and there is evidence that even these
observations have not resolved all the fine structure
\cite{Rouppe:etal:2004}. In the cores of spectral lines formed in the
middle and upper photosphere this brightness modulation appears to be
washed out or, at least, to be restricted to larger spatial scales
\cite{Wiehr:Stellmacher:1989}.

\subsection{Sizes and lifetimes}
\label{subsec_size}

Sunspots exhibit a considerable range of sizes, which is well
approximated by a lognormal size distribution \cite{Bogdan:etal:1988,
Baumann:Solanki:2005}.  Very large sunspots can occasionally reach
diameters of $60\,000$~km, but are relatively rare. Sunspots smaller
than 3000~km in diameter are also rare.  Smaller photospheric magnetic
structures usually manifest themselves as pores or magnetic elements
(see Sect.~\ref{sec_small-scale}).

Small sunspots live for hours, the largest ones for months. The
lifetime, $T$, increases linearly with maximum area, ${A_0}$:
$A_0 = WT$, where $W \simeq 3.1\,10^7\,$km$^2\,$day$^{-1}$
\cite{Gnevyshev:1938,Waldmeier:1955}.  Sunspots decay steadily soon
after they reach their maximum size. The decay is thought to be driven
by turbulent diffusion of the magnetic field
\cite{Meyer:etal:1974,Petrovay:MorenoInsertis:1997}.

\subsection{Magnetic structure}
\label{subsec magnetic}

\begin{figure}
\centering
\includegraphics[width=0.8\hsize]{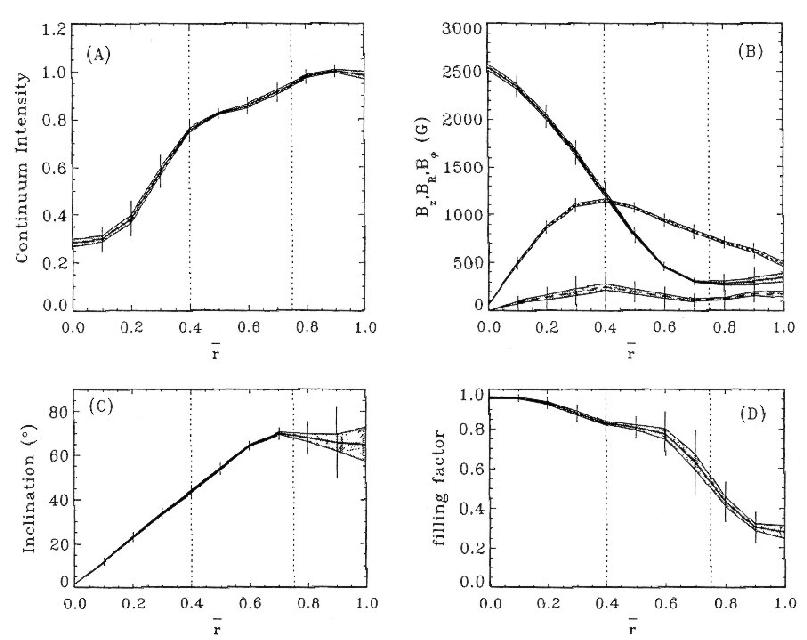}
\caption{\label{fig:3}Intensity and magnetic parameters vs. 
normalized radial distance, $\overline{r}$, from sunspot centre, as
determined from 16 observations of sunspots. Vertical dotted lines
indicate the umbra (left) and the penumbra (right) boundaries. Plotted
are the continuum intensity in Panel A, vertical ($B_z$, solid curve),
radial ($B_r$, dotted curve) and azimuthal ($B_\phi$, dashed curve)
components of the magnetic field in Panel B, magnetic inclination in
Panel C, and magnetic filling factor in Panel D.  $\overline{r}$ is
normalized to a radius well outside the visible sunspot. (from
\cite{Keppens:MartinezPillet:1996}).}
\end{figure}

Sunspots have a field strength of $B$ = 2500--3500~G in their darkest
portions and 700--1000~G at their outer edges. At the `centre' of a
spot (i.e., at the location of the largest field strength) the field
is vertical (i.e., the zenith angle of the magnetic vector, $\zeta$,
is close to zero) while it is inclined by
$\zeta\simeq70^\circ\dots80^\circ$ to the vertical at the visible
sunspot boundary. Measurements of the radial profiles of
magnetic field and inclination angle in almost circular sunspots are
shown in Figure~\ref{fig:3} \cite{Keppens:MartinezPillet:1996}. Other
observations made in the visible and infrared spectral ranges
\cite{Adam:1990,Lites:Skumanich:1990, Solanki:etal:1992, 
Mathew:etal:2003,Hewagama:etal:1993} give similar, although not
completely identical, results. Such regular, isolated sunspots do not
appear to show significant global azimuthal twist of the field
\cite{Landolfi:Deglinnocenti:1982,Lites:Skumanich:1990,
LandiDeglinnocenti:1979,Keppens:MartinezPillet:1996,
WestendorpPlaza:etal:2001a}.  Fits to such data assuming a `buried'
dipole are reasonable, but not perfect. This suggests that the global
structure of the field inside regular sunspots is close to but not
exactly potential. The observations also indicate that sunspot magnetic
fields are bounded by current sheets, i.e., at the sunspot boundary
$B$ falls off rapidly within a radial distance that is small compared
to the size of the sunspot \cite{Solanki:Schmidt:1993}.

From the measured profiles $B(r)$ and $\zeta(r)$ it is possible to
determine the relative amounts of magnetic flux emerging through the
umbra and the penumbra \cite{Schmidt:1991,Solanki:Schmidt:1993}. The
observations imply that over half of the magnetic flux emerges in the
penumbra.  Consequently, sunspot penumbrae are deep, i.e.\ the lower
boundary of the field in the penumbra is inclined and lies
significantly below the solar surface, at least in the inner penumbra.
The outer part of the penumbra could be shallow, however
\cite{Balthasar:Collados:2005}. 

Above the visible solar surface, the field lines of a sunspot fan out
rapidly, so that the magnetic field forms an almost horizontal
{\em canopy,} which overlies the nearly field-free plasma below.  The
interface between these two regions (the canopy base) is found to lie
in the photosphere \cite{Giovanelli:1980, Giovanelli:Jones:1982,
Solanki:etal:1992, Solanki:etal:1999}.  The field strength above the
canopy base decreases steadily outward.

Modern techniques allow the 3-D structure of the magnetic vector in
the lower solar atmosphere to be determined. Within the visible
outline of the sunspot, the field strength decreases with height. At
photospheric levels, $\partial B/\partial z\approx 0.5$--$3$ G
km$^{-1}$ \cite{Balthasar:Schmidt:1993, Bruls:etal:1995,
Eibe:etal:2002}. When averaged over a height range of 2000 km or more,
values around $\partial B/\partial z\approx 0.3$--$0.6$ G km$^{-1}$ are
found in the umbra
\cite{Henze:etal:1982,Lee:etal:1993,Penn:Kuhn:1995,Balthasar:Collados:2005}.
These values are in rough agreement with simple theoretical predictions
of 0.5--1~G km$^{-1}$ \cite{Yun:1972}. More recently, discrepancies have
been found between gradients deduced in the penumbra from visible and
infrared spectral lines, respectively
\cite{WestendorpPlaza:etal:2001, Mathew:etal:2003}. 
Such seeming inconsistencies can be reconciled if the fine structure
of the penumbra is taken into account, in particular, the presence of
horizontal flux tubes interspersed with the inclined field, as
described below
\cite{MartinezPillet:2000,Borrero:etal:2004}.

\begin{figure}
\centering
\includegraphics[width=0.7\hsize]{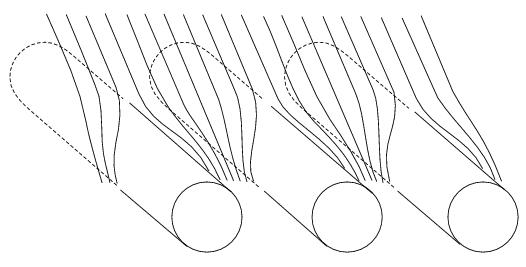}
\caption{\label{fig:4}Detail of the fine-scale magnetic structure of 
the penumbra, as derived from observations. Small-scale horizontal
flux tubes are surrounded by inclined field lines (adapted from
\cite{Solanki:Montavon:1993}).}
\end{figure}

At small scales, the penumbral magnetic field is filamented into
two components, an inclined component and a horizontal component
\cite{Degenhardt:Wiehr:1991,Schmidt:etal:1992,Title:etal:1993}. 
The filaments associated with each component are found to run
uninterrupted across the entire width of the penumbra 
\cite{Langhans:etal:2005}. 
The horizontal magnetic component is restricted in height and is well
described by horizontal flux tubes that are embedded in an inclined
field (\cite{Solanki:Montavon:1993}, see Figure~\ref{fig:4}).  However,
there is still considerable controversy regarding the exact magnetic
structure. One controversial point is the actual width of the horizontal
flux tubes, with proposed values ranging between very slender,
completely optically thin flux tubes and structures that are a few
hundred km thick \cite{SanchezAlmeida:2001,MartinezPillet:2001,
Borrero:etal:2005}.  Recently, is has also been proposed that the
magnetic structure is quite different, with field-free gas intruding
into (or nearly into) the photosphere and the horizontal field overlying
this field-free material \cite{Spruit:Scharmer:2005}.

\begin{figure}
\centering
\includegraphics[height=8cm]{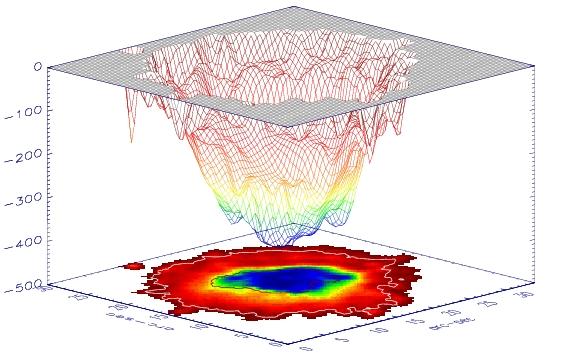}
\caption{\label{fig:5}Wilson depression ($Z$, downward shift of the
layer of optical depth unity) of a regular sunspot obtained from
spectro-polarimetric observations in the infrared. The inner and outer
contours overplotted on the horizontal plane represent the umbral and
penumbral boundaries obtained from the continuum image, respectively
(from \cite{Mathew:etal:2004}).}
\end{figure}

\subsection{Dynamic structure}
\label{subsec_dynamic}

The dominant signature of dynamics in the photospheric layers of
sunspots is the Evershed effect, named after its discoverer
J. Evershed. It is composed of a shift and asymmetry of spectral
lines, with opposite signs on the limbward and centreward sides of the
penumbra\footnote[1]{As a sunspot moves over the visible solar disc
owing to the rotation of the Sun, the line of sight towards the
observer changes its inclination with respect to the locally vertical
(radial) direction: when the sunspot is near the center of the disc,
the spot is observed vertically, while sunspots near the limb are seen
under a large inclination. It is important to take these geometrical
effects into account when it comes to measuring quantities (like
Doppler shift) which reflect the line-of-sight component of a vector.} 
\cite{StJohn:1913, Maltby:1964, Ichimoto:1987, Wiehr:1995,
WestendorpPlaza:etal:2001, Rouppe:2002}.  This observational signature
is generally interpreted in terms of a nearly horizontal, radial
outflow of material. Depending on the spectral line used speeds of up
to 6--9 km~s$^{-1}$ have been inferred \cite{Penn:etal:2003}.  There
has been considerable controversy about whether the Evershed effect
continues beyond the outer penumbral boundary, or not
\cite{Wiehr:Degenhardt:1992,Dere:etal:1990,Borner:Kneer:1992}. 
It is now accepted that the Evershed effect does continue outside the
boundary, but mainly above the base of the magnetic canopy
\cite{Solanki:etal:1994}. The mass flux in the penumbra is found to be
larger than in the canopy region beyond the visible penumbra
\cite{Solanki:etal:1994,Solanki:etal:1999}, so that some of the mass carried
by the Evershed flow must return into the solar interior within the confines
of the sunspots. The problem raised by
this finding was solved by the discovery of downflows
near the penumbral edge \cite{WestendorpPlaza:etal:1997}, which suggests that
the excess mass flows again into the solar interior there. The flow is
also found to be structured in the sense that the outflow follows the
nearly horizontal component of the magnetic field
\cite{Title:etal:1993,BellotRubio:etal:2003}.

The Evershed effect is restricted in height. The line shifts decrease
rapidly with height of line formation (e.g.\
\cite{StJohn:1913,Borner:Kneer:1992}) and at sufficiently large heights
(above the temperature minimum) even change sign, so that the flow in the
chromosphere  is directed inwards (inverse Evershed
effect, e.g., \cite{StJohn:1913,Dere:etal:1990,Tsiropoula:2000}). This
inflow is seen at temperatures up to a few 10$^5$~K, i.e.\ in
both chromospheric and transition region gas. The transition region flow
is both more vertical and faster than the flow at chromospheric
temperatures, and is also seen as an almost vertical flow above the
umbra. In spite of the larger flow speeds, the mass flux transported by
the inverse Evershed effect is only a few percent of the mass flux
flowing out from the umbra in the much denser photosphere.

With the help of local helioseismic techniques it is now starting to
become possible to detect flows below the solar surface. Sunspots are
found to have a distinctive signature, with a downflow below them
\cite{Duvall:etal:1996} and an inflow feeding this downflow just below
the surface (this flow pattern is termed a collar flow and was
proposed by Parker \cite{Parker:1992} to help hold sunspots
together). This suggests that the Evershed flow and the adjacent moat
flow (a horizontal outflow surrounding the sunspot) do not extend more
than a couple of Mm into the solar interior.

Besides such steady flows, sunspots are also rich sources of
oscillations and waves. The most prominent are oscillations with a
period of about three minutes, which are measured in chromospheric
layers above umbrae \cite{Lites:1992}. Above penumbrae, waves
traveling horizontally outwards (so called running penumbral waves)
are dominant \cite{Zirin:1972}. Finally, the oscillations in the
photospheric layers are found to have a period of five minutes with
similar properties as the acoustic oscillations of the quiet Sun. They
are identified as magnetoacoustic oscillations \cite{Staude:1999}.

\subsection{The Wilson depression}

For sunspots near the solar limb, the umbra and often the part of the
penumbra nearer to the centre of the visible solar disk are no longer
visible. This effect is called the Wilson effect after its discoverer,
A. Wilson.  The Wilson effect is best explained if the visible light
within the umbra arises 400--800~km deeper than in the quiet Sun
\cite{Suzuki:1967,Gokhale:Zwaan:1972,Collados:etal:1987}.  The
presence of such a {\em Wilson depression} implies that the gas
pressure within the sunspot must be significantly smaller at the same
geometric level than outside, in accordance with the idea that the
magnetic field is confined mainly by horizontal balance of total (gas
plus magnetic) pressure.

The Wilson depression may also be estimated by introducing measured
values of magnetic field strength and temperature into the horizontal
component of the magnetohydrostatic force-balance equation 
\cite{Alfven:1943,MartinezPillet:Vazquez:1993,Solanki:etal:1993}.  
Comparison of the Wilson depression estimated in this manner with
observation indicates that the magnetic curvature forces within the
sunspot are significant, but do not dominate over pressure gradient
forces. Spatial maps show that the Wilson
depression is largest in the darkest parts of the umbra, where $B$ is
largest \cite{Mathew:etal:2004}. Also, there is a relatively abrupt
change of the Wilson depression at the umbral boundary. The Wilson
depression surface of a regular sunspot obtained in this manner is
plotted in Figure~\ref{fig:5}.

\subsection{Theoretical models of the magnetic structure}

The magnetic structure of sunspots is generally modelled assuming
axial symmetry for reasons of theoretical tractability, although most
sunspots possess an irregular shape. The field is confined by
the external gas  pressure and the magnetohydrostatic equilibrium is
described by

$$ \mu_0^{-1}\, {\rm curl}\ {\bf B}\times{\bf B}=\nabla p-\rho{\bf g} \ ,$$

\noindent
where {\bf B} is the magnetic vector, $p$ is the gas pressure, $\rho$
the gas density and {\bf g} denotes gravitational acceleration.
Usually significant additional assumptions have to be made since
otherwise the computation of the magnetic configuration would require
the treatment of all relevant spatial and temporal scales (which span
very wide ranges, e.g., spatially between 10~km and 10$^5$~km). A
complete theoretical sunspot model would also require the simultaneous
and consistent solution for the magnetic and thermodynamic structures,
i.e., a solution of a complete energy equation (with radiative
transfer and a consistent treatment of magneto-convection) in addition
to the force balance.

The most comprehensive existing models of the magnetic structure of
sunspots \cite{Jahn:1989,Jahn:Schmidt:1994} have a threefold structure
that consists of the umbra, the penumbra, and the surrounding
nonmagnetic stratification, each separated by current sheets.  Such
models, which also could include electrical body currents (leading to
deviations from a potential magnetic field) provide acceptable fits to
the observations of the global magnetic structure of sunspots and are
the most promising for future study.  One basic assumption underlying
all attempts to quantitatively model the global magnetic structure of
sunspots is the assumption that the sunspot is monolithic below the
solar surface.  Recent helioseismic work, however, suggests that this
assumption may not be valid (see below).

\begin{figure}
\centering
\includegraphics[height=9cm,angle=-90]{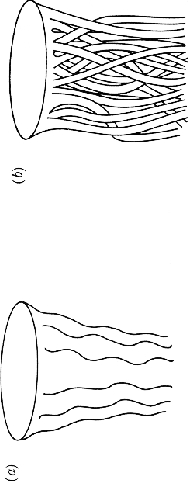}
\caption{\label{fig:6}Sketch of the monolithic (a) and cluster (b) models
of the subsurface structure of sunspot magnetic fields (from
\cite{Thomas:Weiss:1992}).}
\end{figure}

Parker \cite{Parker:1979a,Parker:1979b,Parker:1979c} proposed that
just below the surface the magnetic field of a sunspot breaks up into
many small flux tubes due to the fluting (interchange) instability
\cite{Parker:1975b}. Later is was found, however, that magnetic
buoyancy protects sunspots with magnetic flux $\phi >10^{20}$ Mx from
this instability, at least in the first 5--10~Mm depth
\cite{Meyer:etal:1977}. Nevertheless, a cluster model of sunspots 
(see Figure~\ref{fig:6}) can readily explain
umbral dots as field-free intrusions into the sunspot from below
\cite{Parker:1979c,Choudhuri:1986}, thus providing the high thermal
energy flux of the umbra, which cannot be transported by radiation
alone. The strongest (but still indirect) support for a cluster model
comes from measurements made with the technique of local
helioseismology, which can (with some restrictions) image the
subsurface thermal and velocity structure
\cite{Duvall:etal:1996}. At least in one case, signs of a
nearly horizontal flow that passes a few Mm underneath a sunspot has
been found \cite{Zhao:etal:2001}.  Since the magnetic field is nearly
frozen in the plasma, the field lines at the surface are nearly
vertical, and the spot is stable, such a flow cannot occur if the spot
is monolithic below the surface.

One idea concerning the nature of the small-scale magnetic structure of
the penumbra considers it to be dynamic and its complexity to result
from the convective exchange of flux tubes
\cite{Spruit:1981c,Schmidt:1991}. In the framework of this scenario, a
flux tube at the outer boundary of the penumbra (the magnetopause) is
heated by the field-free convecting external plasma.  The subsurface
section of the heated tube becomes buoyant and rises, eventually coming
to the surface near the outer penumbral edge.  Later, portions of the
tube closer to the umbra reach the solar surface. At the surface, the
tube cools by radiation, loses its buoyancy, becomes more horizontal,
and eventually sinks down again, so that the cycle can be repeated. Part
of this scenario has been studied by numerical simulation
\cite{Schlichenmaier:etal:1998a,Schlichenmaier:etal:1998b}:  a flux
tube lying at the magnetopause does indeed rise to the
surface until it lies there horizontally through much of the penumbra.
Hot material continues to rise along the flux tube and flows
 horizontally outward along the flux
tube. This flow reproduces some properties of the Evershed effect,
suggesting that the magnetic and velocity structure are intimately
connected. Evidence supporting the presence of magnetic structure
consistent with these simulations has been provided 
\cite{Schlichenmaier:Schmidt:2000,Schlichenmaier:Collados:2002}.
However, the simulations carried out so far do not reproduce the second
half of the convective transport cycle: the horizontal flux tubes do not
sink back down again, but remain at the surface. In this respect, the
simulations are in accordance with observed time series of magnetograms
that display little change in the magnetic structure with time
\cite{Solanki:Ruedi:2003}. An alternative proposal assumes that the
fluted structure of the penumbral field is produced by upwellings of
hot, field-free material, which parts the penumbral field lines
approximately like two curtains and reaches very close to the
surface. Above these intrusions, the field is nearly horizontal, beside
them more vertical \cite{Spruit:Scharmer:2005}.

\subsection{Models of the brightness and thermal structure}
\label{theory bright}

L. Biermann \cite{Biermann:1941} first proposed that the reduced
brightness of sunspots is due to the inhibition of convective motions
by the magnetic field, whereby the criterion for the onset
of convection is modified by the presence of the magnetic field
\cite{Proctor:Weiss:1982}.  Since convection dominates energy
transport below the observable layers, quenching of convection turns a
sunspot into an obstacle to the outward heat flux through the
convection zone. This leads to a diversion of energy away from the
sunspot, which reduces the energy flux through the spot and produces a
darkening. The diverted energy flux is mainly stored in the convection
zone and released very slowly over its thermal time scale of $10^5$
years, much longer than the lifetime of a sunspot
\cite{Spruit:1977,Spruit:1982a,Spruit:1982b}. The recent rediscovery
of a (very faint) bright ring around large sunspots
(\cite{Rast:etal:2001}; see Waldmeier \cite{Waldmeier:1939} for the
original discovery) has only served to show just how efficiently the
blocked energy is redistributed within the convection zone. Since only
a few percent of the blocked energy flux immediately re-emerges in the
bright ring, the presence of sunspots at the solar surface leads to a
global darkening of the Sun.

A complete quenching of convection so efficiently reduces the heat flux
that the question to ask is not why umbrae and penumbrae are so dark,
but rather why they are so bright, in particular the penumbrae. The observed
brightness can only be explained if efficient mechanisms of heat transport
act within sunspots in spite of the strong magnetic field.

Of the approaches that have been taken to solve the problem basically
two are still considered viable.
The first is to form the sunspot out of a cluster of small flux tubes
(see Section 6.6). Owing  to the tapered shape of each small
flux tube, larger amounts of field-free gas are present between them at
increasing depth. Consequently, below the sunspot, convection can
penetrate relatively unhindered until close to the surface. In addition,
for a sunspot composed of $N$ small flux tubes the surface area of the
side walls of the flux tubes, over which the convective gas can radiate
into the magnetized gas, is $\sqrt N$ times larger than for a simple
monolithic sunspot. A larger side-wall surface compared to the
horizontal cross-sectional area leads to a more efficient heating of the
tubes. Note that, instead of flux tubes, marrow flux sheets als lead to
an enhanced energy flux at the surface.

The second proposal considers convective transport within a monolithic
sunspot umbra.  Magnetoconvection is a vast subject in itself and we
refer the reader to reviews devoted specifically to this topic (e.g.,
\cite{Weiss:1997,Weiss:2002,Proctor:1992}. Here we only mention two
results. The character of the convection below a sunspot umbra depends
largely on the parameter $\zeta=\eta/\kappa$, where $\eta$ is the
magnetic diffusivity and $\kappa$ is the (radiative) thermal
diffusivity. For $\zeta<1$ the convection is oscillatory, while for
$\zeta>1$ overturning convection is the preferred mode
\cite{Meyer:etal:1974}.
Since $\zeta\approx 10^{-3}\ll 1$ at the solar surface, but increases
rapidly with depth, the current picture is that oscillatory convection
dominates in the first 2~Mm below the surface, while overturning convection
takes over in the deeper layers. In any case, such modes of
magneto-convection lead only to moderate variations of the magnetic
field strength between the upflow and the downflow regions
\cite{Weiss:etal:1990}. 

Recent 3-D simulations exhibit cases of filamentation. For magnetic
fluxes appropriate to sunspot umbrae, islands of upwelling hot gas
with a relatively weak magnetic field were found, surrounded by cooler,
strong-field material \cite{Tao:etal:1998,Cattaneo:etal:2003}. These
hot upwellings are reminiscent of umbral dots. Unfortunately,
such magnetoconvection simulations cannot as yet
be used to decide between the monolithic and the cluster model. This
is partly because the parameters and geometries for which these
computations are carried out are still significantly removed from real
sunspots.


The problem of heating the umbra pales in comparison with that of
heating the penumbra with its nearly 4 times higher radiative
flux. One possibility would be a `flat penumbra', such that the
sunspot magnetopause were to lie within a few photon mean-free-paths
of the solar surface under the whole penumbra
\cite{Schmidt:etal:1986}. However, the measured magnetic field
structure excludes such a model, since it does not account for the
amount of magnetic flux emerging in the penumbra
\cite{Solanki:Schmidt:1993}. Most other options, such as
interchange convection (see previous subsection), the excess heat
brought to the surface by the gas feeding the Evershed flow, or heating
by magnetic reconnection between the inclined and the horizontal field,
face problems. For example, the first runs afoul of observations
suggesting that the penumbral magnetic structure does not change
sufficiently rapidly to provide the necessary heat flux
\cite{Schlichenmaier:Solanki:2003,Solanki:Ruedi:2003}. 
A steady outflow of matter as the source of the heat could work, but only
if the flow dips below the solar surface multiple times during its
passage across the penumbra. Another possibility is a special mode of
magneto-convection in an inclined magnetic field (e.g.,
\cite{Danielson:1961}, or the intrusion of field-free material between
the penumbral field lines, making the penumbra locally shallow
\cite{Spruit:Scharmer:2005}. 

\subsection{Models for the Evershed effect}
\label{subsec_theory}

The simplest interpretation of the Evershed effect is that it is
produced by a steady, almost radial outflow. This scenario was put on
a solid physical footing by Meyer and Schmidt
\cite{Meyer:Schmidt:1968a,Meyer:Schmidt:1968b}, who presented a siphon
flow model of the Evershed effect. The field strength in the outer
penumbra (700--900~G) is smaller than in typical small-scale magnetic
elements (1000--1500~G), so that the gas pressure is expected to be
larger in the outer penumbra. If the field lines from the outer penumbra
are connected to elements of concentrated magnetic flux outside the
sunspot (see Sect.\ref{sec_small-scale}), then the gas pressure
difference at equal geometrical height drives a flow from the penumbra
to the magnetic elements. At the same time, this model also explains the
inverse Evershed effect. The field strength in the umbra (2000--3000~G)
is larger than in small external magnetic elements, so that a field line
connecting the umbra with a magnetic element should support an inflow
into the umbra. The relative simplicity and intuitive appeal of this
model has led to a number of investigations (e.g.,
\cite{Degenhardt:1993,Montesinos:Thomas:1997}).


Observations indicate that the outflow associated with the Evershed
effect is largely restricted to the penumbra, with downflows present
in the outer penumbra \cite{WestendorpPlaza:etal:1997}, so that only a
small fraction of the mass continues into the superpenumbral canopy
\cite{Solanki:etal:1994}. Since the (average) field strength decreases by 
roughly a factor of two from the inner to the outer penumbra, this would
lead to a siphon flow directed towards the umbra, opposite to the
observed flow. Only a difference in the heights (i.e., Wilson
depression) to which the field strength observations correspond leaves a
possibility to obtain the correct flow direction
\cite{Montesinos:Thomas:1997}.  However, another solution is also
possible. The horizontal component of the field supporting the
Evershed flow changes very little in strength between umbral boundary
and outer spot boundary \cite{Ruedi:etal:1999, Borrero:etal:2004}
while the more vertical background component of the field displays a
strong outward decrease. The balance of total pressure between these
two components leads to a strong radial gas pressure gradient in the
flow-carrying horizontal flux tube, which can drive an outward flow.


\section{Small-scale magnetic structure}
\label{sec_small-scale}

\subsection{Introduction}
\label{subsec_introduc}

The magnetic field in the photospheric layers is concentrated in active
regions and in a network distributed over the whole Sun. In active
regions (outside sunspots) the magnetic field is concentrated into more
or less discrete features that together form faculae or plage
regions.\footnote[1]{Historically, a plage is defined as an extended
bright region seen in cores of stronger chromospheric spectral
lines. Plages are roughly co-spatial with photospheric faculae, bright
areas seen near the solar limb in the continuum or in photospheric
spectral lines. Both, plages and faculae, are caused by a large density
of small-scale elements of concentrated magnetic flux, so that often the
terms plage or faculae are used to denote areas densely populated by such
magnetic elements.}  In the quiet Sun (i.e., outside active regions) the
magnetic flux elements form a network outlining the borders of
supergranular cells with a length scale of 20--40 Mm. Another type of
magnetic feature in the quiet Sun are the internetwork elements, which
are located in the interiors of supergranular cells.  On a more
localized scale, the magnetic elements forming faculae and the network,
and very likely also those in the internetwork, are located in the
downflow lanes between granules
\cite{Title:etal:1987,Solanki:1989}.

There is a whole spectrum of magnetic features having very different
sizes and properties on the solar surface. Sunspots
(Sect.~\ref{sec_sunspots}) are the largest and rarest.  Even at times
of greatest solar activity they cover only a fraction of a percent of
the solar surface. On average smaller than sunspots are pores,
although the largest pores can be bigger than the smallest
sunspots. Pores typically have diameters of a couple of thousand km
and are dark. They are distinguished from sunspots by the fact that
they have no penumbra, appearing like isolated umbrae
\cite{Keppens:2001}. Even smaller, and far more
common, are magnetic elements, bright structures with diameters
smaller than a few hundred km.  High resolution observations show
magnetic features with sizes right at the achievable spatial
resolution of 150~km \cite{Keller:1992}. In the quiet Sun, indirect
estimates suggest the existence of features of the internetwork field
with a diameter of at the most 50~km \cite{Lin:1995}. Finally,
observations indicate an omnipresent turbulent field in photospheric
layers.

The current picture of the magnetic elements is that they have many
features in common with sunspots (although at first sight the
differences may appear to dominate). In particular, both structures,
as well as pores, are thought to be manifestations of intense
concentrations of magnetic flux and are usually described by the
theory of magnetic flux tubes. Overviews of the properties of magnetic
elements can be found in \cite{Spruit:etal:1991,Solanki:1993}.

The magnetic structure in both active regions and in the quiet Sun
network is very similar: the magnetic field is concentrated into more
or less discrete elements of magnetic flux separated by regions with
comparatively little magnetic flux (to first approximation they are
considered to be field free, although they probably harbor a
considerable turbulent magnetic field; see
Sect.~\ref{subsec_internetwork}). Hence the measured intensity from a
part of the Sun can, employing a simple 2-component model, be written
as:

\begin{equation}
\label{eq:7.1}
I_{\rm obs} = f I_{\rm m} + (1-f)I_{\rm b},
\end{equation}

\noindent
where $I_{\rm obs}$ is the observed intensity, $I_{\rm m}$ the
intensity of light from the magnetic feature, $I_{\rm b}$ from the
``field-free'' background and $f$ is the magnetic filling factor,
i.e.\ the fraction of the observed part of the solar surface covered
by intense magnetic fields. Equation~\ref{eq:7.1} is based on the
assumption that the spectra of magnetic elements and background are
the same everywhere, or, alternatively, it describes average
properties. One reason why Equation~\ref{eq:7.1} has been widely used is
that many magnetic features are not resolved by spectropolarimetric
observations, which usually do not reach the spatial resolution
achievable in broad-band images. Although $I_{\rm m}$ and $I_{\rm b}$
do to a certain extent depend on $f$, to first order on different
parts of the Sun $I_{\rm obs}$ can be determined simply by changing
$f$, while leaving $I_{\rm m}$ and $I_{\rm b}$ unchanged. When
considering the net polarization (as described by the Stokes
parameters $Q$ and $U$ for net linear polarization and Stokes $V$ for
net circular polarization) produced by the Zeeman effect to first
order

\begin{equation}
\label{eq:7.2}
S_{\rm obs}=f S_{\rm m}, \,\,\,\,\,\,{\rm where}\,\,\,\,\,\, S=Q, U, {\rm or}\,\,\,
V.
\end{equation}

In reality, the relationship is more complex, but Equation~\ref{eq:7.2}
does indicate
one major advantage of using the polarized Stokes parameters when
studying small-scale magnetic elements, specially when they are
spatially unresolved: to first order the field-free atmosphere does not
contribute to the polarized Stokes profiles.

\subsection{Field strength measurements}
\label{subsec_field}

The first evidence for magnetic fields outside sunspots was found by
G.E.\ Hale \cite{Hale:1922} in what he termed `invisible
sunspots'. Such features were first mapped with the photoelectric
magnetograph by Babcock and Babcock
\cite{Babcock:Babcock:1955}. Although the field strength averaged over
the spatial resolution element of the magnetograph (typically a few Mm
on the Sun) was on the order of tens to a few hundred G, evidence kept
accumulating that the intrinsic field strength of these features is
much larger
\cite{Sheeley:1966,Beckers:Schroter:1968,Frazier:Stenflo:1972}.  The
difference between the apparent and intrinsic field strength can be
explained if the features are spatially not resolved, i.e., $f\ll 1$
in Equations~(\ref{eq:7.1}) and (\ref{eq:7.2}). Finally, a technique
involving the ratio of a carefully chosen line pair was used by
Stenflo \cite{Stenflo:1973} to show that magnetic elements
intrinsically have kG field strength (cf.\
\cite{Wiehr:1978,Stenflo:Harvey:1985,Solanki:etal:1987}).  The strong
fields have been confirmed by a number of other techniques and
diagnostics in the meantime, most prominently the 
spectral lines of neutral iron (Fe I) at 1.56 $\mu$m, which are
sufficiently Zeeman-sensitive to directly reveal the splitting by the
kG field \cite{Rabin:1992a,Rabin:1992b,Ruedi:etal:1992}. More
recently, line profile inversions, which allow the magnetic field to
be determined consistently with the other atmospheric parameters have
confirmed and refined the results obtained with other techniques
\cite{MartinezPillet:etal:1997,BellotRubio:etal:2000,Frutiger:Solanki:2001}.
At the same time, the spatial resolution of the observations has also
been increasing, in some cases allowing magnetic elements to be nearly
resolved \cite{Keller:1992}, although simulations of magnetoconvection
show many features that lie below the best currently achievable
spatial resolution (see Sect.~\ref{subsec_models}).

\begin{figure}
\centering
\includegraphics[width=0.8\hsize]{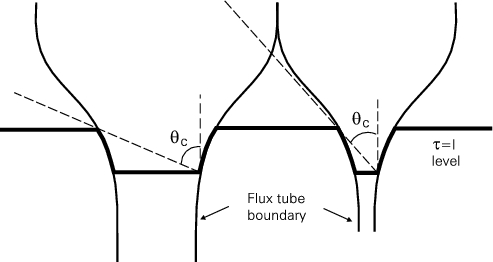}
\caption{\label{fig:7}
`Wine glass model' of magnetic elements. 
The bounding field lines (current sheets) of two neighbouring magnetic
elements are indicated by the curved, nearly vertical lines, the solar
surface (i.e.\ the continuum unit optical depth, $\tau=1$, level) by
the thick line.  Also indicated are the angles $\theta_c$ at which the
bright walls of the flux tubes are best visible. Note that $\theta_c$
is larger for the thicker flux tube.}
\end{figure}

The magnetic field strength drops rapidly with height. In the deep
photosphere, the field strength is around 1500--1700~G
\cite{Rabin:1992a,Rabin:1992b, Ruedi:etal:1992}), dropping to roughly
1000--1200~G in the middle photosphere \cite{Stenflo:Harvey:1985}),
and to 200--500~G near the temperature minimum
\cite{Zirin:Popp:1989,Bruls:Solanki:1995}). How strongly the field strength
drops with height in still higher layers depends on the filling
factor. Owing to magnetic flux conservation, the magnetic features
expand with decreasing field strength (increasing height) until
neighboring flux tubes merge, at least in the case when they have the
same magnetic polarity, which is typical of active regions. This case is
illustrated in Figure~\ref{fig:7}.  Above the merging height, the field
strength is expected to drop much more slowly. Observations have also
provided evidence for the expansion of the flux tubes and their merging
\cite{Bruls:Solanki:1995,Briand:Solanki:1998}. In a mixed polarity
environment like the quiet Sun, the situation is more complex since
field lines can bend over and return to the solar surface relatively
close by.

\begin{figure}
\centering
\includegraphics[width=0.7\hsize]{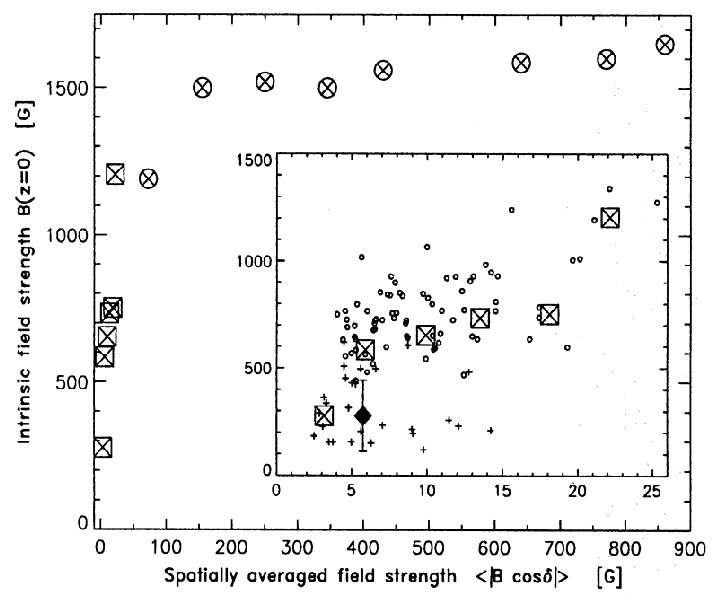}
\caption{\label{fig:8}
Intrinsic magnetic field strength at the solar
surface, $B(z=0)$, vs.  the unsigned, spatially averaged longitudinal
field strength $\langle |B\cos\delta|\rangle$, where $\delta$ is the
angle between the direction of the magnetic vector and the line of
sight of the observer.  Plotted are binned values resulting from quiet
Sun (squares) and plage (circles) observations.  The inset shows only
the left-most part of the main figure, with the squares being
identical to those in the main figure. The small circles and crosses
represent individual measurements.  The diamond denotes the value of
$B$ derived from the profile averaged over all the magnetic features
denoted by crosses, which individually lie below the noise level (from
\cite{Solanki:etal:1996}).}
\end{figure}

The field strength of magnetic elements increases slowly with increasing
filling factor, $f$, and also with increasing cross-sectional area of
the elements \cite{Ruedi:etal:1992,Zayer:etal:1990}. Thus between the
quiet network and strong active region plage the field strength
increases by roughly 10\% at a given geometrical height, although $f$
can change by nearly an order of magnitude. At very small filling
factors or magnetic fluxes, as found mainly in supergranule cell
interiors, however, the situation is less clear. Observations using
the infrared iron lines at
1.56$\mu$m show that $B$ drops rapidly with decreasing flux
\cite{Lin:1995,Solanki:etal:1996,Khomenko:etal:2003}. High
resolution observations in the visible spectral range, on the other
hand, suggest the presence of strong fields throughout the
supergranule cell interiors
\cite{Dominguez:etal:2003a,Dominguez:etal:2003b}. The dependence of the
intrinsic field strength on the spatially averaged field, as deduced
from 1.56$\mu$m lines, is shown in Figure~\ref{fig:8}.

High resolution images suggest that the average size of magnetic
features increases as the filling factor increases.  This is also
supported by the comparison of spectropolarimetric data with MHD
simulations \cite{GrossmannDoerth:etal:1994}.  These results indicate
that the increase in field strength with filling factor may mainly
reflect a dependence on flux-tube size.

\subsection{Formation and evolution}
\label{subsec_formation}

New magnetic flux emerges into the solar atmosphere in the form of
bipolar magnetic regions.  During emergence, the field initially is
relatively weak (being a few hundred Gauss) and predominantly
horizontal \cite{Lites:etal:1998}. At this point we are seeing the top
of the emerging $\Omega$ loop. As the loop continues to rise, its two
photospheric footpoints move apart, the field gets stronger and nearly
vertical.

Right from the beginning, the field interacts with convection. At the
location of flux emergence, the granulation is heavily disrupted, with
long thin parallel dark and light striations being visible instead of
normal granules \cite{Strous:Zwaan:1999}. This interaction continues
throughout the lifetime of the magnetic flux, although the field and
convection segregate early (such that the magnetic flux then lies in the
intergranular convective downflow lanes). As the granules evolve and
move, the flux located between them is forced to move along, wandering
with the intergranular lanes. Finally, it is probably the turbulence
associated with convection which buffets and twists the field on a small
scale, bringing opposite polarities together and thus allowing for the
dissipation of magnetic energy.

The currently favored physical picture of the formation of magnetic
elements involves a two-stage process. In a first step, the convection
(i.e.\ the granular flow) expels the initially weak and relatively
homogeneously distributed magnetic flux into the intergranular
downflow lanes
\cite{Parker:1963,Weiss:1966,Tao:etal:1998,Hurlburt:Weiss:1987}. This
process is called flux expulsion and can concentrate the field until
it is roughly in equipartition with the flow, i.e., the magnetic
energy density equals the kinetic energy density, $B \simeq
\sqrt{\mu_0\rho}\,v$, where $\rho$ is the gas density and $v$ is the
average horizontal convective flow speed. For the solar photosphere
values of 100--400~G are thus obtained, depending on the height to
which they refer ($\rho$ drops exponentially with a scale height of
$\approx$ 100~km) and on the value of the convective velocity assumed.
Of considerable importance for the further concentration of the
magnetic flux is an instability, the so-called convective collapse,
which is driven by the cooling, through radiative losses at the solar
surface, of the gas trapped between the field lines
\cite{Parker:1978,GrossmannDoerth:etal:1998}.  As the gas cools, its
density increases, so that it sinks and settles in deeper layers,
leading to low pressure in the upper layers of the flux
concentration. The pressure of the surrounding gas pushes the field
lines together, thus strengthening the field to kG values at the solar
surface.  A field of this strength quenches convection and stops the
instability from growing further. Radiative heating (mainly through
the side walls of the flux concentration; see
Sect.~\ref{subsec_models}) keeps the gas from cooling further and thus
allows the magnetic element to approach a stable magneto-static state
\cite{Spruit:1979}. Since the lateral influx of heat is more efficient
for more slender flux tubes (which have a larger ratio of wall area to
volume and which are less optically thick), the less flux in a
small-scale magnetic feature, the lower the final field strength
\cite{Venkatakrishnan:1986}.  Observations confirm this theoretical
prediction
\cite{Solanki:etal:1996}.

In reality, the various stages of the above multistep process
(emergence, brightening, flux expulsion and convective collapse) seem
to run in parallel \cite{Schuessler:1990}.  The emerging field already
has equipartition field strength and is very rapidly swept into the
intergranular downflow lanes, where is becomes intensified by the
convective collapse process.  As the field becomes stronger, the
buoyancy force increases $\propto B^2$, so that an initially inclined
field quickly becomes nearly vertical \cite{Schussler:1986}. Indeed, a
strong correlation is found between inclination and the strength of
the field in an emerging flux region \cite{Lites:etal:1998} and is
also observed in simulations of magnetoconvection appropriate for
regions of mixed magnetic polarity \cite{Gadun:etal:2001}.

The non-stationary granulation appears to be the major driver of the
evolution of individual magnetic features, although on a larger scale
the evolution of the supergranulation and possibly the mesogranulation
(i.e., a velocity pattern on an intermediate scale of 5--10~Mm) play
the more important role \cite{Dominguez:2003}.  Individual magnetic
features are dragged along as intergranular lanes change and move
\cite{Berger:etal:1998,Choudhuri:1993}.  Polarimetric observations have
so far provided only limited information on the evolution of
individual magnetic elements, with proxies revealing much of what we
know.  Proxy observations (see Sect.~\ref{subsec_proxies}) show, e.g.,
the splitting and merging of bright points \cite{Berger:etal:1998}.
However, a deeper insight is given by MHD simulations (see
Sect.~\ref{subsec_models}).

The most common mode of decay of the magnetic field is through
cancellation with features of opposite magnetic polarity, as has been
revealed from time series of magnetograms
\cite{Martin:etal:1985,Livi:etal:1985}. Again, the spatial resolution
of the observations is not sufficient to learn more about the exact
processes acting during the cancellation. However, after the emergence
of a bipolar loop, its individual footpoints generally keep moving
apart and finally cancel with opposite polarity elements belonging to
another bipole (or merge with field of the same polarity). Hence, the
cancellation must be accompanied by magnetic reconnection of some
form, i.e.\ with some amount of energy release. This process has been
proposed as the source of the observed X-ray bright points
\cite{Priest:1994}.

\subsection{Facular brightening and proxies of the magnetic field}
\label{subsec_proxies}

The measurement of the magnetic field often suffers from a low
signal-to-noise ratio, in particular when considering regions with low
magnetic flux (or filling factor). Since net circularly polarized
light needs to be recorded at high spectral resolution, the low signal
level necessitates longer integrations, which results in a loss of
spatial resolution caused by smearing due to fluctuations of the
refractivity index in the turbulent terrestrial atmosphere. This has
led to the use of more easily accessible proxies, generally the
intensity in some (narrower or broader) spectral band in which the
magnetic features exhibit a particularly large contrast. These proxies
are useful since they are more rapidly recorded, permit a simpler
instrument setup and do not suffer from polarization cross-talk.
Magnetic elements are generally visible as bright structures.  Their
visibility thus depends on the amount of excess heating which they
undergo.

\begin{figure}
\centering
\includegraphics[height=12cm]{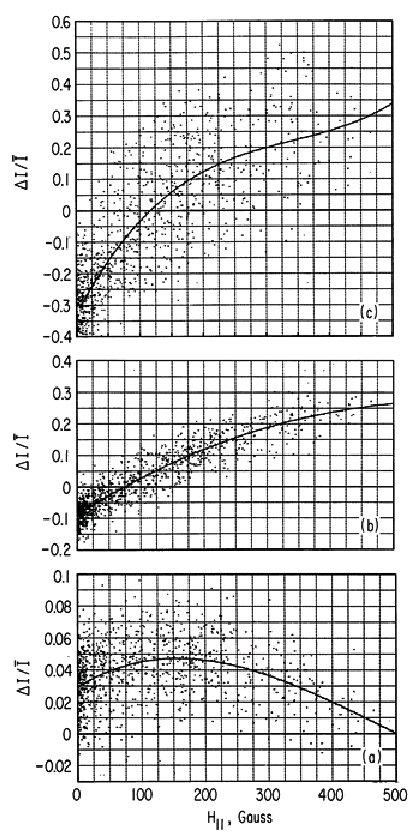}
\caption{\label{fig:9}Brightness contrast $\Delta I/I$ of, from bottom to top, (a) the continuum,
(b) the core of Fe I 525.02 nm, and (c) Ca II K vs. the magnetic field
strength averaged over the spatial resolution element of the
observations. The observations refer to active regions (from
\cite{Frazier:1971}).}
\end{figure}

The intrinsic brightness of magnetic features depends on their size.
In general, the intrinsic brightness of a magnetic feature located
near the centre of the visible solar disc (so that the observational
line of sight is vertical) increases steadily with decreasing
cross-sectional area or magnetic flux per feature. Such a relationship
also holds if instead of size the magnetic flux per pixel, or filling
factor $f$, is considered \cite{Topka:etal:1992,Ortiz:etal:2002}.
However, unless the spatial resolution is very high or a region
located near the solar limb is observed, magnetic elements cannot be
recognized in continuum radiation. This behavior can be understood in
terms of a flux-tube model (see below). The dependence on magnetic
flux itself depends on the height from which the radiation originates
(e.g., \cite{Frazier:1971}). Thus, pores are dark in the visible
continuum, but bright in radiation coming from the chromosphere (e.g.,
emitted in the core of the strong spectral line of ionized calcium, Ca
II K). Quite generally, the contrast of the magnetic features located
near solar disc centre increases with increasing height of emission or
absorption of the radiation (within the photosphere and chromosphere).
Hence the best wavelengths at which to study magnetic elements are
those for which the radiation is emitted in the mid-photosphere or
higher and where the intensity is particularly temperature
sensitive. This is illustrated by Figure \ref{fig:9}, which shows the
large difference between the contrast of magnetic features measured in
the visible continuum (bottom), and the cores of spectral
lines. Clearly, the contrast saturates and decreases again in the
continuum for increasing magnetic flux (formation of pores), but
continues to increase in the line cores.

Recent G-band observations with a resolution of 100~km or better have
unveiled new aspects of the morphology and dynamics of small-scale
magnetic features. Their morphology strongly depends on the amount of
magnetic flux that they harbour. In regions of low magnetic flux (quiet
Sun) the features are more point-like, while in strong plage (active
regions) they reveal are more ribbon-like appearance, often exhibiting a
rather convolved shape. The features continuously evolve along with the
granulation, changing their form, buckling at their edges, breaking up
or coalescing \cite{Berger:etal:2004,Rouppe:etal:2005}.

Quite distinct is the behavior of the brightness contrast (with
respect to the quiet Sun) of magnetic elements as a function of their
distance from the centre of the solar disk (centre-to-limb variation
of contrast due to an increasing angle between the line of sight and
the local vertical direction on the Sun).  In white light or at
continuum wavelengths, faculae are often almost invisible near the
centre of the solar disk, but become bright near the limb even at
relatively low spatial resolution.  This dependence on limb distance
also is a strong function of the amount of magnetic flux or filling
factor \cite{Topka:etal:1997,Ortiz:etal:2002}.  The steep increase in
contrast towards the limb is typical of regions with a high flux
density, with features that are dark at the centre of the disk
appearing bright near the limb. At small $f$, the contrast peaks
closer to the centre of the disk, although it is comparatively flat as
a function of limb distance.  The contrast of chromospheric plage and
the chromospheric network is much higher and is roughly independent of
position on the solar disk. This suggests that different mechanisms
are responsible for the excess brightness of magnetic elements in the
lower photosphere than in higher layers.

Although broad-band brightness is the most easily recorded quantity,
it is not the best proxy of the magnetic field. The continuum contrast
of magnetic elements is small, being a couple of percent under typical
to good observational conditions. This is similar to the contrast of
the granulation pattern, so that it is not straightforward to identify
magnetic features except near the limb or at very high spatial
resolution.  The contrast can be enhanced if the observations are
carried out in the cores of spectral lines, in particular if these
lines are temperature sensitive. Most commonly used are the
chromospheric cores of the Ca II H and K lines and wavelength bands
dominated by lines of diatomic molecules, whose relatively low
dissociation potential makes their spectra extremely
temperature-sensitive. Examples are the G-band at around 430.5~nm
dominated by spectral lines of the CH molecule or the CN band head near
388~nm \cite{Steiner:etal:2001,SanchezAlmeida:etal:2001,Berdyugina:etal:2003}.
Observations in these wavelength bands allow higher resolution to be
obtained than with magnetograms since they can be carried out with
wider filters (0.1--3~nm wide) rather than in the narrow bands
necessary for magnetograms ($<$ 0.01~nm), so that shorter exposure
times are possible.  The higher contrasts in spectral line cores
suggest that the temperature enhancement in the upper photosphere and
the chromosphere in magnetic elements is much stronger than in the lower
photosphere.

\begin{figure}
\centering
\includegraphics[width=0.65\hsize]{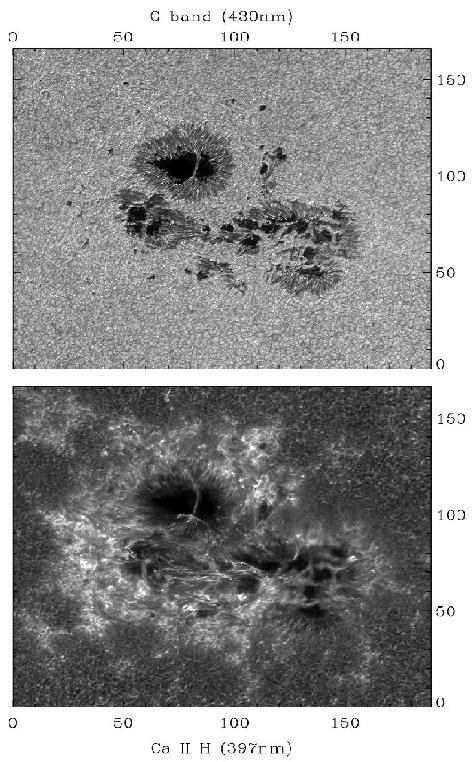}
  \caption{\label{fig:10}AR 10375 imaged with the Dutch Open Telescope
  in the G-band (upper frame) and the core of the Ca II H line (lower
  frame). Figure courtesy of P.\ S\"utterlin and R.J.\ Rutten.}

\end{figure}

The brightness in the Ca II K line core has been used as a proxy of
magnetic brightening since the invention of the spectroheliograph,
showing both the magnetic network and the active regions with high
contrast. Recently, very high resolution Ca II K line observations
have been made with the Dutch Open Telescope
\cite{Rutten:etal:2004}. An example is shown (along with an image
in the G band, which originates in the photosphere) in
Figure~\ref{fig:10}.  The sunspots are seen in both photospheric and
chromospheric radiation, but are less readily visible in the
latter. The plage, on the other hand, appears particularly bright in
the chromosphere, and the network structure is also clearly
discernible. Note that it is not always the same features that are
most clearly visible in the chromosphere and in the photosphere.

With the advent of observations in the ultraviolet spectral range,
further proxies for the magnetic field became available. These include
in particular the He II line at 30.4~nm, which is recorded by the EUV
Imaging Telescope (EIT) on board the SOHO spacecraft
\cite{Delaboudiniere:etal:1995}.

\begin{figure}
\centering
\includegraphics[width=0.65\hsize,angle=-90]{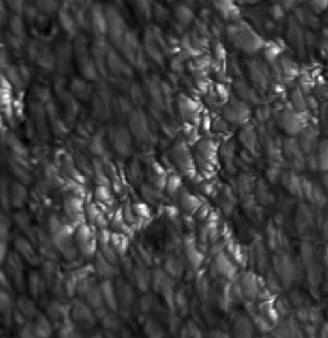}

\caption{\label{fig:11}G-band image of a plage region observed at an
inclination of $\cos\theta\approx 0.5$ with a spatial resolution of
about 70~km (a tenth of a second of arc) at the Swedish Solar
Telescope on La Palma (figure courtesy of V.~Zakharov).}
\end{figure}

Besides on the inclination, $\mu=\cos\theta$ (where $\theta$ is the
angle between the line of sight and the local vertical direction on the
Sun), on the size of magnetic features and on wavelength, the exact
value of the intensity contrast depends strongly on the spatial
resolution of the observations. At high resolution, (e.g., in the
G-band) the smaller magnetic features in the faculae break up into
strings of small bright points or elongated features lying in the dark
intergranular lanes. Moreover, at very high resolution, the size and
shape of the bright structures changes with $\mu$, so that it is even
unclear if the same magnetic features are seen as bright structures at
different $\mu$ values \cite{Lites:etal:2004}. There are indications
that bright features near the limb do not represent the same magnetic
structures as bright points at larger $\mu$ values
\cite{Ortiz:etal:2002,Topka:etal:1997}.  High-resolution observations
reveal that the tiny bright points disappear closer to the limb and are
replaced by larger bright `fans' that appear as if the walls of the
granules behind the magnetic feature had brightened
\cite{Lites:etal:2004,Hirzberger:Wiehr:2005} 
as shown in Figure~\ref{fig:11}. This figure also
suggests a 3-D structure of the visible solar surface, with the granules
lying higher than the intergranular lanes. The bright fan appearing
behind magnetic elements results from the larger transparency of the
latter in comparison to the non-magnetic atmosphere, so that the
radiation comes from slightly deeper (and thus hotter) layers of the
granule behind the magnetic element (see Sect.~\ref{subsec_models}).

\subsection{Models and numerical simulations}
\label{subsec_models}

Almost all models of magnetic elements or, more generally, magnetic
flux concentrations consider them to be some form of magnetic flux
tube. Strictly speaking, this term is restricted to axially symmetric
structures, but it is often used for magnetic flux concentrations in a
more general sense. An exception are magnetic features displaying
translational symmetry, which are sometimes also referred to as flux
slabs. In the following we will refer to magnetic flux concentrations
as flux tubes irrespective of their exact geometry. Basically, flux
tubes are bundles of nearly parallel field lines whose cross section
is bounded by a topologically simple closed curve. Often the field
lines are considered to lie almost parallel to each other (and almost
parallel to the `axis' of the tube), although structures with a
twisted field (sometimes called magnetic flux ropes) have also been
considered in many investigations.

Flux-tube models of many types are known, differing in their level of
generality.  An adequate working description of magnetic elements
that satisfies many currently available observations is provided by
the thin-tube approximation
\cite{Defouw:1976,Roberts:Webb:1979,Spruit:1981}.
A thin-tube model can be considered to be the solution that satisfies
the MHD equations to lowest order in an expansion in radius (distance
from the axis). Thus, the thin-tube approximation should be a good
assumption for the smallest magnetic elements, but an increasingly
poor representation of larger magnetic features.  The approximation
basically neglects magnetic curvature forces and the internal
structure of the flux tube in the horizontal direction. The lateral
force equilibrium is simply described by a balance of total pressure:
$p_{\rm e} = p_{\rm i} + B^2/2\mu_0$, with the external gas pressure,
$p_{\rm e}$, being equal to the sum of the gas pressure in the flux
tube, $p_{\rm i}$, and the magnetic pressure. Consequently, the
internal gas pressure of a flux tube with kG magnetic field in the
solar atmosphere reaches only 10--30~\% of the value in its
surroundings at the same height.

In its simplest form the thin-tube approximation neglects temporal
variations and, in addition to lateral pressure balance, assumes
hydrostatic equilibrium along the field lines. Consequently, only the
internal temperature profile as a function of height remains to be
specified (apart from the total magnetic flux and the value of the
field strength at an arbitrary height) in order to obtain a uniquely
defined model of a thin flux tube.  If the radial expansion is carried
to second order, the approximation remains valid for somewhat thicker
tubes \cite{Pneuman:etal:1986,Ferriz-Mas:etal:1989}. Other approaches
include combinations of a potential field with a boundary current
sheet \cite{Simon:etal:1983} or a full solution of the magnetostatic
equations for an axially symmetric flux tube \cite{Steiner:etal:1986}.

Comprehensive 3-D radiation MHD simulations (see below) suggest that
horizontal pressure balance is satisfied in the strong-field magnetic
features to a high degree, so that for many purposes a thin tube (or
slender slab) is a good approximation of magnetic elements. In
addition, models based upon the thin-tube approximation also satisfy a
number of observational constraints.  A major success has been the
explanation of the centre-to-limb variation of the continuum contrast
of faculae (at least as measured at a spatial resolution of the order
of a Mm on the Sun).  Owing to the strong evacuation of the tube, the
surface of optical depth unity (the visible surface) drops by
100--200~ km within the flux tube, similar to the Wilson depression in
sunspots. The temperature of the solar gas increases rapidly with
depth, so that the `side walls' of the flux tube between the external
and the internal levels of optical depth unity tend to be rather hot
and thus bright.  Since the depression of the visible surface is
largely independent of the flux tube size, the ratio of width to
depth of the depression depends strongly on the diameter of the
magnetic feature. This is illustrated in Figure~\ref{fig:7}. The hot
walls of narrow flux tubes are only seen if they are located
relatively close to the centre of the solar disc. At larger
inclinations, the backward wall gets hidden. For broader tubes this
happens only much closer to the solar limb
\cite{Spruit:1976,Knolker:Schussler:1988}. This simple geometrical
concept qualitatively explains the centre-to-limb variation of the
continuum contrast \cite{Ortiz:etal:2002,Topka:etal:1997}. The details
seen in the highest resolution data, such as the bright fan behind
magnetic elements near the limb require more detailed models
\cite{Keller:etal:2004,Carlsson:etal:2004,Steiner:2005}.

A number spectro-polarimetric observations can also be explained by
relatively simple models. As the gas pressure drops with height
approximately at an exponential rate, the magnetic field strength
drops with $\sqrt p$ and, to allow magnetic flux conservation, the
cross sectional area increases as $1/\sqrt p$. Observations of Stokes
$V$ profiles of Zeeman-sensitive spectral lines in the infrared and of
chromospheric lines are consistent with such gradients of the field
strength \cite{Briand:Solanki:1995}.  The field strengths observed in
lines formed at different heights also support this model
\cite{Ruedi:etal:1992,Bruls:Solanki:1995}.  The expansion of the flux
tubes, combined with their location in downflow lanes also provides a
natural explanation of the combination of asymmetric and unshifted
Stokes $V$ profiles \cite{GrossmannDoerth:etal:1988,Solanki:1989}.

The models discussed so far do not include an energy equation and thus
do not possess any predictive power for the thermal or dynamic
structure of magnetic elements.  In order to reproduce such
observations (e.g. line weakening or G-band contrasts) and to obtain a
better understanding of the physical processes acting in and around
magnetic elements, one has to carry out MHD simulations including a
proper description of radiative transfer.  Similarly, static or
stationary models have no predictive power regarding the high field
strengths of the concentrated magnetic features. In order to
understand the large observed values the formation process of
flux tubes must be modeled.

Closer to reality than the thin-tube approximation are 2-D MHD models of
flux slabs, either with \cite{Deinzer:etal:1984b,Deinzer:etal:1984a}
or without imposed mirror symmetry \cite{Steiner:etal:1998}. In the
former case, the solution quickly reaches a nearly stationary state,
in the latter case it turns out to be very dynamic, with the flux tube
swaying to and for as it is buffeted by the nearby granules.

3-D simulations of solar convection have been very successfully carried out
\cite{Stein:Nordlund:2000,Spruit:etal:1990}. These simulations
are fully compressible and take into account non-gray radiative transfer and
a realistic equation of state. They reproduce a wide variety of observations
\cite{Nordlund:1984a,Asplund:etal:2000}. Similarly realistic 3-D MHD 
simulations of magneto-convection have also been run for some time
\cite{Nordlund:Stein:1990,Bercik:etal:1998}. Such
simulations now  reproduce critical observations in detail 
\cite{Bercik:etal:2003,Voegler:etal:2003,Voegler:Schuessler:2003,
Vogler:etal:2005}.

\begin{figure}
\resizebox{1.0\hsize}{!}
{\includegraphics[width=0.3\hsize]{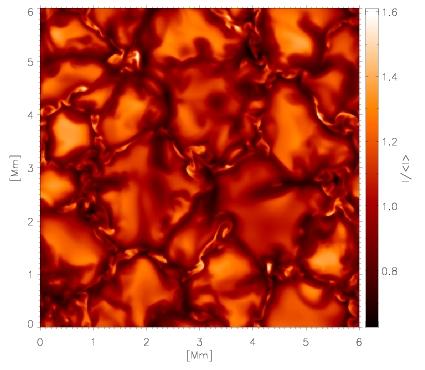}
 \hglue 0mm \includegraphics[width=0.3\hsize]{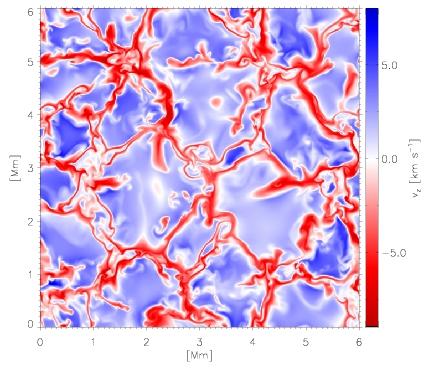}}
\resizebox{1.0\hsize}{!}
{\includegraphics[width=0.3\hsize]{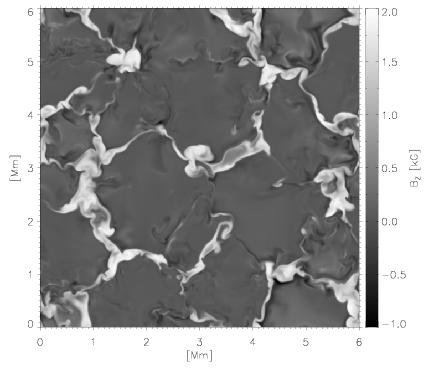}
 \hglue 0mm \includegraphics[width=0.3\hsize]{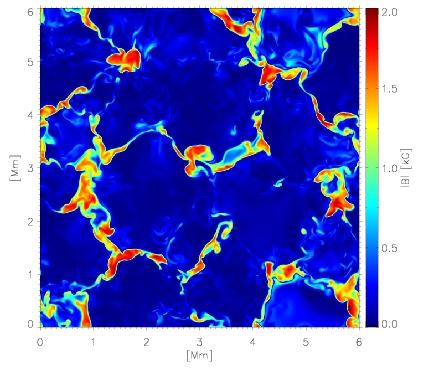}}
\caption{\label{fig:12}Snapshot of brightness (upper left),
vertical component of the velocity field (upper right), vertical
component of the magnetic field (lower left), and magnetic field
strength (lower right) at the height of the (corrugated) solar surface
deduced from a 3-D radiation MHD simulation covering a region of
6$\times$6~Mm on the solar surface
\cite{Vogler:etal:2005}.  The horizontally averaged vertical magnetic
field strength is 200 G. The correspondence between the color
scale and the physical quantity it represents can be deduced from the
colour bars on the right of each panel.}
\end{figure}

Figure~\ref{fig:12} shows a snapshot of a simulation which was started
with an initially homogeneous and unipolar field of 200~G
\cite{Voegler:etal:2003,Vogler:etal:2005}. The magnetic field rapidly
becomes concentrated in the dark downflowing intergranular lanes within
the convective turnover time of the granules. Values of over 2000 G are
commonly reached in the flux concentrations, in good agreement with the
observations. In addition to these strong fields, a weak field is also
found everywhere in the simulation box.  The probability distribution
function of the field strength has a peak at around zero (very weak are
present mainly above the granules) and drops off towards higher field
strengths.  If the magnetic flux in the domain is small (e.g., an
initial field of 10 G), then the field strength drops almost
exponentially (in agreement with observations in the infrared
\cite{Khomenko:etal:2003}), although flux concentrations with kG field
elements are still present. This corresponds to the situation in the
quiet Sun, which is best reproduced by a mixed-polarity field with about
20 G average vertical field strength \cite{Khomenko:etal:2005}.  As the
vertical magnetic flux in the simulation box is increased, a hump
appears in the distribution function of magnetic field strength at kG
field strengths. This indicates an increasingly large proportion of
strong fields, which are mainly found in horizontally elongated
structures located along the downflow lanes (see Figure~\ref{fig:12}).

At values of the average flux density at the solar surface beyond
roughly 200 G, the properties of the convection are strongly affected
by the magnetic field.  The granules become smaller and more stable
(compare with \cite{Title:etal:1989}), the downflow within the
magnetic features is quenched, with downflows now being concentrated
at their edges. This quenching of the downflow is clearly visible in
Figure~\ref{fig:12} at locations of strong field. Also well visible is
that the magnetic elements brighten, producing thin, bright ribbons
surrounded by the darker intergranular gas, very similar to what is
observed in high resolution continuum images.  This enhanced
brightness is caused by radiation flowing into the evacuated magnetic
feature from the sides (through the hot walls).  In realistic
simulations, the flow of the radiative flux is followed
\cite{Vogler:etal:2005} and shown to behave according to the hot-wall
model \cite{Spruit:1976}.

The (Wilson) depression of the visible surface in small magnetic flux
concentrations leads to an increase in the solar surface area, so that
the elements act like leaks in the solar surface and lead to a
brightening of the Sun as a whole. This excess radiation extracts
energy from the gas in the convection zone. Owing to the high
effective thermal conductivity brought about by convective mixing, a
part of the deficit is distributed through the whole convection zone,
so that a net increase in the total solar energy flux results
\cite{Spruit:1982a,Spruit:1982b}.

The gas in the magnetic features is not entirely optically thin and a
part of the strong influx of radiation into a magnetic element is
absorbed, especially in the mid-photospheric layers. The absorbed
radiation heats up the gas in the magnetic element, so that at equal
optical depth the magnetic features become hotter than the
surroundings, leading to the ionization of neutral species (i.e. to
the weakening of lines of neutral atomic species) and the dissociation
of molecules.  This qualitatively explains the enhanced brightness of
magnetic elements at particular wavelengths (e.g.\ in spectral line
cores) and in particular spectral bands (G-band and other spectral
bands populated by molecular spectral lines).

Quantitative and stringent tests of the simulations can be made by
computing spectral lines or whole spectral regions for each horizontal
grid point of the simulation domain. In the case of the high contrast
G-band, approximately 300 CH and atomic lines are computed for each
horizontal grid point. The similarity with properties of observed
features is striking \cite{Schussler:etal:2003,Shelyag:etal:2004}. The
bright fan-like structures seen in facular regions near the limb are
also well reproduced by such simulations
\cite{Keller:etal:2004,Carlsson:etal:2004}. This 
suggests that the simulations contain much of the relevant
physics necessary to describe the thermal structure of the
photospheric layers of magnetic elements.  However, more tests, i.e.\
comparisons with observational data, in particular with sensitive
spectro-polarimetric diagnostics, are needed in order to probe other
aspects of the simulations.

The simulations show the ceaseless movement of the magnetic features as
they are buffeted around by the granulation. The concentrated
magnetic flux moves along the downflow lanes, carrying out something
close to a random walk as granules are born, evolve, and disappear. It
becomes difficult to identify individual magnetic features and, in
particular, to determine lifetimes, since there is a constant transfer
of flux between them as concentrations of magnetic flux wax and
wane. If features of opposite polarity are close together, these
dynamics leads to flux cancellation and to a steady reduction of the
magnetic flux in the simulation box, if no additional flux enters it
from outside.

Considering that the photospheric magnetic field forms the footpoints
of the coronal loops, one of the effects of this permanent motion is
that the field lines in the corona, which are pulled along by the
dynamic photospheric field, become braided. As a consequence,
tangential discontinuities and current sheets form, where the stored
magnetic energy  can be released or dissipated
\cite{Parker:1983b,Parker:1988,Gudiksen:Nordlund:2002} (see
Sect.~\ref{sec_corona}).  It is evident from the simulations that
spatial scales below even the best resolution reachable by current
observations play an important role in this process.

\subsection{Oscillations and waves}

Magnetic elements support various wave modes, most of which have
mainly been studied theoretically.  The observational evidence
for flux-tube waves has been limited to low amplitude oscillations,
mainly at a period of five minutes
\cite{Giovanelli:etal:1978,Fleck:etal:1993}, although in one case a shorter
period oscillation has been reported \cite{Volkmer:etal:1995}. It has
been argued that the observational evidence is so limited because the
magnetic features are spatially unresolved. If multiple magnetic
features, each supporting an oscillation at a different phase, are
present in the resolution element, then the various oscillation
signals at least partly cancel.  The interest in the study of such
waves is driven by their potential contribution to chromospheric and
coronal heating.

Basically, there are three wave modes in a slender flux tube: a slow
magnetosonic mode called the tube wave, a transversal kink wave, and a
torsional Alfv\'{e}n wave. In a flux tube with sound speed $c_{\rm s}$,
Alfv\'{e}n speed $c_{\rm A}=B/\sqrt{\mu_0\rho_{\rm i}}$, internal gas
density $\rho_{\rm i}$, and external density $\rho_{\rm e}$, the three
modes travel at the tube speed $c_{\rm t}$, the kink speed, $c_{\rm k}$,
and the Alfv\'{e}n speed
\cite{Defouw:1976,Spruit:1981},
respectively, with

\begin{equation}
c_{\rm t}^2 = {{{c_{\rm s}^2 c_{\rm A}^2}}\over 
              {c_{\rm s}^2 + c_{\rm A}^2}},\quad
{c_{\rm k}^2} = 
  \left( {{\rho_{\rm i}}\over {\rho_i +\rho_{\rm e}}}\right) c_{\rm A}^2\;.
\end{equation}

The Alfv\'{e}n wave has no cutoff, so that any torsional oscillation
excited in a flux tube propagates while kink and longitudinal modes
propagate only at frequencies above their respective cutoff values
\cite{Spruit:1981,Roberts:Ulmschneider:1997}. Studies of the
excitation of such waves generally concentrate on the effect of
stochastic perturbations in the surroundings of the flux tube
\cite{Musielak:Ulmschneider:2002,Musielak:Rosner:1987}, which is
similar to the excitation mechanism for the solar global oscillations.
In both cases, the perturbations are due to convection. The waves
differ in their propagation characteristics. The longitudinal tube
waves are affected by radiative damping \cite{Webb:Roberts:1980} and
easily form shocks with energy dissipation
\cite{Herbold:etal:1985,Fawzy:etal:2002a}. The kink modes are far less
susceptible to damping and dissipation, while Alfv\'{e}n waves are not
damped and easily reach the corona.  Most accessible to observations are
the longitudinal waves
\cite{Solanki:Roberts:1992,Ploner:Solanki:1997,Ploner:Solanki:1999}
and these are the only ones for which positive observational
detections exist
\cite{Volkmer:etal:1995,Fleck:etal:1993,Giovanelli:etal:1978}. In
order to detect the other wave modes, the flux tubes need to be
resolved near the solar limb, which is even harder than at disk centre
since due to foreshortening even more magnetic elements are present in
a given angular resolution element.  In the case of torsional waves,
individual magnetic elements need to be completely resolved.

\subsection{Internetwork and turbulent fields}
\label{subsec_internetwork}

Magnetic fields in the interiors of supergranules outside active
regions were dubbed internetwork fields, in order to distinguish them
from the more readily detectable network fields
\cite{Livingston:Harvey:1975,Livi:etal:1985,Zirin:1985}.
The magnetic flux of a typical internetwork feature is about
10$^{16}$~Mx, which is 1--2 orders of magnitude smaller than the flux
of a typical magnetic element in the network. However, the emergence
rate of internetwork elements is so large that the total magnetic flux
emerging in the form of the internetwork field per unit of time is
estimated to be two orders of magnitude larger than the rate of flux
emergence in the small (ephemeral) bipolar regions that regenerate the
network \cite{Hagenaar:etal:2003}.  Internetwork elements emerge inside
the supergranules and move with the horizontal flow to the downflow
lanes, where they merge or cancel with other internetwork features or
with network elements
\cite{Zhang:etal:1998}. They have an estimated average lifetime of
around 2 hours, which is an order of magnitude lower than the
estimated lifetime of the network flux.

A controversy surrounds the intrinsic field strength of internetwork
features.  Whereas observations of infrared lines indicate low
intrinsic field strength of below 500~G
\cite{Lin:1995,Khomenko:etal:2003}, recent measurements using visible
lines \cite{Dominguez:etal:2003a,Dominguez:etal:2003b} suggest that the
field strengths are mostly in the kG range or are a mixture of weak and
strong fields \cite{Socas:etal:2004}.  It is at present unclear whether
both types of observations are simply showing fields populating
different parts of a single, broad field-strength distribution
\cite{BellotRubio:Collados:2003}, or if there is a problem with one of
the diagnostics.

Besides the relatively well studied internetwork fields a truly
turbulent component of the field has long been proposed. Field lines
belonging to this component are tangled on a very small scale,
reaching to below the spatial resolution of current
observations. While initial searches using the Zeeman effect remained
unsuccessful in detecting such a field, they did provide upper limits
\cite{Stenflo:Lindegren:1977}. More
recent work has concentrated on using the Hanle effect, which can detect
weak fields even in the presence of an isotropic distribution of
magnetic field directions \cite{Stenflo:1982,
Faurobert-Scholl:etal:1995, Faurobert:etal:2001, Stenflo:etal:1998,
Berdyugina:Fluri:2004}.  These measurements suggest that a turbulent
field of a few tens of Gauss may be present, although the exact value
(lying between 10 and 60 G) is still unclear. Recently, it has been
proposed on the basis of the Hanle effect of atomic and molecular lines
that even much more flux is hidden in the form of a turbulent field
\cite{TrujilloBueno:etal:2004}.

The origin of this turbulent field is at present completely open, with
the same mechanisms being discussed as for the internetwork field. The
MHD simulations indicate that the distinction into different types of
magnetic features is artificial and they all are part of a single
distribution of field strengths. As the average field strength
increases, the spatial distribution and associated properties of the
field gradually change, from a turbulent field to flux tubes.

\subsection{Influence of magnetic features on solar irradiance}
\label{subsec_influence}

\begin{figure}
\centering
\includegraphics[height=8cm]{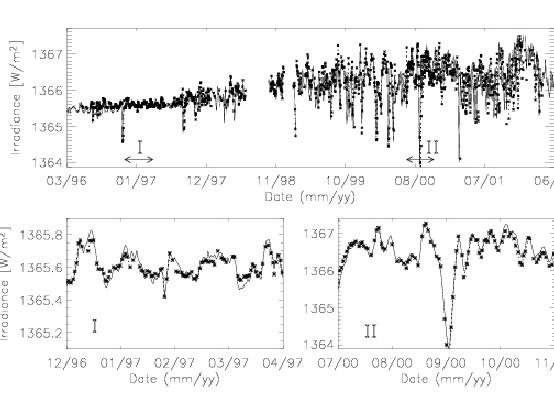}
\caption{\label{fig:13}Total solar irradiance measured by the 
VIRGO instrument on SOHO (solid line) and synthesized from models of
solar magnetic features and MDI magnetograms (stars).  The top panel
shows the period between 1996 and 2002, while the lower panels show
shorter stretches on an expanded scale (from \cite{Krivova:etal:2003}.)}
\end{figure}

The total irradiance of the Sun (i.e., its radiative flux integrated
over wavelength, as measured above the Earth's atmosphere), is known
to vary on all time scales observed so far \cite{Frohlich:2003}. At
the short end of the scale, at around minutes to hours, the
variability is dominated by the convection (mainly granulation) and
the acoustic oscillations, which contribute mostly at around 5
minutes.  At longer time scales reaching up to the solar cycle
different causes of the irradiance have been proposed. These include
changes in the Sun's internal thermal structure, $R$-modes
\cite{Wolff:Hickey:1987}, a toroidal, subsurface field
\cite{Kuhn:Stein:1996,Sofia:Li:2000} and  the magnetic field
at the solar surface \cite{Chapman:1987,Foukal:1992}. The variability
of the surface magnetic features as the dominant source of the
observed irradiance variations has been steadily gaining ground and is
now well established. The most developed models include the effects of
both dark sunspots (and pores) as well as of bright magnetic elements.
The spectra resulting from empirical model atmospheres describing each
type of feature are computed and the spatial locations of the features
(taken from magnetograms and continuum images) are also employed to
synthesize the irradiance.  The resulting synthetic irradiance agrees
very well with the observed value \cite{Krivova:etal:2003,
Ermolli:etal:2003} (see
Figure~\ref{fig:13}).

These investigations suggest that the surface magnetic field is
responsible for over 90\% of the Sun's total irradiance variations on
time scales up to the solar cycle. Such models also explain the
apparent paradox that the Sun is brightest at activity maximum when
the number of dark sunspots its surface is largest. The magnetic
elements forming faculae, each of which is rather inconspicuous in
white light compared to sunspots, together cause a larger and, in
particular, longer lasting brightening than sunspots do.  This has
partly to do with the fact that faculae cover a considerably larger
area than sunspots. Also, the magnetic field composing sunspots decays
into smaller elements, producing bright facular regions, which can
survive considerably longer than the original sunspot. Finally, much
of the excess emission from magnetic elements comes from the upper
layers of the photosphere (see Sect.~\ref{subsec_proxies}) and thus
mainly contributes in the UV spectral range. Hence a significant
fraction of the facular contribution to total irradiance is not
visible to detectors with a wavelength sensitivity similar to the
human eye.

\section{The magnetic field in chromosphere and corona}
\label{sec_corona}

%

The phenomena observed above the solar surface are so rich in
variety (and so are the names given to them) that here we can only
mention the most common and most spectacular. Due to its
dominance, the magnetic field in almost all cases is the key to the
physics behind them.

The study of the magnetic field in the atmospheric layers above the
photosphere however suffers from the fact that although it here exhibits a
dominant influence it is difficult to measure.
We will first review briefly from which techniques we draw our present
knowledge about the chromospheric and coronal magnetic field, we then
give a phenomenological overview of some typical structures inferred
from these observations and discuss at the end some more fundamental
aspects.

\subsection{Measurement and modeling of the magnetic field}
\label{sec:measure+model}

Imprints of the magnetic field above the solar surface are usually faint
and can sometimes be observed only indirectly.

At radio frequencies below about 3 GHz thermal bremsstrahlung from
free electrons in the upper chromosphere and lower corona can be
observed by ground-based radio antennas.
This radiation propagates in the source region at a frequency above
the local plasma frequency in one of the two natural electromagnetic
wave modes, the o- or x-mode, depending on its sense of circular
polarization with respect to the local magnetic field direction.
As the two modes are differently absorbed, the radiation which finally
escapes the birefringent corona retains a finite polarization.
The degree of polarization is to lowest order proportional to the
magnetic field strength along the propagation direction in the
source region \cite{Gelfreikh:1994}.
Above active regions and sunspots with a strong magnetic field, the
3$^{rd}$ (occasionally even the 2$^{nd}$) harmonic of the electron
gyrofrequency exceeds the local plasma frequency and the corresponding
gyro emission can escape the corona. The observed gyro harmonic
frequencies are typically above 1\;GHz and are directly proportional to
the magnetic field strength in the source region (e.g.,
\cite{White:Kundu:1997}).

At optical frequencies the emission lines from coronal ions are
usually very faint because the plasma density is low and the
emissivity roughly scales with the square of the density. In addition,
the lines are extremely broadened due to a coronal temperature beyond
10$^6$\;K for the emitting ions.
The Zeeman effect is therefore difficult to detect
and to be analyzed quantitatively to yield estimates for the
magnetic field strength.
The conditions are more favorable at chromospheric heights (e.g.,
\cite{Metcalf:etal:1995}) or in prominences (e.g.,
\cite{Bommier:Leroy:1998}) where the plasma is cooler and more dense.

The possibilities to resolve the Zeeman line shift also improve at infrared
wavelengths because the line splitting increases with $\lambda^2$
while the temperature broadening increases only linearly with the
observing wavelength $\lambda$.
For example, combined Zeeman and Hanle effect measurements of Ca I
emissions at $\lambda$ = 850 and 854\;nm have been made above sunspots
\cite{SocasNavarro:etal:2000} and of the He I
line at $\lambda$ = 1083\;nm above active regions
\cite{Solanki:etal:2003,Lagg:etal:2004}.
Form these spectro-polarimetric measurements the full magnetic field
vector could be retrieved in the low chromosphere and near the base of
the corona, respectively.

Successful attempts to detect the magnetic field in the hot corona
above the limb have for a long time been restricted to the observation
of resonance scattering of the forbidden line $\lambda$ = 530\;nm
emitted by Fe XIV (e.g., \cite{Arnaud:1982a}).
The radiation is excited at the transition from an excited state
with a life time which greatly exceeds the Larmor period imposed
by the excited state's magnetic dipole moment.
The observed orientation of the polarization axis of the scattered radiation
then relates to the orientation of the coronal magnetic field perpendicular
to the line-of-sight.

By a careful analysis of the Hanle effect together with Doppler shifts
experienced by solar wind ions on the resonantly scattered O VI line
at $\lambda$ = 103.2\;nm
the magnetic field above the pole can be constrained to within a few G
\cite{Raouafi:etal:2002}.
The longitudinal Zeeman effect has only recently been successfully
detected in the corona, 0.15\;R$_{\odot}$ above the limb for infrared
emission lines from Fe XIII at $\lambda$ = 1075 and 1080\;nm
\cite{LinH:etal:2000}. 
Other infrared lines which are useful for coronal Zeeman and Hanle
effect observations are currently investigated
\cite{Judge:1998,Kuhn:etal:1999a,Judge:etal:2002}.

In spite of this recent progress our present knowledge of the magnetic
field is still largely based on extrapolations from photospheric
magnetograms where Zeeman and Hanle effect measurements are often
routinely made from ground based observatories (e.g.,
the Wilcox Solar Observatory, WSO, near Stanford,
the Synoptic Optical Long-term Investigation of the Sun, SOLIS, at Kitt Peak,
the Vacuum Tower Telescope, VTT, and the
Heliographic Telescope for Solar Magnetic and Instability Studies, THEMIS,
on Teneriffe)
and from space
(Michelson Doppler Imager, MDI, on board the SoHO spacecraft and
the facilities on the upcoming Solar-B satellite).

Quasi-static extrapolations of the photospheric field into the corona
are usually justified by the drastic change of the role the magnetic field
plays in the dynamics of the photosphere on the one hand and in the
corona on the other hand.
The magnetic flux which emerges from the solar surface is
tightly anchored in the highly conducting, massive and high $\beta$
photosphere and is there pushed around at a typical speed of 1\;kms$^{-1}$.
In the low $\beta$ corona magnetic forces dominate and force
imbalances are transmitted along the magnetic field with an Alfv\'en
speed of about 1000\;kms$^{-1}$.
The line-tied coronal field therefore immediately adjusts to the
boundary flux imposed in the photosphere while the photosphere hardly
reacts due to its large inertia to occasional rapid processes in the
corona.
However the height region sandwiched between photosphere and corona,
the chromosphere and transition region is much more complex than many
of the extrapolation models suggest and the impact of the Sun's lower
atmosphere on these models has not yet been fully assessed.

To accomplish the extrapolation, different approximations to the full
set of MHD equations governing the magnetic field in the corona are used.
In its simplest form, the corona is assumed to be current-free and Gauss'
theorem is used to obtain a potential field (or Laplace field) approximation
of the coronal magnetic field which matches the observed normal component
on the solar surface and some reasonable outflow boundary condition which
mimics the solar wind impact on the field at a distance of
1-2\;R$_{\odot}$ above the solar surface (e.g., \cite{Altschuler:Newkirk:1969})

In the low corona, the small value of $\beta$ allows us to neglect
pressure and gravity forces to lowest order, so that in a stationary corona
the Lorentz-force $\vect{j}\!\times\!\vect{B}$ must approximately
vanish. This requires $\vect{j}=\alpha\vect{B}$ for some yet unknown
scalar $\alpha$.
The vanishing divergence of the current density $\vect{j}$ has the
consequence that $\alpha$ must be constant along any given field line,
$(\vect{B}\cdot\vect{\nabla})\,\alpha$ = 0.

The coefficient $\alpha$ has the dimension of an inverse length and
$1/\alpha$ can be visualized as the distance over which coronal currents
induce perturbations in $\vect B$ of the same order as the potential field
strength.
Values of $|\alpha|\lesssim$10\;R$_{\odot}^{-1}$ have been inferred
from, e.g., vector magnetograph measurements \cite{Hagyard:Pevtsov:1999,%
BaoS:ZhangH:1998}.

The force-free approximation however it is intrinsically nonlinear and if
based on boundary information alone, its calculation is a highly
ill-posed problem (see discussion in, e.g., \cite{Demoulin:etal:1992,%
McClymont:etal:1997}).
For this reason, a ``light'' force-free version is often employed
where $\alpha$ is reduced to a global constant (e.g., \cite{Seehafer:1978};
so-called linear force-free (or Taylor field, \cite{Taylor:1974}) which
has its own physical significance as will be discussed in chapter
\ref{sec:energy+helicity} below).
Besides, it is not quite obvious which boundary data is necessary and
sufficient for a unique nonlinear force-free field extrapolation.
The invariance of $\alpha$ along the field lines, for example, has
the consequence that the overall flux conservation
$\int_S B_r d^2x$ = 0 normal to the solar
surface $S$ is replaced by the much stricter requirement that for every
value of $\alpha$ the differential flux $\int_{dS(\alpha)} B_r d^2x$
must vanish. Here $dS(\alpha)$ is the differential subsurface of $S$
where $\alpha$ assumes a given value.
Even more constraints exist (see, e.g., \cite{Aly:1989,Sakurai:1989})
and various numerical methods have been proposed to solve the force-free
field boundary value problem \cite{Aly:Seehafer:1993,Roumeliotis:1996,%
Amari:etal:1997,Amari:etal:1999a,Wheatland:etal:2000,Yan:Sakurai:2000,%
LiJQ:etal:2003}.

Full MHD models are required if the outer parts of the corona are to
be included where $\beta$ approaches unity. Quasistationary solutions
can either be obtained by relaxation \cite{Linker:etal:1999,%
Mikic:etal:1999} or from solving a variational problem
(e.g., \cite{Wiegelmann:Inhester:2003}).

\begin{figure}[t]
\centering
\includegraphics[width=0.7\hsize]{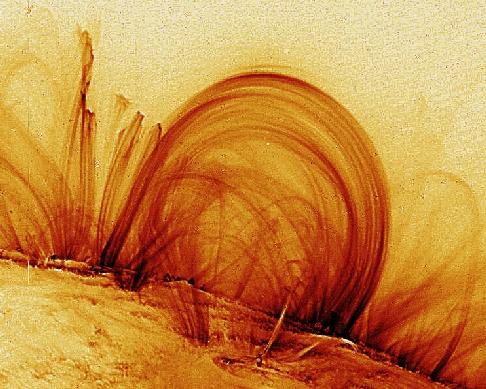}
\caption{\label{Fig:loops}
  Highly structured loops above the limb taken in EUV from the TRACE
  satellite. The loops could be viewed as materialized magnetic field lines.
  The TRACE telescope today yields the highest spatial resolution for
  these type of images.}
\end{figure}

With the advent of space missions like Yokhoh, the Solar and
Heliospheric Observatory (SoHO) and the Transition Region and Coronal
Explorer (Trace) high resolution images with high cadence rates became
available from coronagraphs, UV and X-ray cameras.
Their data show the radiation from plasma which has risen from hot spots
in the chromosphere upwards along individual field lines and thus trace
out thin magnetic flux tubes in the corona . As an example Fig.~\ref{Fig:loops}
displays loops above the limb observed by the TRACE satellite.
The images of these loops therefore tell us much about the topology of
the coronal field and the changes it undergoes by quasi-static
line-tying and also when the field configuration becomes eruptive (see
the recent review \cite{Aschwanden:etal:2001a}).
The image data now are available with both high spatial and temporal
resolution so that quantitative 3D reconstructions from the observed
2D projections are being attacked to verify magnetic field
extrapolations from surface magnetograms.

\subsection{Chromospheric network, canopy and carpet}

\begin{figure}[h]
\centering
\includegraphics[width=0.7\hsize]
                {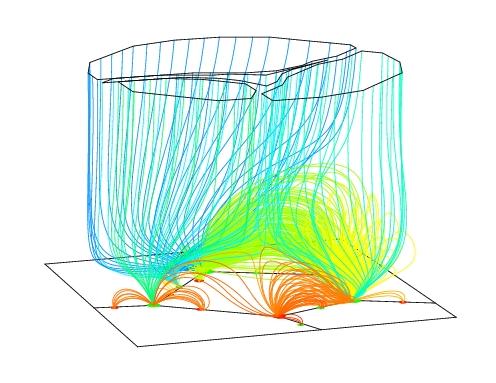}
\caption{\label{Fig:canopy} Magnetic field above a set of magnetic
  elements located at the edge of supergranule cells (lines on the
  solar surface). Part of the field lines open to the corona while
  the carpet field connects elements of different polarity flux on
  the surface. The field lines are calculated from a potential
  field model.}
\end{figure}

As described in chapters \ref{sec_largescale} and
\ref{sec_small-scale}, the magnetic flux which emerges from the
photosphere is not distributed smoothly over the solar surface but
highly concentrated in small flux elements.
The upper photosphere and chromosphere form a relatively cool layer up
to a height of several 1000\;km above the visible surface where
the plasma density and pressure decrease with a scale
height of only a few 100\;km.
As a result the flux elements in turn widen their horizontal cross
section (see Fig.~\ref{fig:8})to keep the pressure balance with the
surrounding plasma. Eventually, individual flux tubes are bound to
merge with flux of equal polarity or bend into magnetic arches to
connect to flux elements of opposite polarity as sketched in
Fig.~\ref{Fig:canopy}.

The horizontal fields at the bottom of the `wine glasses' and
of the lowest arches and formed this way above the
supergranulation cells constitute the magnetic canopy. Observations
have shown that in the quiet chromosphere the flux tube expansion is
relatively gentle, the filling factor is small so that the flux tubes
merge only in the chromosphere and form a canopy base at about 1000\;km
above the photosphere.
In active regions the magnetic filling factor is enhanced
beyond 50\% in the upper photosphere \cite{Giovanelli:Jones:1982}.
Consequently, neighboring flux tubes merge already in the photosphere
\cite{Bruls:Solanki:1995}.
In sunspots, this expansion is so rapid, that the
formation of a layer of horizontal fields occurs already in the mid
photosphere.

Some observations yield evidence
for a particularly rapid expansion of the magnetic field in the lower
chromosphere leading to the formation of a nearly horizontal field
canopy at a height of roughly 800\;km above $\tau$=1, cf.
\cite{Jones:Giovanelli:1983,FaurobertScholl:1994,Bianda:etal:1998,%
Bianda:etal:1999}.
It has been shown that magnetic canopies
at these heights can be produced if the chromosphere is very
inhomogeneous in temperature, with hot flux tubes and cool gas in
between them \cite{Solanki:Steiner:1990}.
The necessary inhomogeneity of the chromospheric plasma
was proposed by \cite{Ayres:Testermann:1981,%
Ayres:etal:1986} and {\cite{Solanki:1994}.
Recent observations of CO molecular emissions in the infrared
seem to confirm this highly structured lower chromosphere \cite{Ayres:2002}.
The plasma $\beta$ is just at the verge between values $\gg$ 1
in the photosphere to $\ll$ 1 in the upper chromosphere. Hence,
the canopy fields may have some influence to thermally insulate
the plasma and to sustain the temperature inhomogeneity.

Magnetic arches above the canopy up to the height of the transition
region typically span larger horizontal distances of several 10$^4$\;km and
more. These field structures make up the magnetic carpet
\cite{Title:Schrijver:1998}.
Again, temperature and plasma density and flow vary enormously among
these flux tubes. Cool chromospheric material and hot coronal plasma may
lie side by side with neighboring magnetic arches separated only by
some 10$^3$\;km \cite{BergerTE:etal:1999}. 
Images taken in H$\alpha$ disclose the horizontal direction of these arches
and hence of the predominant horizontal field direction in this height
by the orientation of fibrils at the edges of plages and in the
neighborhood of filaments. In these images fibrils are visible as elongated
dark threads (see Fig.~\ref{Fig:Hafibrils}).

The distinction between the lower chromospheric
internetwork plasma, the canopy as the lowest magnetic layer and the
carpet field above is not very precise and has been critizised
\cite{Schrijver:Title:2003} since even the
internetwork region is not entirely field free (see chapter
\ref{subsec_internetwork}). Hence, with smaller flux elements
being taken account of, Fig.~\ref{Fig:canopy} would have to be extended
by a large number of low flux magnetic connections to
the surface area inside the supergranulation cells.

\begin{figure}
\centering
\includegraphics[width=0.7\hsize]
                {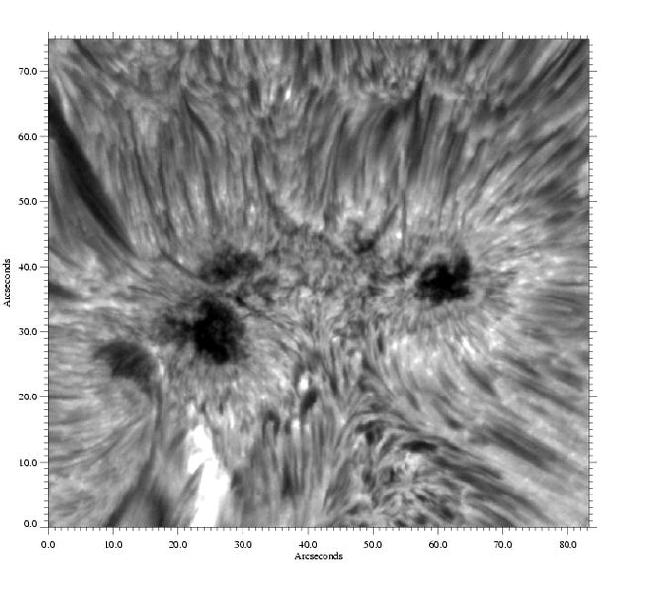}
\caption{\label{Fig:Hafibrils} H$\alpha$ image taken by
   B. de Pontieu with the Swedish Vacuum Solar Telescope (SVST).
   The dark elongated fibrils are radially arranged around a group
   of small sunspots indicating strongly diverging field lines
   emanating from the sunspots}
\end{figure}


\subsection{Coronal holes and plumes}

\begin{figure}[t]
\centering
\includegraphics[width=0.7\hsize]
                {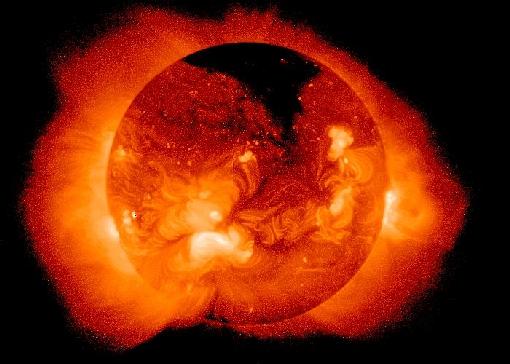}
\caption{\label{Fig:XraySun}
Bright magnetically closed coronal regions in an X-ray image of the Sun from
the YOHKOH spacecraft. The dark polar regions are coronal holes from which
the magnetic field escapes into the heliosphere.}
\end{figure}

Coronagraph observations reveal a striking inhomogeneity of the
Sun's corona and demonstrate the enormous influence the magnetic
field has in structuring the Sun's atmosphere. An extended part of
the solar surface, about 20-30\%, is covered by coronal holes.
These are surface regions which are magnetically connected to the
outer heliosphere \cite{Altschuler:etal:1972,Munro:Withbroe:1972,%
Cranmer:etal:1999,Belenko:2001,WangYM:Sheeley:2002}.
A local dominance of photospheric flux elements of one polarity
provides a net field of about 10 G or more at the coronal base which
feeds the open flux.

There is no containment of plasma possible on these open field lines
but the plasma is blown away from the Sun along the open field lines
to feed the fast solar wind \cite{WangYM:etal:1996}. Already down at
the coronal base blue shifts of as much as $\simeq$ 3\;kms$^{-1}$ have been
observed \cite{Wilhelm:etal:2000}.
The plasma density on these field lines is therefore reduced by almost
an order of magnitude compared to regions in the corona which are
permeated by magnetically closed field lines \cite{Sittler:Guhathakurta:1999}.
The reduction in density reaches down to the base of the corona. Since
emission intensities in X-ray and EUV lines roughly scale with the
plasma density squared, coronal holes are clearly distinguishable by
the lack of X-ray and EUV emission (see Fig.~\ref{Fig:XraySun}).

The flux imbalance in a coronal hole not only controls the net
magnetic flux that can effectively penetrate the chromosphere and
transition zone
but also influences the height, length and hence temperature
of the resulting closed field arcades.
The larger and hotter loops are found strongly underrepresented
in the coronal hole relative to the quiet Sun, while shorter and cooler
loops are almost equally present in both regions. This additionally
contributes to the darkening of coronal hole areas
\cite{Wiegelmann:Solanki:2004}.
Only for emission lines formed at temperatures below about 7\;10$^5$\;K,
the intensity difference between coronal holes and the closed-field quiet
Sun regions appear less pronounced \cite{Stucki:etal:2000,Stucki:etal:2002}.

Another indicator of coronal hole regions on the solar surface are
observations of the chromospheric He~I 1083\;nm line made regularly at
Kitt Peak Observatory. This line appears brighter in coronal holes than
on closed field line regions. The physics behind this radiation process
is, however, very complex. 

The shape and distribution of coronal holes on the Sun's surface
changes drastically during the 11-year's activity cycle. At activity
minimum coronal holes are concentrated at high latitudes and cover,
with opposite magnetic polarity, the polar caps of the Sun almost entirely down
to about a latitude of 50$^{\circ}$ - 60$^{\circ}$. The coronal whole
then occupies about 20\% of the solar surface with an average magnetic
flux density of 5-10 G \cite{WangYM:Sheeley:2002}. The latitudinal
average flux density below 50$^{\circ}$ cancels to less than a Gauss
\cite{WangYM:Sheeley:1995a}.

At activity maximum, coronal holes are distributed in patches all over
the solar surface, sometimes linked by ``narrow coronal hole
channels'' to one of the polarities of a larger bipolar region.
These channels sometimes extend over a good fraction of $R_{\odot}$
and may also cross the equator.
Relations between the coronal hole shape and the lowest multipole moments of
the magnetic surface field during solar maximum have been derived by
\cite{Sanderson:etal:2003} 

A large fraction of the open magnetic flux at these times escapes from
active regions and their neighborhood. The enhanced field strength in
these areas is compensated by a smaller overall coronal hole area of
only 5-10\% of the solar surface \cite{WangYM:Sheeley:2002}, so that the
open flux is remarkable stable over the solar cycle.

\begin{figure}[t]
\centering
\includegraphics[width=0.8\hsize]
                {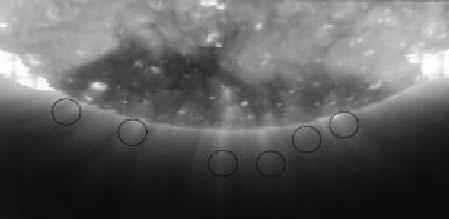}
\caption{\label{Fig:plumes}
  Polar plumes over the Sun's southern coronal hole during times of
  low solar activity. The almost radially extended density striations
  outline the direction of the open magnetic field from the south pole.
  (from \cite{DeForest:etal:1997}).}
\end{figure}

At solar minimum some magnetic flux tubes which emerge out of a
coronal hole are visible against the dark background sky
in coronagraphs or EUV images as polar plumes. These are
occasional field aligned plasma striations with a life time of several
days. The photospheric origin of plumes is newly emerged small bipolar
flux \cite{WangYM:Sheeley:1995b} which reconnects with the background coronal
hole field. The reconnection seems to provide the heating needed to support
the enhanced density inside the plume flux tube.
Enhancements by a factor 5 and more with respect to the coronal hole
background density have been observed \cite{Wilhelm:etal:1998}.
The plasma outflow speed into the solar wind from plumes, however, 
seems to be reduced \cite{DelZanna:etal:1997,Wilhelm:etal:2000}.

\subsection{Active regions and loops}

\begin{figure}
\centering
\includegraphics[width=0.65\hsize]
                {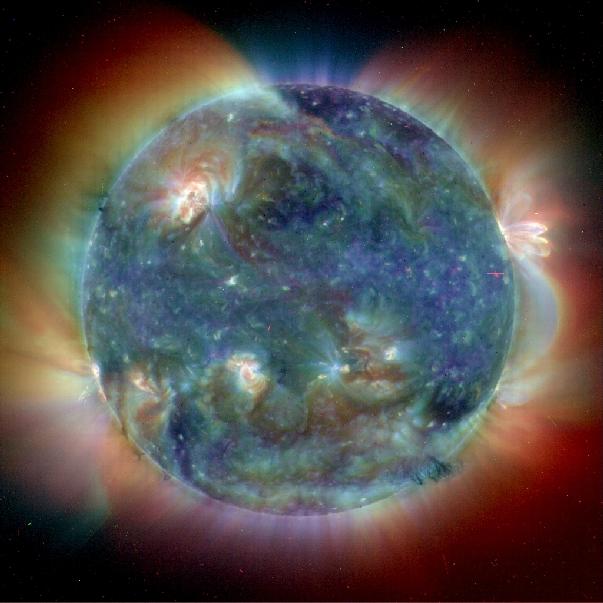}
\caption{\label{Fig:eit_sun} Bright active regions in the EUV image
  taken by the EIT instrument of the SoHO spacecraft. The image is
  actually a composite of images taken at three different wavelengths
  (17.1, 19.5 and 28.4 nm) and superposed in blue, green and red.
  The solar surface background appears dark because the continuum
  radiation from the cool surface at these wavelengths is well below
  the emission from the hot coronal loops. On the Sun's limb active
  regions loops are distinctly visible.}
\vspace{7mm}
\centering
\includegraphics[width=0.75\hsize]
                {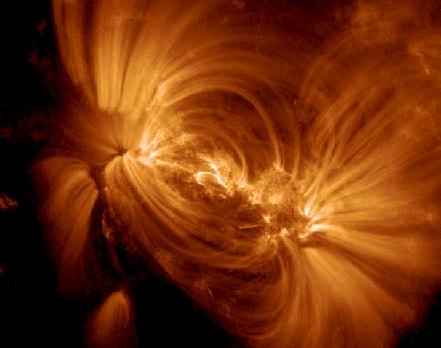}
\caption{\label{Fig:active} Details of bright loops above an active
  bipolar region seen in EUV wavelengths by the TRACE spacecraft.}

\end{figure}

The brightest objects in EUV and soft X-rays images of the Sun
are active regions. They are associated with strong magnetic
bipolar regions on the Sun's surface.
These regions form within a few days as massive amounts of magnetic
flux break through the solar surface. They are irregularly shaped,
sometimes arranged in groups but they are always arranged in the
same longitudinal order (``Hale's Law'', see \ref{subsubsec_rules}).
Magnetic flux densities of up to 100~G and more can be reached in the lower
corona above a strong bipolar region. The lifetime of active regions
may extend over many solar rotations though they start to diffuse
after one or two weeks which gently reduces the flux density and the
overall (absolute) flux that can build up in the lower corona.
Differential rotation together with the diffusion also causes a
characteristic deformation of the active region towards the end
of its life time (see Fig.~\ref{fig:magnetograms}).

Flux tubes emerging from an active region are filled with plasma which
is markedly hotter than elsewhere in the Sun's atmosphere and makes them
distinctly visible in EUV images (Fig.~\ref{Fig:active}) as loops which
magnetically connect the opposite polarities of the bipolar region.
Which physical mechanism provides the heating energy is still an active area
of research. The enhanced density scale height on flux tubes rooted in
active regions makes them distinctly visible in EUV and soft X-rays and
also in optical emissions from ions which form at temperatures well
above 10$^6$\;K.
An example is the forbidden green corona line at $\lambda$ = 530\;nm
emitted from Fe XIV. Positive correlations between the emissivity at
various wavelengths and the field strengths at the foot points of the
respective field line have been derived by several authors
\cite{Schrijver:1991,Benevolenskaya:etal:2002,WangYM:etal:1997}.
Hence there is strong evidence that the magnetic field is an important
ingredient in the heating mechanism for coronal loops.

Comparisons of extrapolation models (see chapter~\ref{sec:measure+model})
of the coronal magnetic field with the
observed shapes of field lines in EUV images help to estimate the
electric current strengths which flows along these field lines.
Especially in active regions these currents may be appreciable and
lead to marked deviations of the shape of observed loops and their
models based on a potential field
\cite{Green:etal:2002,Demoulin:etal:2002b,Bleybel:etal:2002,%
Regnier:etal:2002,Marsch:etal:2004}. A substantial current was also
found by \cite{Wiegelmann:etal:2005} when they tried
to match their field extrapolation to the loops observed by
\cite{Solanki:etal:2003} near a young active region.

Loops visible in EUV images which emerge from active regions appear 
astonishingly thin even though simple field extrapolation predicts a
widening of the flux tube's cross section with height (see
Fig.~\ref{Fig:loops} and discussion by \cite{Klimchuk:etal:2000}).
An explanation for this observation might be that at lower heights the
emitting flux tubes are so thin in diameter that they cannot be
resolved and the widening of the cross section therefore remains
undetected.
Also, unresolved fragmented bundles of emitting loops (``multithreads'')
of different temperature and density have been proposed to explain the
non-isothermal emission observed from these loops
\cite{Reale:Peres:2000,Aschwanden:etal:2000a,Aschwanden:Schrijver:2002}.
As an alternative explanation a twist of the thin flux tubes has been
suggested \cite{Mikic:etal:1990,VanHoven:etal:1995,Klimchuk:etal:2000}.
This requires a field-aligned current which induces a
toroidal magnetic field and a magnetic curvature force which then
keeps the flux tube concentrated on a small cross section like in a
plasma pinch device in the laboratory.
Simulations have shown that as the current increases, the loop axis also rises,
and eventually becomes distorted off the loop plane into an S-shape.
Finally, Mikic et al. found that the loop is driven kink unstable if the
number of turns over the whole loop length exceeds about 2.4
\cite{Mikic:etal:1990,VanHoven:etal:1995}.

In soft X-rays and less pronounced in EUV, the expected S-shape
deformation of bright loops is indeed observed
(``sigmoids''; see, e.g., \cite{Pevtsov:2002, Gibson:etal:2004}).
The sense of the S reflects the sign of $\alpha$ or of the current
helicity density $\vect{j}\cdot\vect{B}$ \cite{Seehafer:1990} on the
respective magnetic flux rope.
This sign has been found to be almost unique on either hemisphere
but reversed with respect to the other hemisphere (``hemispheric helicity
law''; \cite{Rust:Kumar:1994,Rust:Kumar:1996}). Why the S-shape for EUV loops
is not as apparent as for X-ray loops is not clear. The former are
usually lower, shorter and therefore may be more stable against an
S-shape distortion than the higher reaching X-ray loops.

\subsection{Filaments and helmet streamers}


\begin{figure}[t]
\centering
\includegraphics[width=0.75\hsize]
                {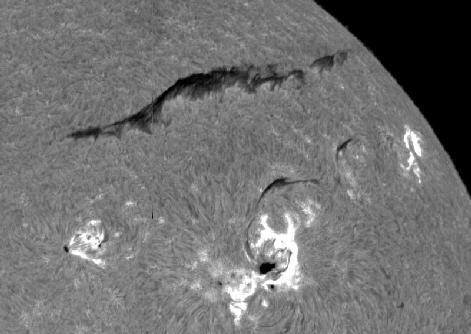}
\caption{\label{Fig:Hafilam} Dark filament lying above the surface
   of the Sun in an image taken at the emission line of H$\alpha$ by the
   Big Bear Solar Observatory (BBSO). The cool filament material absorbs
   the H$\alpha$ radiation from the surface so that they appear dark on
   the solar disk. Closer to the equator, a bright active region with
   smaller filaments in the neighborhood}
\end{figure}

Regions of opposite polarity of the radial magnetic field in
the lower corona are separated by a neutral line where the vertical
magnetic flux at the coronal base changes sign. Often, especially in the
vicinity of active regions and sunspots, filaments are stretched out
above the neutral line. A filament is a vertical sheet of cool and
high density plasma with temperature of about 10$^4$\;K hovering some
10$^4$ to 10$^5$\;km above the photosphere. Often they stretch
out horizontally over 10$^5$\;km and more.
Due to their high density, the filament material is optically thick. 
On the disk, filaments therefore absorb the solar surface emissions
so that they appear dark (Fig.~\ref{Fig:Hafilam}). When observed above
the limb, their dense plasma intensively scatters the solar radiation
so that they are seen as bright prominences.
A good account on the observational facts about filaments can be found in
\cite{TandbergHanssen:1995} and \cite{Martin:1998}.
Depending on the activity of the neighborhood, filaments also show some
activity but in general they are remarkably stable and resistant to
nearby perturbations.
Filaments originally formed above the neutral line between the
poles of a bipolar active region often survive the active
region as a quiescent prominence by weeks.
The more astonishing is that many filaments sometimes disappear after
days and weeks of quiescence in a sudden eruption within a few hours.

Closer inspection in the wings of H$\alpha$ and in EUV
Doppler shifts reveal a considerable, presumably field-aligned 
mass flow. The fact that the filament mass does not simply rain down
onto the solar surface can only be attributed to magnetic fields which
suspend the plasma against gravity. Also, their apparent stability on
the one hand and the vigor of their eruption on the other hand are
evidence of an energetic magnetic field system which extends far beyond the
visible filament.

Hanle effect measurements of the magnetic field that permeates the dense,
cool prominence material reveal flux densities of 5-10~G for quiescent
prominences but may locally reach up to 100~G in more active ones
\cite{Bommier:Leroy:1998,TandbergHanssen:1995}.
The field inside the filament is found to be almost horizontally
oriented with a considerable component elongated along the filament
axis, i.e., along the neutral line underneath.
Leroy, Bommier and coworkers found an average angle of the filament
magnetic field with respect to the neutral line of 40$\pm$30 degrees
\cite{Bommier:Leroy:1998}.
The component perpendicular to the neutral line was in the vast
majority of cases found in opposition to the polarity of photosphere
below (``inverse polarity law'').
Also the orientation of the field component along the neutral line was
found to be systematic (``chirality law''): if $\hat{\vect{p}}$ denotes
the horizontal unit vector normal to the neutral line in normal-polarity
direction (i.e., from magnetic ``$+$'' to ``$-$'') and
$\hat{\vect{z}}$ the vertical unit vector then the field component
along the neutral line is always in direction of
$\hat{\vect{p}}\times\hat{\vect{z}}$ in the northern hemisphere
(``dextral'') and in $-\hat{\vect{p}}\times\hat{\vect{z}}$ direction
in the southern hemisphere (''sinistral'').
The rules for two components of the field inside the filament seem to
hold independent of the solar cycle and can be expressed in brief as
$\vect{B}_{\mathrm{filament}}$ = $a\hat{\vect{p}}$ +
$b\hat{\vect{p}}\times\hat{\vect{z}}$ where $a$ and $b$ have the same
order of magnitude with $a<0$ (``polarity law'') and $b>0$ or $<0$ on
the northern or southern hemisphere, respectively.


\begin{figure}[h]
\centering
\includegraphics[height=9cm]
                {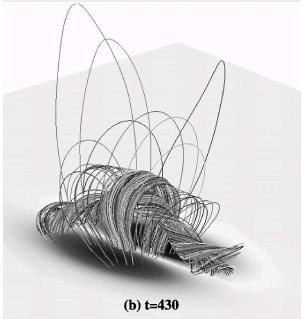}
\caption{\label{Fig:promimod} Field lines of the numerical model for a
  filament \cite{Amari:etal:2000}. The shading on the ground denotes the
  surface field strength. Note the central helical flux tube enclosed by
  an outer arcade field.}
\end{figure}

The simplest field topology which properly connects the photospheric
field and the field observed inside the filament is originally due to
Kuperus and Raadu \cite{Kuperus:Raadu:1974} and has been further
extended to 3D by numerous authors ( e.g., 
\cite{VanBallegooijen:1999,Amari:etal:2000,DeVore:Antiochos:2000}).
Fig.~\ref{Fig:promimod} displays the field configuration assumed to exist
in the vicinity of a filament.
It involves a twisted magnetic flux rope, stretched along the neutral
line, which gives the necessary support to the cool filament plasma in
the upwardly curved pockets of the helical field \cite{Lionello:etal:2002}.
In these heights, the horizontal $\hat{\vect{p}}$ component of the
magnetic field, belonging to the flux rope, is directed in inverse
polarity direction.
At the bottom of the flux rope closer to the surface, this field component
eventually has switch to normal polarity and the field components in a plane
normal to the neutral line describe a saddle point. The location of these
saddle points for all planes along the flux rope are referred to as
the ``x-line''.
Hence the filament material is thought to reside above the x-line at
the bottom of the flux tube in a height where the magnetic field is
about horizontal and is directed in inverse polarity direction, as
observed.

The twist of the flux rope relates the observed geometrical chirality
to the more physical current helicity density $\vect{j}\cdot\vect{B}$.
With the above model, dextral filaments would have a flux rope with
negative current helicity, sinistral filaments need positive current
helicity. This way, the chirality law also agrees with the hemispheric
helicity rule for the sigmoid distortion of X-ray loops mentioned in
the previous section.
In fact, the surface current helicity density observed by vector magnetographs
in active regions and sunspots where the field intensity is strong enough
to be reliably measured shows a clear dominance of the sign expected
from the helicity rule on either hemisphere independent of the solar cycle
\cite{Pevtsov:etal:1995,BaoS:ZhangH:1998}.

Also other observed features like the orientation of the filament
barbs (see the filament `legs' in Fig.~\ref{Fig:Hafilam}) and the
horizontal field orientation derived from fibrils
(Fig.~\ref{Fig:Hafibrils}) in the filament channel comply with the
above general hemispheric chirality or helicity rule (e.g.,
\cite{Rust:Kumar:1994,Zirker:etal:1997}).

\begin{figure}[t]
\centering
\includegraphics[viewport=35 33 585 433, clip,width=6.1cm]
                {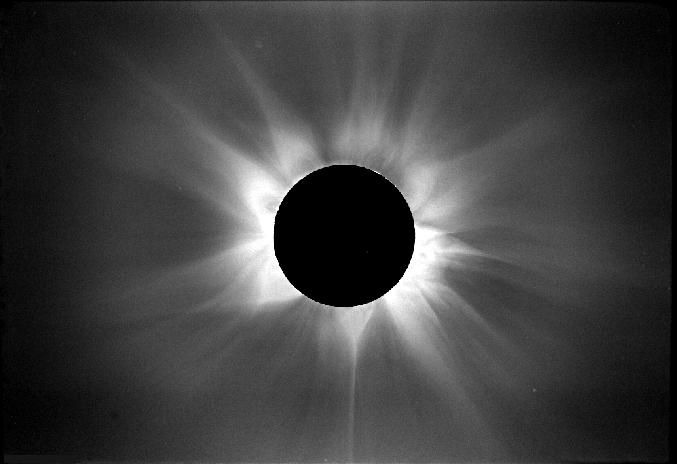}
  \hspace{3mm}
\includegraphics[viewport=65 25 615 425, clip,width=6.1cm]
                {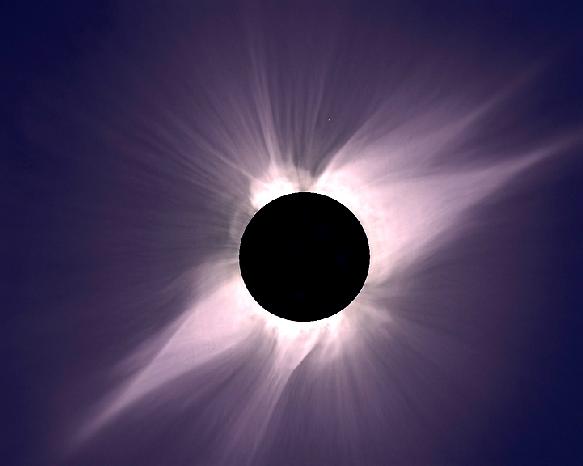}
\caption{\label{Fig:eclipse} White light coronagraph images of
  the solar eclipses 1980 (left) and 1991 (right).
  In the right image helmet
  streamers extend out to beyond three R$_{\odot}$ in all directions
  at times of high solar activity 1980 while they are more concentrated
  at lower latitudes during times of more moderate activity as on the
  left example (images courtesy of the High Altitude Observatory).}
\end{figure}

The filament magnetic field model also complies with the general
orientation found for coronal X-ray arcades which span the neutral line at
heights well above a filament \cite{Martin:McAllister:1996}.
The magnetic arcade field often continues outwards to
more than a solar radius. Limb observations with coronagraphs of the corona
above a neutral line reveal, if seen head-on, a helmet streamer
above the polarity inversion separated from the filament by a cavity
of depleted plasma density (see Fig.~\ref{Fig:eclipse}).
The cavity may reach out above the Sun's surface for a fraction of
$R_{\odot}$ while the tip of the helmet streamer extends to 2
$R_{\odot}$ or more. The cavity and the inner edge of the helmet
streamer show the concave shape which we expect from the magnetic
field lines of an arcade above a neutral line.
The outer edge of the helmet streamer, in contrast, is straightened out
into interplanetary space and joins into the heliospheric current
sheet. Coronagraph observations with Lasco/SoHO show that these cusps
may be the source region for the dense, slow solar wind
\cite{Sheeley:etal:1997}.
At times of small coronal activity, the streamers extend almost
two-dimensionally in azimuthal direction and cover the lower
heliomagnetic latitudes (``streamer belt'').
Whether the helmet streamer magnetic field possesses a significant
longitudinal component along the neutral line underneath like the
filament, a feature which seems to be present in according MHD models
\cite{Linker:etal:1999}, has not yet been observed.

While the Kuperus-Raadu flux rope magnetic field topology for the
filament is widely accepted, there is still a debate about
how the flux rope is formed,
what gives this configuration its stability and
how the stable field configuration may turn into a filament eruption
within hours.
For the genesis of filaments shear motion of the field line foot
points along the neutral line with either flux cancellation (e.g.,
\cite{VanBallegooijen:Martens:1989,VanBallegooijen:Martens:1990,%
Inhester:etal:1992,VanBallegooijen:1999,Amari:etal:1999b})
or emergence of properly wound flux (e.g., \cite{Rust:Kumar:1994,Low:1996,
Mackay:etal:1998,Priest:1998,Gibson:etal:2004}) have been considered.
Least understood, however, is the eruption process.
It often starts with a slow rise of the filament and the arcade field
a few hours before both are rapidly lifted upwards (see
Fig.~\ref{Fig:filerupt}).
Following the eruption, a set of loops across the neutral line
are left behind which brighten in EUV and X-rays (``post-flare loops'',
see Fig.~\ref{Fig:arcade}).  The chromosphere at their feet on
either side of the neutral line starts to radiate intensively in H$\alpha$
(``two-ribbon flare'').
More observational details and also conceptual ideas about filament
eruptions have been collected by, e.g.,
\cite{Hudson:Cliver:2001}.

Tens of Minutes to an hour after the eruption coronagraphs reveal
large parts of the overlying helmet being blown away into the
heliosphere as a coronal mass ejection (CME). In many cases the
quiescent three-part structure, i.e., filament - cavity - helmet, is
maintained in the ejecta.
Often the filament material, distinctly visible in coronagraph
observations, displays the helically twisted structure of the original
filament flux rope (see Figs.~\ref{Fig:promirupt} and \ref{Fig:cme})
The imprint of this helical structure can still be observed in-situ
at 1 AU in the magnetic clouds into which the CME evolves as it propagates
into the heliosphere \cite{Rust:1999,Bothmer:Schwenn:1998,Kumar:Rust:1996,
Plunkett:etal:2000}.

\begin{figure}[t]
\centering
\includegraphics[width=0.75\hsize]
                {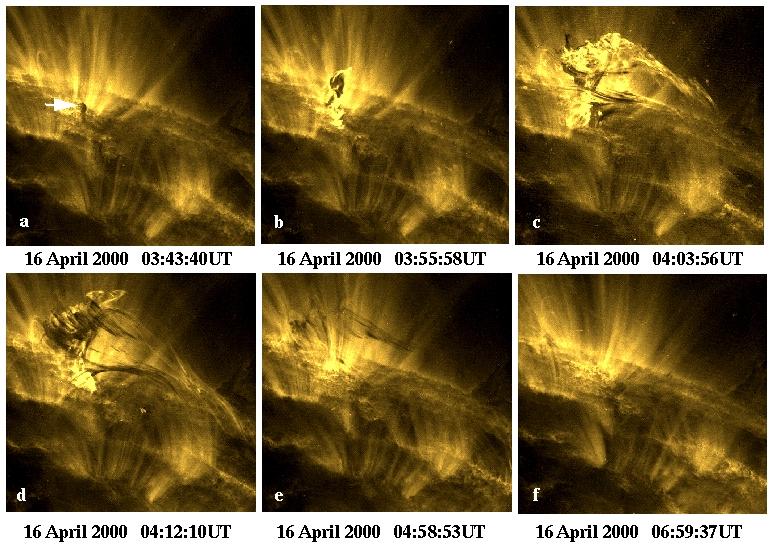}
\caption{\label{Fig:filerupt}
  Eruption of a small filament observed by TRACE. The initial location
  of the filament is at the tip of the arrow in the first panel.}
\vspace{5mm}
\includegraphics[height=8cm]
                {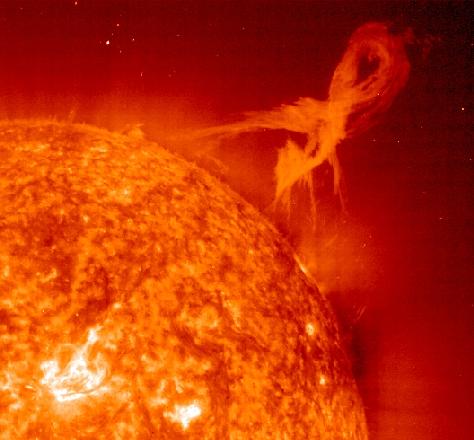}
\caption{\label{Fig:promirupt} Cool filament material just after
  eruption carried high up into the corona by magnetic forces as
  observed by the EIT instrument on board SoHO.}
\end{figure}

\begin{figure}
\centering
\includegraphics[width=0.6\hsize]
                {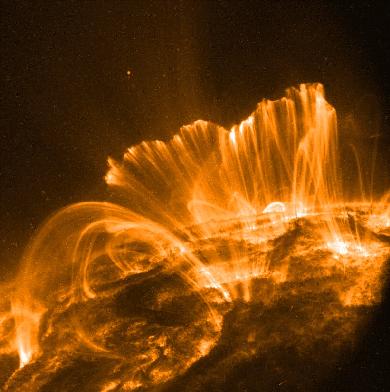}
\caption{\label{Fig:arcade}
Post-flare loops arranged in an arcade crossing the magnetic neutral line.
The EUV image was taken by the TRACE spacecraft}
\end{figure}

\begin{figure}
\centering
\includegraphics[width=0.8\hsize]
                {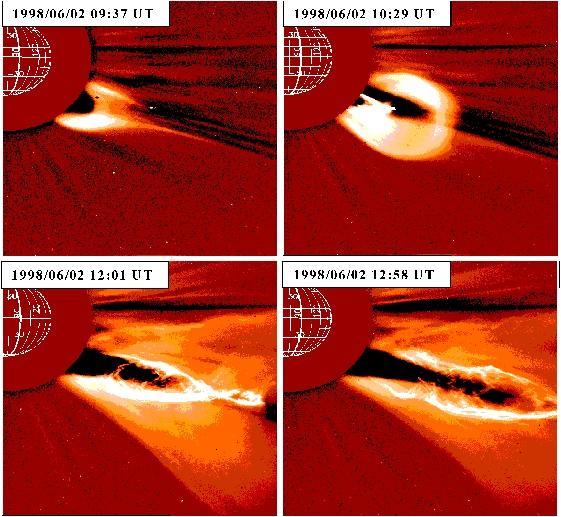}
\caption{\label{Fig:cme} Sequence of images showing the
  eruption of coronal mass into the heliosphere observed by
  the LASCO coronagraph on board the SoHO spacecraft.} 
\end{figure}

The few phenomenological facts presented so far convincingly
demonstrate the overwhelming role of the magnetic field in the Sun's
atmosphere. For the remaining part of this chapter we would like to
turn the logic around: given the dominance of the magnetic field in
the corona, one can derive certain characteristics or constraints of
coronal processes. Again due to lack of space we can touch only few of
the issues discussed in the present literature.

\subsection{Magnetic surfaces and null points}

The magnetic field induces a structure into space: the high mobility
of charged particles parallel to the field compared to their
perpendicular mobility brings points on the same field line into a
physical neighborhood even though they may geometrically be far
apart.

In laboratory plasma devices with some symmetry imposed by the
boundaries of the plasma chamber, field lines may ergodically cover
whole surfaces. These magnetic surfaces are then spanned locally by
the divergence-free fields $\vect{B}$ and $\vect{j}$ and can be
labeled by the local plasma pressure $p$ since $\vect{\nabla}p$ =
$-\vect{j}\times\vect{B}$ is normal to these surfaces.
In the force-free limit when the pressure $p$ becomes negligible,
surfaces $\alpha$ = const can play a similar role. Since $\alpha$ is
constant along a field line, any $\alpha$ contour on the solar surface
maps to a surface in a force-free field corona. These nested surfaces
are an important concept if, e.g., we want to define what exactly we
mean by a flux tube or flux rope.
It is possible however at least for some regions of the corona, that the flux
distribution on the solar surface is so heterogeneous and the related
magnetic surfaces are so eroded that constant-$\alpha$ contours and
surfaces are no longer useful.

Another concept to order magnetic-field-permeated, 3D space is by topological
connectivity. Provided we can separate the surface flux in an enumerable
set of localized flux elements, field lines which connect the same source
elements may be defined a flux tube. Obviously, the surfaces of these flux
regions are particularly prone to reconnection.
Especially those points, where the magnetic field vanishes have been found
to be critical for the stability of the magnetic configuration
\cite{Priest:Schrijver:1999}.
In a 2D geometry, the x-lines as they exist below a filament flux rope, are
the preferred site of reconnection. In a 3D magnetic field configuration,
however, it turns out that lines of vanishing magnetic field strength are 
structurally unstable: a slight random change in the field components
dissolves a $|B|$=0 line into a number of individual $|B|$=0 points.
Only these $|B|$=0 points (``null points'') are stable since small
changes in any field component make the point move but do not destroy it.
Magnetic null-points have therefore attracted quite some attention recently
\cite{Lau:Finn:1990,Bungey:etal:1996,Brown:Priest:2001,Solanki:1994}.
  
Obviously, the way we defined flux tubes above, a null-point has to lie
on the surface of a flux tube.
The field in its neighborhood can be approximated by the first term
of its Taylor expansion $\vect{B}$=
$(\partial\vect{B}/\partial\vect{x}) \cdot (\vect{x}-\vect{x}_0)$ and
is therefore topologically determined by the Jacobian matrix
$(\partial\vect{B}/\partial\vect{x})$.
Its eigenvalues and associated eigenvectors determine the orientation of
a fan plane and an orthogonal spine direction formed by field lines which
pass close to the null point as sketched in Fig.~\ref{Fig:skeleton}
\cite{Lau:Finn:1990}.
At larger distances from the null point, the field lines on the fan
extend to a surface (``separatrix'', as this surface separates
differently connected flux regions), and the spine extends into a
singular field line.
The whole set of interwoven null points, fan surfaces and spines
form what is often referred to as the skeleton of a magnetic field.


In the region near the null point energetic plasma may be locally trapped
and isotropized because the mirror force prevents it from escaping both
along the spine and, somewhat less effectively, along the fan.
It has been found that the intersection line of fans of related null points,
the so called separator, is particularly prone to reconnection
(e.g., \cite{Priest:Schrijver:1999,Longcope:VanBallegooijen:2002,%
Craig:Watson:2000}).
Priest and coworkers investigated systematically the changes of the
skeleton of a potential field with respect to changes of surface magnetic
elements (e.g., \cite{Beveridge:etal:2002}). These topological
changes indicate the state towards which a real plasma might want to
relax. If this state is topologically different ( i.e., has a different
skeleton) from the original state, it can only be reached by
reconnection taking place.

\begin{figure}[h]
\centering
\includegraphics[width=8cm]
                {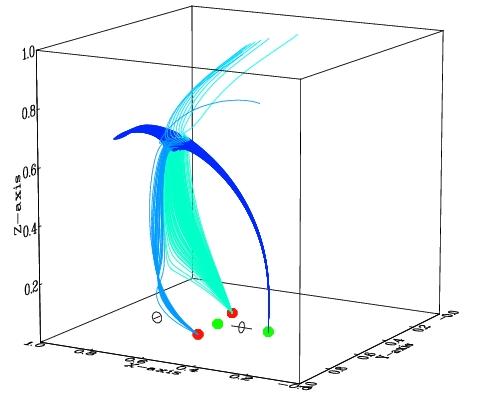}
\caption{\label{Fig:skeleton} The magnetic field lines through the
  vicinity of a coronal null point (at $z\sim$0.6) of the potential
  field due to four magnetic surface elements (full dots on the surface,
  the magnetic sign of the surface elements is color coded in red and green).
  The spine connects to the element on the right side and leaves the box
  on its left end. The two reversed polarity elements on the front and rear
  side lie on the curved fan surface.
  Two further null points in the photosphere are indicated by a small
  stick/disk which shows the orientation of their respective spine/fan.
  A fourth null point is located underneath the photospheric surface.}
\end{figure}

\subsection{Energy and helicity of the coronal magnetic field}
\label{sec:energy+helicity}

The energy output from the chromosphere and corona is considerable:
about 1000 and 300\;Wm$^{-2}$ are the estimated mean outward energy
flux density from the respective atmosphere layer of the quiet Sun
\cite{Withbroe:Noyes:1977,Aschwanden:etal:2000b} into EUV and X-ray
radiation, electron heat conduction and into the solar wind.
Individual active regions and bright loops radiate locally
up to 10$^4$\;Wm$^{-2}$ (e.g., \cite{Browning:1991,Aschwanden:etal:2001b}).
Large CMEs have been estimated to release up to 10$^{26}$\;J
within minutes \cite{Hundhausen:etal:1994} which is comparable to
the radiative loss of the whole corona during one or two days.

According to a concept promoted by Parker in a series of papers,
(e.g., \cite{Parker:1972,Parker:1988}), large flares and CMEs are
just the bottom end of eruptive processes on all scales which convert
magnetic energy into heat and bulk plasma motion.
The source of this energy is assumed to be the continuous motion of the
foot points of magnetic field lines which leads to a stretching and
braiding of the coronal field.
The energy flux induced by this surface motion is easily estimated if
the frozen-in condition $\vect{E}$ = $\vect{v}\times\vect{B}$ is used
in the Pointing flux vector component normal to the solar surface.
The resulting power transported through the solar surface is
(here and below, subscripts $h$ and $r$ refer to the horizontal and
radial components on the solar surface, respectively)
\begin{equation}
P_E = \frac{1}{\mu_0}
     \hspace*{-1em}\int\limits_{\mathrm{photosphere}}\hspace*{-1em}
     \big(B_h^2 v_r - (\vect{B}_h\cdot\vect{v}_h) B_r \big) \;d^2x
\label{Eflux_photosph}\end{equation}
$P_E$ has two terms which are the contributions from flux
emergence/cancellation and from work done by line-tying against
magnetic stress.
Both terms are highly fluctuating and their averages are not straightforward
to estimate.
Parker argues that the line-tying term alone can yield the power
necessary to explain the radiative output of the corona while the
term due to flux emergence and cancellation rapidly changes sign
and gives only a minor contribution to the energy balance.
The contribution of the line-tying term can be estimated
if we regroup the respective surface integration measure $B_r\;d^2x$
pairwise into the flux elements at the conjugate foot points of the
closed flux tubes. With $B$ $\simeq$ 1-10\;G, and a mean foot
point separation velocity of $\simeq$ 1\;kms$^{-1}$ we obtain the
order of magnitude of the quiet Sun energy flux density.

Parker's key paradigm to explain the conversion of this energy input
into coronal heat is the conjecture that even if the foot point motion
has only moderate gradients, the magnetic field driven by it will
eventually develop arbitrarily thin current sheets. These will
probably dissipate intermittently in small reconnection events.
Field lines which encounter a null point may play an important role
in this development of current sheets. Their foot points are close to
each other at one end but far separated at the other so that they may
well experience a distinctly different shear.

The emergence of new flux from below the photosphere (first term in the
integrand of eq.~\ref{Eflux_photosph}) provides another
source of current sheet formation. Indeed, many flares are observed
to occur near young active regions \cite{Feynman:Martin:1995,%
WangYM:Sheeley:1999}. First direct evidence for the existence of a strong
current sheet in the vicinity of newly emerged flux was obtained
only recently \cite{Solanki:etal:2003}.

Indeed short duration flares and explosive events are observed with a
broad range of energies in EUV and X-rays. Their statistics shows an
energy dependent occurrence rate of $f(E)$ $\propto$ $E^{-\eta}$ with $\eta$
$\simeq$ 1.8 - 2.6 \cite{Aschwanden:etal:2001a,Benz:Krucker:2002,%
Brkovic:etal:2001,Winebarger:2004} down to energies of $E$ = 10$^{17}$\;J.
An exponent $\eta$ larger than 2 implies that the small energy flares
contribute most to the heating of the corona. The total energy input
into the corona would in this case depend on the minimum possible flare
energy.

It is still controversial, however, whether these frequent, small-energy
flares by themselves yield enough power to explain the heating of the
corona entirely.
Alternative and additional heating mechanisms involve continuous ohmic
heating by DC currents (e.g., \cite{Heyvaerts:Priest:1984,%
Galsgaard:Nordlund:1996a,Gudiksen:Nordlund:2002,Solanki:etal:2003}) and
dissipation of waves (e.g., \cite{Grossmann:Smith:1988},chapter~3 of
\cite{Ulmschneider:etal:1991},\cite{Goossens:etal:2002,%
VanDoorsselaere:etal:2004a}).
Some fraction of the free energy available in small scale flare events
will probably be released by exciting one or more of the MHD wave modes.
Wave activity and oscillations are observed in the chromosphere and
corona in a broad range of frequencies. Spectral line broadening beyond
the expected thermal width is often interpreted as due to high-frequency
turbulent or wave induced plasma motion. Oscillations of the super-granular
plasma with a 3-min period are a prominent feature in chromospheric Doppler
shifts. In the corona, transverse Alfv\'enic loop oscillations at periods of
200-300\;s (e.g., \cite{Nakariakov:etal:1999,Schrijver:etal:2002}) and
slow mode type compressional oscillations at even longer frequencies
\cite{WangTJ:etal:2003} have been observed.
Their damping and hence their contribution to coronal heating is, however,
still a matter of research \cite{DeMoortel:Hood:2003}.
Besides quasiperiodic variations, an increasing, seemingly random, dynamic
variability is observed in EUV images, the finer the observational resolution.
Some authors \cite{Schrijver:etal:1999} take this as evidence that
coronal heating occurs rather erratically and on small time scales.

Less controversial than the minimum flare energy is the maximum
possible energy released in a large flare or CME. These major
eruptions though they are highly dynamic in the corona are not able
to change the magnetic flux distribution in the photosphere. A
lower bound of the magnetic energy the corona will retain after the
eruption therefore is the minimum magnetic energy state which complies
with the surface boundary conditions. This state is given by the
associated potential field. If in addition to the fixed
surface field, we require a certain amount of relative helicity to
remain in the corona, the minimum energy state is that of the
respective constant-$\alpha$ force-free field (Taylor field,
\cite{Taylor:1974}). If in addition to the normal component of the
boundary field we impose a connectivity map to the boundary, the
nonlinear $\alpha$ force-free field which complies with this
connectivity has the least magnetic energy (see \cite{Grad:1964,%
Taylor:1986,Sakurai:1989,Biskamp:1993}).

The amount of energy which can be set free by a large flare is
therefore limited by the energy difference of the field prior to the
flare to the energy of one of the above minimum-energy fields for
the same surface magnetic flux distribution.
There is strong evidence that the eruptions are enabled by field
line reconnection which alters the connectivity of the foot points.
It has further been found that reconnection hardly changes the overall
helicity \cite{Berger:1984,Hornig:Rastaetter:1997}.
Field line reconnection is a rather local process while the helicity,
e.g., of a flux tube, resides in its twist or, equivalently, in the
toroidal flux which in equilibrium is evenly spread out all along the
flux tube.
If flux tubes with different twist reconnect the toroidal flux from
each of the reconnected ends redistributes along the newly formed
tubes by means of Alfv\'en waves. The total helicity is almost
conserved, but due to the mixing of toroidal flux the difference in
helicity per unit length between the new flux tubes is less than the
difference was between the old ones.
Hence reconnection tends to equalize the helicity among the reconnecting
flux tubes \cite{Hornig:Rastaetter:1997} and drives the plasma towards
a constant-$\alpha$ state.
The lowest energy state of the field above the solar surface we can
expect on these grounds after a large flare will therefore be the
linear force-free field. But even this is probably an underestimate because
$\alpha$ will only be leveled out on those flux tubes which were
involved in a reconnection process and not in the whole of the corona.

The above minimum energy estimates of the potential or linear
force-free field can be calculated from the observed surface field $B_r$,
using the extrapolation methods mentioned in chapter \ref{sec:measure+model}.
An energy estimate of the actual field before the flare is much more of a
problem. In principle it should be possible to obtain such an estimate
by the virial $\int \vect{r}\cdot\vect{f} d^3x$ (e.g., \cite{Sakurai:1989})
over the total force density $\vect{f}$ in the plasma.
In the MHD approximation and conventional notation,
\begin{equation}
\vect{f} = \frac{1}{\mu_0}\vect{j}\times\vect{B}-\vect{\nabla}p
          -\rho\frac{GM_{\odot}}{r^2}\hat{\vect{r}}
          -\rho(\vect{v}\cdot\vect{\nabla})\vect{v}
\end{equation}
Inside a stationary corona, $\vect{f}$ should vanish and so should the virial,
independent of where we place the origin of $\vect{r}$.
With $\vect{r}$ at the Sun's center we obtain after partial integration
\begin{eqnarray}
  & \frac{1}{2\mu_0} \int\limits_{\mathrm{corona}} B^2\;d^3x
  \hspace*{1em} = \int\limits_{\mathrm{corona}}
     \big( \rho\frac{GM}{r} -3p -\rho v^2 \big)\;d^3x &
 \nonumber\\
 &+ \hspace*{-1em}\int\limits_{\mathrm{coronal~boundaries}}\hspace*{-1.5em}
      \big( \frac{1}{2\mu_0}B^2\vect{r}
           -\frac{\vect{B}}{\mu_0}(\vect{B}\cdot\vect{r})+p\vect{r}
            +\rho(\vect{v}\cdot\vect{r})\vect{v}\big)\cdot\vect{n}\;d^2x &
\label{MHDvirial1}\end{eqnarray}
The inner boundary of the corona is chosen as the coronal base at
$r$ $\simeq$ $R_{\odot}$ where the non-magnetic forces are negligible
and the surface normal $\vect{n}$ points towards the Sun's center.
The contribution to the surface integral in (\ref{MHDvirial1})  
from the outer boundary of the corona is usually neglected.
With $\vect{B}$ $sim$ $\hat{\vect{r}}$ beyond $r$ $sim$ 2.5 $R_{\odot}$,
the integrand of the surface integral decreases in magnitude and changes
sign as the solar wind $\vect{v}$ $sim$ $v_r\hat{\vect{r}}$
increases beyond about the Alfv\'en speed.
With these simplifications
\begin{eqnarray}
  \frac{1}{2\mu_0} \int\limits_{\mathrm{corona}} B^2\;d^3x
  &=& \int\limits_{\mathrm{corona}}
     \big( \rho\frac{GM}{r} -3p -\rho v^2 \big)\;d^3x
 \nonumber\\ &+& \frac{R_{\odot}}{2\mu_0}
      \hspace*{-1em}\int\limits_{\mathrm{coronal~base}}\hspace*{-1em}
              (B_r^2-B_h^2)\;d^2x
\label{MHDvirial2}\end{eqnarray}
remains which yields the magnetic energy in terms of the surface field
and some minor non-magnetic contributions which may be neglected in a first
approach.

The problem with eq.~(\ref{MHDvirial2}) is, however, that the magnetic
surface field is mostly observed in the photosphere where the
force-free condition may not yet hold and not at the coronal base
\cite{Metcalf:etal:1995,McClymont:etal:1997}.
How errors in the surface observations, especially in $B_h$ may seriously
affect eq.~(\ref{MHDvirial2}) is discussed in \cite{Klimchuk:etal:1992}.
Yet it is instructive to consider some consequences of eq.~(\ref{MHDvirial2})
more closely.

If the coronal magnetic energy is enhanced by line-tying as in
eq.~(\ref{Eflux_photosph}) we cannot expect a systematic change of $B_r^2$
to account for the increase in magnetic energy on the right hand side of
eq.~(\ref{MHDvirial2}). 
Consequently, $B_h^2$ in eq.~(\ref{MHDvirial2}) will have to decrease which
we indeed observe as the rising of coronal loops with increasing shear.
The maximum energy that can possibly be achieved is when $B_h$ $\simeq$ 0,
i.e., when all field lines are stretched out almost radially into the
heliosphere. However, this configuration is never reached because
as $B_h$ decreases with increasing energy, and so does the capability
to pump further energy into the corona by line-tying according to
eq.~(\ref{Eflux_photosph}).

Hence, besides the lower energy bound given by the linear force-free field
there is also an upper bound for the energy of a force-free coronal magnetic
field which is represented by a field with all field lines open to the
heliosphere (``Aly's conjecture''; \cite{Aly:1984,Aly:1991,Sturrock:1991})
A simple calculation shows that the energy of a field
$\vect{B}$ = $B_r(R_{\odot})R_{\odot}^2\vect{r}/r^3$
(though this is not force-free; see, e.g. \cite{Aly:1991}, for an
exact construction of a force-free ``all-open field lines'' configuration)
with a dipole boundary $B_r(R_{\odot})$ has twice the energy
of the Laplacian dipole field (similarly there is a factor $n$+1 for
general multipoles).
Roughly, we therefore expect a factor of about 2 or more
between the upper and lower energy bounds.
Depending on how close the actual magnetic energy approaches this upper
limit an appreciable amount of the coronal magnetic field energy
can in principle be released in large flares.
It was pointed out by \cite{Low:1996} that the gravitational forces on
the filament and the helmet enhances the energy of the suspending
magnetic field in eq.~(\ref{MHDvirial2}).
If only part of the filament and helmet mass is expelled in a CME, the
mass loading prior to a CME could act like the lid of a jack-in-the-box.

Estimates of the magnetic energy of active regions before and after
a CME have been made by several authors for limited regions of the corona.
\cite{Regnier:etal:2002} found a magnetic energy of
6.5\;10$^{24}$\;J for an active region with an equivalent potential field
energy of about 3.8\;10$^{24}$\;J and an Aly-limit of 7.6\;10$^{24}$\;J.
\cite{Bleybel:etal:2002} found the magnetic energy of an
active region reduced from 8.4\;10$^{25}$\;J to 5.7\;10$^{25}$\;J after an
eruption.

Besides energy, we have seen that helicity is another invariant of ideal MHD
(see chapter~\ref{subsec_helicity}).
It also plays an important role in structuring the coronal magnetic field.
Its original definition eq.~(\ref{eq:helicity}), however,
is useful only for a plasma bounded by magnetic surfaces. Only for these
configurations is $H$ independent of the gauge of the vector potential
$\vect{A}$. 
For open systems such as the corona, the definition of helicity
was generalized by \cite{Berger:Field:1984} and \cite{Finn:Antonsen:1985}
through the introduction of the relative helicity
\begin{equation}
H_r=\int (\vect{A}\cdot\vect{B} - \vect{A}_0\cdot \vect{B}_0)\; d^3x
\end{equation}
which measures the helicity relative to a reference field
$\vect{B}_0 = \vect{\nabla} \times \vect{A}_0$. For $H_r$ to be gauge
invariant, $\vect{B}_0$ must satisfy the same normal boundary conditions
as $\vect{B}$ and $\vect{A}_0$ the same tangential boundaries.

A conventional choice for $\vect{B}_0$ is the potential field
for the boundary conditions given along with the Coulomb gauge
for $\vect{A}_0$ and $(\vect{A}_0)_r$ = 0 on the solar surface.
With the frozen-in $\vect{E}$ = $\vect{v}\times\vect{B}$ employed,
a simple expression then results for the flow of relative
helicity through the solar surface (see, e.g., \cite{Berger:Field:1984,%
Demoulin:Berger:2003})
\begin{equation}
P_H = 2 \hspace*{-1em}\int\limits_{\mathrm{photosphere}}\hspace*{-1em}
        \big((\vect{A}_0\cdot\vect{B}_h) v_r
           - (\vect{A}_0\cdot\vect{v}_h) B_r\big) \;\,d^2x
\label{Hflux_photosph}\end{equation}
As in eq.~(\ref{Eflux_photosph}) there are two separate contributions,
one due to vertical motions and the second due to horizontal line-tying.

In a number of analytical and numerical investigations
\cite{DeVore:2000,Berger:Ruzmaikin:2000,Demoulin:etal:2002a,%
VanDrielGesztelyi:etal:2003} it was shown that random surface motion
$\vect{v}_h$ is much less effective for the helicity transport in
eq.~(\ref{Hflux_photosph}) than it is for the energy transport in
eq.~(\ref{Eflux_photosph}).
Some amount of helicity is produced by the steady rotation of the Sun
alone and is stored on the open field lines in the deformation of the
interplanetary field into Parker's spiral \cite{Bieber:etal:1987}.
The ordered, differential rotation of the Sun is capable of generating
only a small part of the helicity expected to reside in the closed
coronal field \cite{DeVore:2000} and seems insufficient to explain
the total helicity flux which is eventually released in CMEs
\cite{Demoulin:etal:2002b,Rust:1999}.
The only candidate left in eq.~(\ref{Hflux_photosph}) for an effective
production of relative helicity is the vertical convection term, i.e.,
the emergence of properly twisted flux.

In \cite{Berger:Ruzmaikin:2000} the total helicity produced by the Sun during
a solar cycle was estimated to be about 10$^{47}$\;Mx$^2$.
Observational evidence that random line-tying has to be ruled out for
helicity production can be found in the hemispheric asymmetry of the
observed current helicity \cite{Pevtsov:etal:1995,BaoS:ZhangH:1998}, the
chirality law of filaments (e.g., \cite{Zirker:etal:1997,Martin:1998})
and the sigmoid X-ray structures above active regions
\cite{Pevtsov:2002,Gibson:etal:2004}.
Hence a major source of the coronal helicity has to be sought in the twist
of emerging flux represented by the first term of the integrand in
eq.~(\ref{Eflux_photosph}) and partly in the motion of active regions which
concentrate the majority of the surface flux
\cite{DeVore:2000,LopezFuentes:etal:2003}.

As the helicity cannot be dissipated by reconnection like the magnetic
energy, the closed field line region of the corona can only keep its
helicity in quasi-steady balance if helicity of opposite sign is
exchanged across the hemispheres or if helicity is transported away by
the magnetic clouds of CMEs.
Estimates of the release of relative helicity during the major flare
mentioned above range from 1.3\;10$^{42}$ to 0.7\;10$^{42}$\;Mx$^2$
\cite{Bleybel:etal:2002}. An active region was continuously observed
and its helicity budget monitored for five solar rotations
\cite{Green:etal:2002}. It was found
that a major part of the helicity carried away by a series of CMEs,
about 2\;10$^{42}$\;Mx$^2$ each, must have been provided by uprising
twisted magnetic flux.


\section{The magnetic field in the heliosphere}
\label{sec_heliospheric}

%
%

The heliosphere is the only region where in-situ observations of the
magnetic field have been made. The majority of observations were
carried out by various space crafts in the plane of the ecliptic near 1 AU.
However, the Helios 1 and 2 space crafts provided measurements as close
a 0.29 AU \cite{Mariani:Neubauer:1990}, the Voyager 1 and 2 space
crafts have explored the heliosphere out to almost 100 AU
\cite{Burlaga:etal:2003a} and Ulysses scanned the magnetic field over
the solar poles at a distance between 1.3 and 5.4 AU \cite{Neugebauer:2001}.
These measurements show that in the heliosphere, the magnetic field
has lost its dominant role. Instead, the plasma $\beta$
is back to the order of unity with a few exceptions like the interior of
CME magnetic clouds or the inside of magnetospheres and $\beta$ even
increases further with distance from the Sun.

\subsection{Heliospheric current sheet and Parker's spiral}

The open magnetic flux which emanates from the coronal holes fills the
whole space angle of 4$\pi$ beyond about 3 R$_{\odot}$. As the coronal
holes make up only about 20\% of the solar surface at solar minimum and
even less at maximum activity, the magnetic flux has to overexpand
between the coronal base and $\sim$ 3 R$_{\odot}$.
The expansion of an open flux tube cross section in this height range
compared to an $r^{-2}$ expansion is defined as the flux tube's
expansion factor.
Its value usually exceeds 5 and may be up to 20 or more for magnetic flux
close to the boundaries of a coronal hole \cite{Neugebauer:2001}.

The solar wind rapidly accelerates in this height region and helps to
straighten out the field radially along the wind velocity direction.
This produces the characteristic cusps of helmet streamers visible in
coronagraph images. Empirically, a strong anti-correlation has been
found between the asymptotic solar wind velocity in a flux tube and
the expansion factor of its cross section in the corona
\cite{WangYM:etal:1997a,Arge:Pizzo:2000}. There is therefore evidence
that the magnetic field has some control on the acceleration of
the solar wind.

Beyond about 3 R$_{\odot}$ the inward and outward directed flux
polarities are organized in two simply connected angular domains
separated by an undulating, radially stretched heliospheric current
sheet (HCS).
Back on the Sun, this current sheet connects to the cusps of the
helmet streamers which form above the meandering neutral line on the
solar surface. Comparisons with space craft observations have shown
that the position of the HCS can at most times be quite satisfactorily
be modeled (e.g.,\cite{Hoeksema:1995}) by simple potential field
extrapolations of solar surface magnetograms.
These models often employ a virtual sphere at about $\sim$ 3 R$_{\odot}$
(``source surface'') where as boundary condition a strictly radial
field direction is imposed.

\begin{figure}[t]
\centering
\includegraphics[width=0.6\hsize]
                {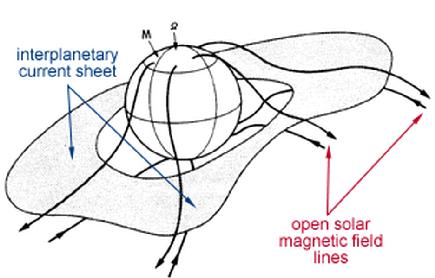}
\caption{\label{Fig:currsheet} Sketch of the current sheet and how it is
  magnetically connected to the coronal helmet streamers.
  The cartoon is from J.T. Hoeksema, Stanford University)}
\end{figure}

At solar activity minimum, the majority of the open solar magnetic flux
emanates from polar coronal holes and the HCS within about 1AU is confined
close to the solar equatorial plane.
At activity maximum, the neutral line on the solar surface is rather
erratic, open flux regions are spread out over the whole solar surface
which causes the HCS to be considerably displaced from the
heliographic equatorial plane.
From Ulysses observations, it was concluded that the HCS at solar
maximum is highly inclined and oriented almost perpendicular to the
equatorial plane \cite{Balogh:Smith:2001,Jones:Balogh:2003,Smith:etal:2003}.
The enhanced ratio of quadrupole to dipole moment indicates that the
HCS is also considerably warped \cite{Sanderson:etal:2003}.
Some authors \cite{WangYM:etal:2002} even argue for a more complex HCS
geometry at solar maximum. They claim that at least occasionally
even isolated reversed polarity cones may exist which would cause a second
detached current sheet on the surface of the radial cone.

The straightening effect of the solar wind ensures, however, that the
current sheet well below 1AU only warps azimuthally and is
straightened in radial direction so that (within about 1 AU) it appears
like a ``ballerina skirt'' (see Fig.~\ref{Fig:currsheet}), a term coined
by Alfv\'en \cite{Alfven:1977}.
During late declining and minimum solar activity, the HCS is often
approximated within less than 1 AU by a plane inclined with respect to
the heliographic equator.
Due to this inclination, the HCS intersects the ecliptic at least
twice. A space craft in the ecliptic plane should therefore see two
(or an even number of) sector boundaries during one solar rotation
where the heliospheric radial magnetic field component changes sign
\cite{Burlaga:1995}.
During maximum activity, HCS encounters were observed by
Ulysses up the the highest latitudes of its orbit \cite{Balogh:Smith:2001}
(see Fig.~\ref{Fig:parkerangle}).

In radial directions pointing well away from the HCS, the solar wind
is accelerated within a fraction of 1 AU to a final speed of about 800
km s$^{-1}$ (``fast solar wind''), more than 5 times the
Alfv\'en velocity at 1 AU \cite{DeKeyser:etal:2001}.
Observations of the Ulysses spacecraft show that beyond 1 AU and above
30$^{\circ}$ latitude there is relatively little variation in the fast
solar wind speed.
Close to the HCS, the plasma has a larger density and the final solar
wind speed only reaches about 400 - 600 km s$^{-1}$ (``slow solar wind'')
\cite{Phillips:etal:1995,Woch:etal:1997}.
Density irregularities carried away by the dense, slow wind can be
observed in coronagraph image sequences out to about 30 R$_{\odot}$
\cite{Sheeley:etal:1997,WangYM:etal:2000a} and by radio scintillation
measurements to even larger distances. These observations also show
that the slow solar wind originates from the cusps of the helmet
streamers and perhaps also from their lateral boundaries.

For distances beyond 3 R$_\odot$ and in regions unperturbed by the
HCS the radial magnetic field component $B_r$ must decline
as $r^{-2}$ in order to ensure flux conservation. Similarly,
the radial mass flux $\rho v_r$ must on average decrease as $r^{-2}$
to maintain a stationary solar wind outflow.
However, while their feet are still anchored on the Sun,
heliospheric field lines will eventually be deformed by
the rotation of the Sun and thus create an azimuthal field
component.
Parker \cite{Parker:1958} and Weber and Davis \cite{Weber:Davis:1967}
gave a first description of the stationary, azimuthally symmetric
heliospheric background field which results from this deformation.
In the Sun's rotating frame the solar wind velocity and magnetic field
vectors must be parallel in order to keep the field stationary. Hence,

\begin{equation}
  (v_{\phi} - \Omega r \cos\theta) B_r = v_r B_{\phi}
\label{vparB}\end{equation}
where $\Omega$ is the Sun's rotation rate of $\sim$ $2\pi/$(27 days),
$\theta$ denotes the heliographic latitude and index $\phi$ the azimuthal
component.
The azimuthal velocity $v_{\phi}$ will turn out to be practically
negligible and is determined from the conservation of angular momentum
\begin{equation}
  \frac{\rho v_r}{r}\frac{\partial}{\partial r} r v_{\phi}
 -\frac{B_r}{\mu_0 r}\frac{\partial}{\partial r}  r B_{\phi} = 0
\end{equation}
which can readily be integrated to
\begin{equation}
  r\big( v_{\phi}-\frac{B_r B_{\phi}}{\mu_0\rho v_r} \big) = L(\theta)
\label{Angmom}\end{equation}
where $L$ is a constant along the field line.
Upon elimination of $v_{\phi}$ from (\ref{vparB}) and (\ref{Angmom})
we obtain for the azimuthal component of the magnetic field
\begin{equation}
  B_{\phi} = \frac{B_r}{r v_r}
                 \;\big( L(\theta) - \Omega r^2 \cos\theta \big)
                 \;\big(1 - (\frac{B_r^2}{\mu_0\rho v_r^2})\big)^{-1}
\label{Bphi}\end{equation}

The field line deflection $\phi_B$ = atan$(B_{\phi}/B_r)$ (``Parker
angle'') from the radial direction can only increase smoothly with
distance from the Sun if the singularity $B_r^2$ = $\rho v_r^2$ in
(\ref{Bphi}) is resolved by a simultaneous vanishing of the numerator.
The distance at which this condition is met, i.e., where $v_r$ is the
(radial) Alfv\'en velocity, is $r$ = $r_A$ (``Alfv\'enic critical
distance'').
The critical distance has an important physical meaning: since the
solar wind is superalfv\'enic for $r>r_A$, no inward propagating
Alfv\'en waves can ever pass this barrier.
The condition for smooth field lines at $r_A$ therefore determines the
integration constant $L$ to $\Omega r_A^2 \cos\theta$ and hence the
angular momentum carried away by the solar wind.
In situ measurements of the heliospheric field and the solar wind speed
between 0.3 and 1 AU together with equation (\ref{Angmom}) gave values
for the Alfv\'en critical distance of $r_A$ $\simeq$ 10-14 R$_{\odot}$
\cite{Pizzo:etal:1983,Marsch:Richter:1984a}.
With these parameters, a deflection for the Parker angle of $\phi_B$
$\simeq$ 45$^{\circ}$ is obtained at about 1 AU.

The conservation of the radial invariants of the solar wind parameters
have been tested extensively in in-situ observations because they may be
extrapolated to distances where measurements have not yet been made.
The radial magnetic flux constant $r^2 B_r$ has a typical magnitude of
3 - 4 nT AU$^2$ and is found to be, apart from fluctuations, independent
of latitude except of course for the sign change across the HCS
\cite{Marsch:Richter:1984a,Smith:etal:1997,Smith:Balogh:2003}.
The mean magnetic pressure therefore seems in perfect lateral balance.
Fig.~\ref{Fig:Bradial} shows the sign flip of $r^2 B_r$ at the HCS of
an otherwise flat latitude profile.
The variation of $r^2 B_r$ with solar activity is only about 1 nT AU$^2$
with the largest values at the decline phase of the activity cycle
\cite{Smith:Balogh:2003}.
This temporal stability of $r^2 B_r$, being a direct measure of the
Sun's open magnetic flux, is surprising because the total unsigned flux
through the solar surface varies by as much as 80\% during the solar
cycle \cite{WangYM:Sheeley:2002a}.
An explanation was offered by \cite{Fisk:Schwadron:2001}
which allows magnetic field line reconnection of open flux only with closed
flux tubes. This is sufficient to feed the solar wind and redistribute
open flux during the course of a solar cycle, however, it does not change the
amount of open flux.

Also, the mean mass flux $r^2 \rho v_r$ has been found from Ulysses
observations to be almost independent of heliographic latitude.
A median value of $\sim$ 2.5 10$^{12}$ amu m/s AU$^2$ at all
latitudes and in fast and slow solar wind regions was found from
the Ulysses in situ observations \cite{Phillips:etal:1995}.
Measurements of Ly$\alpha$ radiation scattered at neutral interstellar
H atoms which penetrated into the heliosphere, however, can only be
reconciled with a decrease of the average mass flux by a about 30 \% at
latitudes above 40$^{\circ} $\cite{Bertaux:etal:1996}.

For distances $r$ $\gg$ $r_A$ where $B_r^2/\mu_0\rho$ $\ll$ $v_r^2$
we obtain from (\ref{Bphi}) another approximate constant along
a heliospheric field line, $r v_r B_{\phi}$ which should coincide
with $-\Omega B_r r^2 \cos\theta$. The latter is a constant provided
the field line stays at a constant latitude $\theta$.
For the Parker spiral angle $\phi_B$ this yields
\begin{equation}
 \tan\phi_B = \frac{B_{\phi}}{B_r} = - \frac{\Omega}{v_r} r \cos\theta
\label{Parkerangle}\end{equation}
which is the same expression we obtain from (\ref{vparB}) if the
vanishingly small $v_{\phi}$ component is neglected. Comparisons of
both expressions for $\phi_B$ in (\ref{Parkerangle}) using Ulysses data
showed good average agreement (see Fig.~\ref{Fig:parkerangle},
\cite{Forsyth:etal:1996,Smith:etal:2003}).
The statistics, however, are obscured by fluctuations in the various
field and flow components so that the results depend on the method of
averaging. For example, \cite{Smith:etal:1997} claim to
have observed somewhat less tightly wound field lines at high latitudes
during Ulysses' first polar passage than Parker's model would predict.

\begin{figure}
\centering
\includegraphics[width=0.65\hsize]
                {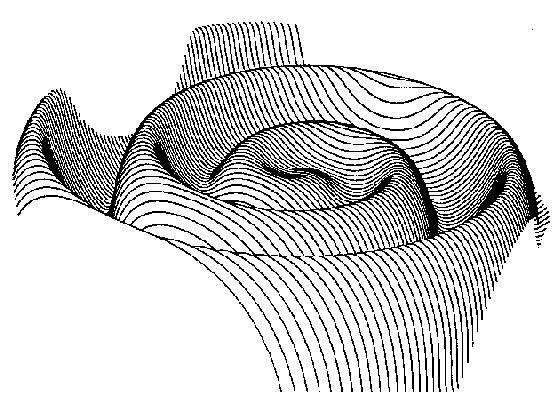}
\caption{\label{Fig:ParkerHCS} Deformation of the HCS up to a distance
  of about 12 AU. Close to the Sun, the HCS is approximated by a plane
  inclined by 15$^{\circ}$ with respect to the heliographic equator.
  With increasing distance from the Sun, the latitude of the inclined HCS
  is preserved along the Parker spiral characteristics
  (from \cite{Jokipii:Thomas:1981}).}
\end{figure}

\begin{figure}
\centering
\includegraphics[width=0.65\hsize]
                {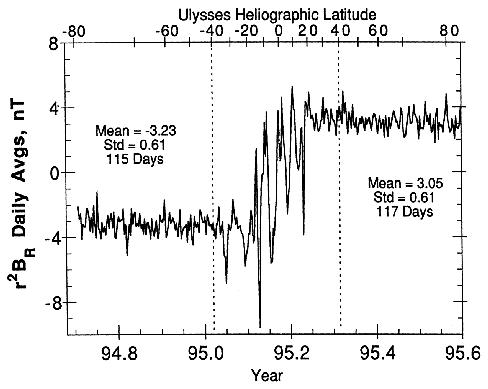}
\caption{\label{Fig:Bradial} Radial heliospheric field component scaled
  to 1AU vs. observation time along the Ulysses trajectory from the south
  to north pole (from \cite{Smith:etal:1997}).
  Between -40 and 40$^{\circ}$ of latitude the HCS was repeatedly
  crossed.}
\end{figure}

\begin{figure}
\centering
\includegraphics[width=0.65\hsize]
                {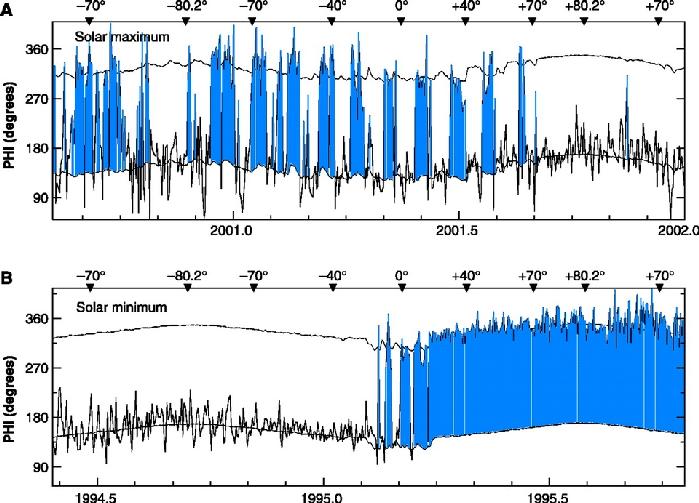}
\caption{\label{Fig:parkerangle} Azimuthal deflection angle of the
  heliospheric magnetic field from the radial direction vs. observation
  time along the Ulysses trajectory (from \cite{Smith:etal:2003}).
  The top and bottom panel show maximum and minimum solar activity
  conditions, respectively. The smooth curves represent the angle
  as derived from Parker's model (rhs of \ref{Parkerangle}), the fluctuating
  curve represents the measured ratio atan $B_{\phi}/B_r$. It flips between
  sectors of inward (angles around 100) and outward directed (angles around
  300)field whenever the heliospheric HCS is crossed.} 
\end{figure}


\subsection{CIRs, transients and turbulence}

Parker's spiral model, however, can only be a first order stationary
approximation to a very dynamic heliospheric field and flow
configuration.
The fact that the HCS is embedded in a slow, dense solar wind compared
to the fast, light wind inside the sector regions will eventually cause to
destabilize the HCS the more it is inclined with respect to the
heliographic equator. The spiral structure of the field causes an
analogous folding of the HCS surface. Fast wind streams will penetrate into
the equatorial plane where the HCS is displaced off to higher
latitudes. As the solar wind flow is practically radial, the fast
wind sector will eventually run into the slow current sheet ahead.

The fact that the solar wind plasma is frozen into the spiraling
magnetic field prevents the fast wind from penetrating the slow wind
region. Instead, the less tightly wound fast solar wind leaves a
rarefaction region behind and compress the warped current sheet ahead,
forming a corotating interaction region (CIR).
The dependence of the CIR formation on the initial coronal and HCS
boundary state has been reviewed by \cite{Balogh:etal:1999},
a general overview can be found in \cite{Forsyth:Gosling:2001}.

Typically, CIRs form at 1 AU and beyond as this is the distance where
the Parker spiral angle becomes significant. Beyond about 3 AU, the
compression of the CIR becomes so strong that a pair of shocks is
formed on either side: a fast mode shock at the front side which
accelerates the slow plasma ahead and a reverse shock on the back side
which decelerates the fast wind from behind. Simulations
show how CIRs are deformed through interaction with the solar wind
\cite{Pizzo:1994b,Pizzo:Gosling:1994}. These calculations
demonstrate also that the wings of the reverse shock can propagate to
much higher latitudes than correspond to the initial tilt of the HCS
thus explaining an initially puzzling observation of the Ulysses
spacecraft \cite{Gosling:Pizzo:1999}.

At even larger distances from the Sun, the CIR surfaces steepen
and merge. Whenever a merged CIR passed by
the Voyager 1 spacecraft observed between 14 an 18 AU 
quasiperiodic magnetic field enhancements by a factor 3-4
with periods around 26-days \cite{Burlaga:etal:1997}.
To the extent that CIRs overtake each other and interact they diminish
in number \cite{Burlaga:etal:2003a} at distances beyond $\sim$ 30 AU.
Beyond this distance, the solar rotation period cannot be found anymore
in the magnetic field fluctuations.

Besides the corotating heliospheric field pattern described so far,
the in-situ field observations display a large amount of transient and
turbulent perturbations.
Most prominent are transient magnetic clouds which originate from
coronal mass ejections. They propagate away from the Sun with
speeds ranging from several 100 to 1200 km/s. In most cases, the
speed of the clouds has at least the speed of the ambient solar wind
\cite{Gosling:etal:1994a}.
If this speed exceeds that of the ambient solar wind by more than
the Alfv\'en velocity, a fast mode shock forms on its front side.
At high latitudes the CME clouds were also observed to rapidly expand
due to excess magnetic pressure inside so that both a forward and
reverse shock may form \cite{Gosling:etal:1994b,Forsyth:etal:2001}.
Another consequence of the cloud's expansion is a drastic reduction
of the particle pressure and plasma $\beta$ in its interior so that we
may expect the field inside the cloud to relax towards a force-free
equilibrium (see, however \cite{Riley:etal:2003}).

A common magnetic field signature of a CME passing a space craft
is the rotation of the meridional field components as expected from
the passage of a giant helical flux rope.
As we pointed out in chapter~\ref{sec_corona} the
sense of rotation of the flux rope is determined by the helicity
of the associated magnetic cloud.
In a number of studies, it was indeed possible to identify the origin of the
CMEs back on the Sun's surface and it was found that in these cases the
sense of rotation corresponds to the helicity of the original filament
and hence on the hemisphere, from where it was ejected
\cite{Bothmer:Schwenn:1994,Kumar:Rust:1996,Bothmer:Schwenn:1998}.

A comprehensive 3D model of CME clouds is still under debate. Field
observations of space crafts only yield point measurements which are
often interpreted under the assumption that the perturbation is
convected radially away from the Sun. It is still unclear, how far the
CME cloud extends azimuthally and whether or how it is magnetically
connected to the Sun. Bi-directional electron streams parallel and
anti-parallel to the magnetic field are often observed inside the
cloud. They are usually taken as evidence that the respective field
line is still connected to the corona at both ends. However, also a
complete lack of electron streams has occasionally been seen, which is
interpreted as the presence of a field line rooted with both feet
in the outer heliosphere \cite{Bothmer:etal:1996,Forsyth:Gosling:2001}

A numerical simulation of a CME and its propagation through the
heliosphere is a formidable task. Purely hydrodynamical computations
already show the complicated interaction of a CME with the fast and
slow solar wind regimes in the background and its subsequent
distortion \cite{Odstrcil:Pizzo:1999a,Odstrcil:Pizzo:1999b}.
Including the magnetic field in these computations poses a real
challenge. Even the background balance between magnetic field and
solar wind are difficult to model numerically \cite{Roussev:etal:2003}.
First steps to follow the CME release process and its propagation have
in an MHD model recently been started (e.g.,
\cite{Groth:etal:2000,Riley:etal:2003,Manchester:etal:2004}).
These calculations reproduce the gross features observed. The magnetic
cloud of fast CMEs in these simulations, however, does not seem to evolve
to a near force-free flux rope but is found to be considerably distorted
by the solar wind plasma it runs into.

Another permanent feature of heliospheric observations are turbulent
magnetic field fluctuations. Large amplitude, almost incompressible
transverse Alfv\'en waves are observed whenever space crafts enter
the fast solar wind \cite{Belcher:Davis:1971}. Ulysses
high latitude observations show that the relative transverse field
variance $<\delta \vect{B}_{\perp}^2>/<B^2>$ increases with latitude
from 0.1 almost linearly to $\sim$ 0.3 at 40$^{\circ}$ latitude
\cite{Smith:etal:1995a}. These waves are therefore well in the
non-linear regime.

The resulting radial evolution of the Alfv\'enic turbulence has been
investigated by a number of authors. A comprehensive review was given
by \cite{Tu:Marsch:1995a}.
The wave power spectra measured by the Helios space crafts between
0.3 and 1 AU show a gradual decline from a power law index of -1
towards Kolmogorov's inertial range index -5/3 with increasing
distance from the Sun \cite{Marsch:Tu:1996}.
The interaction of this turbulence with the solar wind has been
modeled by a number of authors.
\cite{Tu:Marsch:1997} assume an initial Alfv\'enic wave
power spectrum emitted from the solar surface and evolve the spectrum
by a non-linear cascade of wave energy to higher frequencies.
The wave energy is absorbed at the high frequency end of the
cascade by proton cyclotron absorption as proposed by
\cite{Axford:McKenzie:1992}.
This absorption provides the necessary acceleration and
heating to drive fast solar wind.
Other authors consider in their models the solar wind velocity shear
and beyond $\sim$ 10 AU energetic pickup ions (see below) as an
additional source for the Alfv\'enic turbulence
\cite{Matthaeus:etal:1999a,SmithCW:etal:2001}. The observed turbulent
magnetic field fluctuation levels could thus be explained out to
$\sim$ 50 AU


\subsection{The distant heliosphere}

Observations of scattered Ly$\alpha$ radiation originating from
the Sun and nearby stars show that our heliosphere is surrounded by
a partially ionized interstellar cloud of mainly hydrogen gas
(e.g., \cite{Axford:1972,Holzer:1989}).
The interaction of this cloud which constitutes the local interstellar medium
(LISM) of our heliosphere with the solar wind has been thoroughly reviewed
\cite{Zank:1999}.
The LISM cloud drifts relative to the heliosphere causing its ionized
component to be deflected at the heliospheric front side boundary while
the cloud's neutral component can freely enter the heliosphere.
Roughly where the solar wind ram pressure equals the LISM
dynamic and thermal pressure, a reverse termination shock stops the
solar wind and decelerates it into a hot, subsonic flow with increased
density.
Inside the region beyond the termination shock, the heliosheath, the solar
plasma is deflected away into the heliotail and stretched out into the
downstream direction of the LISM flow.
The LISM plasma, on the other hand, has to flow around this obstacle and is
swept away around the outer boundary of the heliosheath into the heliotail.
The contact discontinuity between the heliosheath solar wind plasma on the
inside and LISM plasma on the outside is called the heliopause.

Until a 2004, estimates of the stand-off distance of the termination
shock ranged somewhere between about 90 and 200 AU. Voyager 1,
advancing with $\sim$ 3.5 AU/y in about the LISM upstream direction
passed it in December 2004 at about 94 AU while the decreasing solar
wind pressure in the declining phase of the solar cycle made the shock
front move inwards towards the Sun \cite{Burlaga:etal:2005}.
Unfortunately, the failure of the plasma detector on board Voyager 1
makes it somewhat difficult to observe the plasma compression at the
termination shock directly. However a pile-up of the almost azimuthal
magnetic field, tangential to the shock front, could clearly be
detected from the noisy field measurements. The field enhancement was
from about 0.03 nT, still very close to Parker's predictions, on the
upstram side to about 0.1 nT downstram of the termination shock.
A puzzle in the observations is why the frequent flips of the azimuthal
field polarity, reflecting the transitions between the inward and
outward heliospheric field sectors (see Fig.~\ref{Fig:parkerangle}),
completely ceased after the termination shock had been passed.

Before the passage to the termination shock, our knowledge about the
boundaries of the heliosphere were entirely based on HD and MHD models
which extrapolate the observed solar wind quantities downstream and
vary the LISM parameters within meaningful ranges (see
\cite{Zank:1999} and references therein). These models, however,
suffer greatly the uncertainties of unknown LISM parameters.

The only parameter known with sufficient accuracy is the relative speed
of the LISM cloud of $v_{\mathrm{LISM}}$ $\simeq$ 26 km/s and the direction
of this flow.
This value has both been inferred from the Ly$\alpha$ observations
(e.g., \cite{Lallement:1996}) and from in-situ measurements of the
neutral helium, a minor constituent of the cloud which penetrates the
heliosphere \cite{Witte:etal:1996}.
The temperature estimates of $\sim$ 8000 K are already uncertain so that
it is not quite clear whether the LISM flow is sub- or supersonic. In
the latter case, a bow shock is required in front of the heliopause
to decelerate the LISM plasma to subsonic speeds and allow its
deflection around the heliopause. Depending on the H/He composition
the estimate of 8000 K yields a sound speed of about 10 km/s
and hence a supersonic LISM flow.
A possible cosmic ray component in the LISM plasma could, however,
enhance the sound speed \cite{Zank:1999}. 

In particular the strength and orientation of the local interstellar
magnetic field are rather uncertain.
Typical model assumptions for the field strength are of the order of
0.1-0.3 nT \cite{Barnes:2001}.
A typically assumed H$^+$ density of 0.2 cm$^{-3}$ yields
an Alfv\'en velocity of about 10 km/s. Hence both, the sonic and
the Alfv\'enic Mach numbers of the LISM flow are of the order of
unity so that numerical models show substantial differences if
LISM temperature and magnetic field strength and direction is
varied \cite{Ratkiewicz:etal:1998,Ratkiewicz:McKenzie:2003}.
A stronger field compresses the heliopause further inwards, an oblique
field direction may cause a marked asymmetry of the shape of the heliopause.

Which influence the magnetic field from outside the heliosphere may have
has to await more precise estimates of its strength and direction. The
magnetic field which reaches the heliospheric boundary from the inside
is much better known.
Extrapolating Parker's spiral field to large distances suggests that
the magnetic field regains its dominance again in the outer heliosphere.
The field direction eventually becomes almost entirely azimuthal and the
field pressure decreases as $B^2$ $\simeq$ $B_{\phi}^2$ $\simeq$
$r^{-2}$.
This has indeed been observed out to 80 AU with slight deviations
due to solar cycle and solar wind speed variations but else
in fair agreement with Parker's model \cite{Burlaga:etal:2003b}.
Assuming a constant solar wind speed $v_r$, the conservation of the
radial mass flow causes the plasma density to decay as $\simeq$ $r^{-2}$
and an adiabatic solar wind pressure would then decline with $p^{-2\gamma}$
for a polytropic index $\gamma$.
Hence, $\beta$ $\sim$ $p/B^2$ $\sim$ $r^{-2(\gamma-1)}$ should
decrease with distance $r$ for an index $\gamma$ $>$ 1.

However, with decreasing (thermal) solar wind density a new source of
particles becomes important beyond about 15 AU: pick-up ions ionized
by charge exchange from the interstellar neutral hydrogen of the LISM
cloud which penetrates the heliosphere.
In the direction from the Sun towards the heliospheric stagnation point
these hydrogen atoms acquire a speed of $v_r+v_{\mathrm{\small LISM}}$ 
in the frame of the solar wind. Once ionized, they become frozen to
the solar wind and gyrate about the azimuthal magnetic field direction
with an energy of some keV.
The pickup distribution rapidly isotropizes and the ions pass some of their
energy to the thermal solar wind plasma.
This additional source of heating can be observed by a reduced
temperature decrease beyond about 25 AU \cite{Isenberg:etal:2003}.
The result is an increase rather than a decrease of $\beta$ with
further distance from the Sun to values well above unity
at the termination shock \cite{Zank:1999,Barnes:2001}.

Another consequence of the pickup ions is the mass loading of the
solar wind. Momentum conservation should then decelerate the wind,
and Parker's spiral should wind more tightly according to
(\ref{Parkerangle}). There is, however, also a diamagnetization effect
due to the large magnetic moment of the gyrating pickup ions
\cite{Fahr:Scherer:2004}. The fact that $B_{\phi}$ is found
in reasonable agreement with Parker's model and an undecelerated wind
\cite{Burlaga:etal:2003b} suggests that the two effects may compensate.

The negligible role of the solar wind magnetic field in the distant
heliosphere has led many researchers to neglect it altogether in their
simulations of the heliosphere, with only few exceptions
\cite{Washimi:Tanaka:1996,Linde:etal:1998}.
Upstream of the termination shock the solar wind is decelerated and
heated to be subsonic, the azimuthal magnetic field $B_{\phi}$ piles up,
as is observed by Voyager 1, and is, especially near the stagnation point,
compressed towards the heliopause (``Axford-Cranfield effect'').
This may locally lead to a considerable field strength, however, the
temporal variations of the solar wind and the polarization reversals
with the solar cycle probably destroy this effect to a large degree
\cite{Barnes:2001}.

In addition, the termination shock location is my no means
stationary. Voyager observations suggest that the shock front was
moving inward at the time of the passage. It was shown in model
simulations that the front can move within a solar cycle by as much as
10 AU \cite{WhangYC:etal:1995,Zank:Mueller:2003}.
The remains of CIRs and CMEs will additionally shake the heliosheath
continuously and reconnection of the heliosheath magnetic field with
the reversed polarity solar wind field or with the LISM field may lead to
a further erosion of the magnetic field in the heliosheath.


\section{Outlook}
\label{sec_outlook}


The magnetic field plays a role in almost all parts of solar physics, so
that the study of the Sun's magnetism has wide-ranging
implications. Some of the main results concerning the solar magnetic
field have been presented, or at least mentioned in this review. Many
more have become victims of the limited space and the aim of the authors
to present an overview spiced with a few examples, rather than a
complete compilation. In any case, however important the compilation or
presentation of past achievements may be, for the scientist working at
the coal face of solar magnetic field research, the progress we hope to
make in the near and mid-term future is by far the more interesting. Such
progress is driven not so much by the open, unanswered questions, of
which there are many, including some very basic ones. Rather, the main
driver is the availability of new tools.  These include new space
instruments, ground-based telescopes and associated post-focus
instruments, but also new codes, (numerical) techniques and computers.

In the following we point out some of the requirements for making progress
and list some of the new tools that have either recently become
available, or are planned for the near to mid-term future. We follow
roughly the same order as the sections of this
review, starting with the large-scale structure of the magnetic field and
ending with the magnetic field in the
heliosphere.

The risk of such an endeavour is that unforeseen developments
can change the course of the field very significantly, making even a carefully
written outlook quickly obsolete.

\subsection{Large scale structure of the magnetic field}

The observation of global magnetic field patterns is far less demanding than
the detailed study of individual features in that high spatial resolution is
not a prerequisit. This is offset, however, by the need for long
uninterrupted time series of as many useful quantities as possible.

A point that cannot be overstressed is the need for coherent
synoptic full-disk data sets of constant high quality. Only with the help of
such data can the dynamo be constrained with sufficient accuracy. Such data
often appear routine and it may not seem worth the effort to continue to
gather them over decades and ideally even centuries. However, such data
invariably pay rich scientific dividents in the end. For example, the sunspot
numbers data set, one of the longest running
scientific time series, is one of the most studied time series in any field
of science and even today keeps revealing fresh insights when analysed with
new techniques. In fact, it has become almost standard practice to apply any
newly developed technique for time series analysis to the sunspot number
record.

A negative example are the sunspot area and position measurements
carried out by the Royal Greenwich Observatory. These are important not
just for constraining the solar dynamo, but also for the reconstruction
of solar irradiance. Unfortunately, contributions to this homogeneous
data set ended in 1976, just 2 years before the radiometer on the
NIMBUS-7 satellite revealed the systematic variability of the solar
irradiance. The absence of this overlap still poses problems to
researchers attempting to extend the irradiance record further back in
time.

On a shorter time scale, the regular magnetograms provided by
ground-based observatories, such as the Spectropolarimeter (SPM), on
Kitt Peak and from space by the Michelson Doppler Imager (MDI) on the
Solar and Heliospheric Observatory (SOHO) have made an immense
difference to studies of the large scale magnetic field. However, these
data still suffer from some shortcomings. One is the lack of vector
magnetic field measurements since all these instruments only record the
longitudinal magnetic field component. Another is the insufficient
accuracy and often the relatively low spatial resolution of the
data. Finally, the coverage of the solar poles is extremely poor.

Projects that will address these shortcomings are in different stages of
development. The SOLIS instrument on Kitt Peak, which has recently
started operation, will provide regular vector magnetograms of the full
solar disk in addition to other interesting observables for many
years. The HMI instrument on the Solar Dynamics Observatory (SDO) has a
similar aim, but is planned to provide data at a higher resolution and a
high cadence (but possibly with lower sensitivity). Not just the regular
coverage and the magnetic vector provided by these instruments is
important, but also the enhanced polarimetric sensitivity, needed to
reliably determine the transverse field components. Such an enhanced
sensitivity also allows weak fields (e.g. internetwork fields) to be
followed, which are only imperfectly rendered by current synoptic
magnetograms.  Finally, the solar poles will be studied by the Visible
light Imager and Magnetograph (VIM) on the Solar Orbiter mission of ESA.

\subsection{The solar interior}

Most of the work dealing with the magnetic field in the solar interior
has been theoretical in nature and the basic unanswered question is how
the solar dynamo works, i.e., how the Sun's magnetic field is
generated. Although dynamo theory has made significant progress over the
last decades, some basic questions remain unanswered.

There is no generally accepted model of the solar dynamo.  The relative
importance of dynamo action in the overshoot layer/tachocline and a more
distributed dynamo in the bulk of the convective zone remains unclear,
particularly so concerning the origin of the turbulent magnetic field in
the near-surface layers. While large-scale numerical simulations so far
has not shown solar-like results, there is hope that the rapid progress
in computational power will allow us to reach a more relevant parameter
regime in the next decade or two. Reynolds numbers realistic for the
solar plasma, however, are and will for long be beyond computational
capabilities.

The biggest drawback for dynamo theory, however, is the lack of
observational information about the structure and evolution of the
magnetic field in the solar interior.


Another point that has so far not been adequately addressed in the context
of dynamo theory is the description of the magnetic flux concentrations
with superequipartion fields in the Sun's interior. The interaction of these
flux tubes with sub-surface flows is expected to be markedly different
from that of a more diffuse fields.



The possibility to measure magnetic fields reliably in the solar interior
would be revolutionary.
Local helioseismology has this capability in principle,
but so far it has not been possible to unambiguously distinguish
between magnetic effects on $p$-modes and other influences, such as
thermal inhomogeneities.
The theoretical work needed is demanding and it is not possible to
predict  if and when sufficiently reliable results will be
obtained to run such applications.
In addition, the observational input for the inversion calculations
may also turn out to be difficult to supply: to discriminate between
different effects on wave propagation, data with a low signal-to-noise
ratio obtained after long time integration may be required.

\subsection{The solar photosphere: magnetic elements and sunspots}

Most evident and pressing is the need to study the magnetic field at a higher
spatial resolution than
currently achievable. Even the best images and measurements available today
show structure at the resolution limit.
Furthermore, 3-D numerical MHD simulations of plage or quiet Sun also
predict considerable magnetic fine
structure below the spatial resolution of current observations. Higher spatial
resolution observational data are
needed in order to test whether structures visible in the simulations are also
present on the Sun. A clever choice of spectral diagnostics can overcome a
part of the disadvantages of limited spatial resolution, but cannot replace
high spatial resolution data.

Currently, the highest spatial resolution photospheric magnetic data are
being provided by the Swedish Solar Telescope (SST) on the island of La
Palma, thanks to a careful design and the use of adaptive optics. In the
coming years, a number of new instruments will provide competition,
starting in 2006 with the Solar-B spacecraft. Although the theoretical
resolution of Solar-B is only half that of the SST, the constant
conditions provided by space are a distinct advantage for observations
aiming at, e.g., the evolution of the magnetic field. The Sunrise
balloon-borne observatory is expected to combine the theoretical
resolution of the SST (an even higher resolution will be achieved in the
UV) with the near absence of seeing enjoyed in the stratosphere.  On the
ground, the Gregor telescope on the island of Tenerife, the New Solar
Telescope (NST) at Big Bear Observatory, and in the more distant future
the Advanced Technology Solar Telescope (ATST) should push the spatial
resolution limit even further.

We expect that progress will not come from higher spatial resolution alone.
For the study of weak magnetic fields (features with low magnetic flux), a
higher polarimetric accuracy (lower noise level) is just as important.
In the Sun's photosphere the magnetic field does not only occur in the form
of magnetic flux concentrations, but also in a more irregular or ``turbulent''
state, which is a subject of intense current
studies
Additional work is needed to reliably determine the amount of flux in this
state of the magnetic field, its structure and distribution, evolution and
origin. Since the field is
weak, a high polarimetric sensitivity, as provided, e.g., by the Z\"urich
Imaging Polarimeter (ZIMPOL) is needed, in particular in combination with
high spatial resolution.

In addition, observations in different spectral bands
(or spectral lines) provide the possibility of determining the height
dependence of magnetic parameters, particularly in combination with
state-of-the-art inversion techniques. The determination of the
3-D structure of the magnetic field obtained by combining high spatial
resolution maps with the vertical stratification deduced from spectral
information will, we believe, become a regularly used observational tool
in the future.

Spectacular results are also expected from novel techniques and data, such as
those obtained with the VIM instrument proposed Solar Orbiter, which will
allow stereoscopic measurements of magnetic features when combined with data
obtained in Earth orbit or from the ground.

Theory has taken immense strides to go beyond simple
models aiming mainly to identify physical processes towards realistic
descriptions of the complex magnetic structures found on the Sun.
3-D radiative MHD simulations of quiet Sun and plage regions are
showing a remarkable similarity
with high resolution observations. Detailed comparisons suggest a need for
denser grids (which is not surprising, given that the Reynolds numbers
achieved by the simulations are still orders of magnitude smaller than the
solar values). A denser grid requires larger computational resources, which
are one limiting factor for progress in this field.

Just as
exciting, e.g. for the study of sunspots, are the possibilities opened up
by considering larger domains and running longer time series. In contrast to
magnetic elements, models of sunspots have so far been severely limited.
Models describing complete sunspots are restricted to assuming axial
symmetry and to parameterising many of
the key physical processes acting in them, while models that treat such
processes explicitely are localised to a small region inside a sunspot (e.g.,
a piece of the umbra or penumbra).
Therefore, one of the main aims of an MHD simulation covering a large domain
is to simulate a complete sunspot.
The simulation of small-scale magnetic elements would also greatly benefit
from a larger computational domain, since it would allow their interaction
with meso- and supergranulation to be followed, thus bringing the study of
solar magnetic fine structure a step closer to the study of the global
evolution of the magnetic field.

\subsection{The upper atmosphere: chromosphere and corona}

The very structure of the solar chromosphere is the subject of intense
debate and this uncertainty also extends to the magnetic field. The main
bone of contention is whether or not the quiet chromosphere harbours
low-lying magnetic canopies (a uniquely chromospheric phenomenon), such
as those seen in active regions.  Observations of chromospheric fields
now possible at the Canary Island Observatories will certainly provide
exciting new results. In order to resolve this question, a high
signal-to-noise ratio is expected to play a stronger role than the
highest spatial resolution possible.

None of the planned space instrumentation foreseen for the near future
has a chromospheric polarimetry channel in it. However, there are at
least two new ground based instruments, TIP2 (the upgrade of the
Tenerife Infrared Polarimeter) at the Vacuum Tower Telescope on
Tenerife and SPINOR at the Vacuum Tower on Sacramento Peak, that are
providing such data.

One reason why chromospheric structure still is rather enigmatic has to
do with the difficulties posed by the chromosphere to modellers. A
plasma $\beta$ on the order of unity, the presence of shock waves and
supersonic flows, the need for partial-redistribution NLTE (for typical
chromospheric lines such as the H and K lines of Ca II), ionisation and
recombination time scales larger than dynamical time scales, etc., all
combine to make the life of the modeller difficult. However, there is no
way around extending the successful 3-D simulations from the photosphere
to the chromosphere. First attempts still neglect some of the physics,
but this is one direction in which there is a strong need to go. As far
as the magnetic field is concerned, it is already instructive in a first
step to extrapolate the field from the photosphere (either from
simulations or high spatial resolution observations) into chromospheric
layers.

The main hinderance to more rapid progress in coronal physics is that
the structure, dynamics and thermodynamics of the corona are dominated
by the magnetic field, which, however, can only be measured in a rather
rudimentary way in these layers. Up to now, with the exception of radio
measurements (see below), coronal studies have relied either on
extrapolations of the magnetic field from the photosphere, or (more
commonly) on the use of proxies such as the spatial distribution of
intensity of transition region lines, or loops visible in coronal
radiation.

The premier technique for measuring the coronal magnetic field has so
far been radio observations, which can provide maps of the magnetic
field strength, but do not in general allow the full magnetic vector to
be measured. This is unfortunate, since in the corona the magnetic field
becomes very inhomogeneous in direction, but relatively homogeneous in
strength. The configuration of the field determines whether such
important processes as magnetic reconnection can take place or not.

Maps of the magnetic field strength obtained from radio data have in the
past often suffered from either low spatial and/or temporal resolution,
or a sparse coverage of the Fourier plane (in the case of
interferometrically obtained data).  The introduction of a dedicated
array (FASR) should help to significantly improve coronal magnetic field
measurements with radio techniques.

At the same time, new techniques for coronal magnetic field measurements
are currently being developed and studied, which might overcome some of
the shortcomings of radio measurements. They include longitudinal
magnetograms made in (coronagraphically imaged) infrared coronal lines,
the utilization of the Hanle effect in EUV lines (for a future
application in space), and using the Zeeman effect in the HeI 10830 \AA\
triplet for vector magnetic field measurements near the coronal base.
Each technique has its advantages and disadvantages and none gives a
complete picture of the coronal field. While the Zeeman effect in IR
coronal lines may remain limited to measuring the longitudinal field
owing to the limited signal strength, the HeI 10830 \AA\ triplet does
not sample the hot coronal gas directly, and the Hanle effect in EUV
lines has not been studied sufficiently to determine its full promise or
the pitfalls.

One generic reason why no single technique will be able to provide
anything near a complete picture of the coronal field is that the
coronal gas is optically thin.
However, a tomographic reconstruction of the coronal magnetic field is
probably possible if longitidinal Zeeman and transverse Hanle
observations are combined -- another field where a substantial theoretical
effort is needed.

A further complication is that gas at many different temperatures is
intermingled. Therefore, measurements in a diagnostic line sensitive to gas
within a certain temperature range show only the part of the field
embedded in that gas. Neighbouring field lines may host gas at a very
different temperature that may be accessible only at another
wavelength and/or with another technique. It is likely that, on the
medium to long term, different diagnostics and techniques will have to
be combined to get a proper view of the coronal field. Extrapolations
from photospheric magnetic field measurements will remain an integral
part of the arsenal of techniques for studying coronal fields.
However, in future such extrapolations and proxies
will be increasingly supplemented, augmented and tested by direct
measurements. For example, the extrapolated field can give a hint of
the height to which a certain measurement refers.

Highly dynamic phaenomena in the corona, like eruptions and coronal
mass ejections, which are very probably triggered by the coronal
magnetic field, have attracted great interest in the past and our
understanding of these phaenomena has improved enormously in the
recent decade, both by observations with, e.g., the Yohkoh, SoHO and
TRACE instruments and by improved MHD modelling. Predictions of
individual events are still a challenge though. Again, a problem is
that our knowledge of the magnetic field at the site of the eruption
is still too limited to discern the sequence of physical processes of
an eruption in detail.

\subsection{The heliosphere}

Our knowlegde of the magnetic field in the heliosphere is probably
better than in many other regions of our solar system. From the very
beginning, the importance of the magnetic field was recognized and
almost all space missions were equipped with magnetometers. In-situ
measurements have been complemented with sophisticated MHD simulations,
so that the basic understanding of the magnetic field in the region
between 0.3 and 5 AU is quite advanced.

Owing to the vanishing divergence of the magnetic field, an
extrapolation of the heliospheric field closer to the Sun is possible to
some extent. Yet, the region inward of 0.3 AU, where the corona fades
out into the solar wind, is little understood.  With MHD models of the
corona and inner heliosphere we can roughly map the solar surface
magnetic field to the heliosphere and relate in situ satellite
observations with their sources on the Sun. This magnetic mapping has so
far assumed the field to be stationary and dynamical aspects of the
connection between the coronal and heliospheric magnetic field are not
clear yet.  It is in this region below 0.3 AU where the solar wind is
accelerated to a super-Alfv\'enic speed and CMEs sometimes to speeds
well above 1000 km/s.  The magnetic field almost certainly plays a key
role in these acceleration processes but a widely accepted, physically
sound explanation is still lacking.  The Solar Orbiter mission will
hopefully shed some light onto this region.

The heliosphere is highly dynamic. CMEs and interaction regions are
complex phaenomena which could only partially be disclosed by in-situ
observations. Here, MHD simulations have greatly improved our
understanding of their three-dimensional plasma structure and magnetic
field and how they are shaped in interaction with the background solar
wind while they propagate outwards through the heliosphere.  The STEREO
mission will, for the first time, attempt to follow CMEs by means of
optical obervations out to 1 AU.  Even though only the plasma density
enhancements of the CME cloud will be visible, we will also learn much
about the magnetic field configuration of the cloud.  Certainly, these
observations will give new constraints and inspiration to refined MHD
simulations of the propagation process.

Until recently, our knowledge about the distant heliosphere and its
interaction with the local interstellar medium (LISM) was almost
entirely due to MHD models which use our present knowledge about the
inner heliosphere as boundary values. This situation has drastically
changed since the Voyager 1 probe has crossed the termination shock.
The new and unique in-situ observations of the heliosheath plasma will
considerable constrain future MHD models. As the LISM magnetic field
plays only a minor role in these models, it will have to be seen, how
much the interstellar field value and direction can be constrained by
these observations.
New missions like the Heliopause Probe especially devoted to explore
the outer heliospheric boundaries are considered and may yield more
precise observations than those from the Voyager spacecrafts.
But a lot of patience is required here.
It will take at least 25 years until the next generation of space crafts
will reach the termination shock. 



\section*{References}

\def\solphys{Sol. Phys.}
\def\aap{Astron. Astrophys.}
\def\aapr{Astron. Astrophys. Rev.}
\def\apj{Astrophys. J.}
\def\nat{Nature}
\def\mnras{Monthly Not. Royal Astron. Soc.}
\def\amp{Astron.. Nachrichten}
\def\apjl{Atrophys. J. Lett.}
\def\pasj{Publ. Astron. Soc. Japan}
\def\araa{Ann. Review of Astronomy and Astrophysics}
\def\baas{Bulletin of the American Astronomical Society}
\def\apjs{Astrophysical Journal and Supplement Series}
\def\jgr{J. Geophys. Res.}

\bibliographystyle{unsrt}
\bibliography{solanki_etal.bbl}

\end{document}